\documentclass[pdflatex,sn-mathphys-num]{sn-jnl}

\usepackage{graphicx}%
\usepackage{multirow}%
\usepackage{amsmath,amssymb,amsfonts}%
\usepackage{amsthm}%
\usepackage{mathrsfs}%
\usepackage[title]{appendix}%
\usepackage{textcomp}%
\usepackage{manyfoot}%
\usepackage{booktabs}%
\usepackage{soul} 
\usepackage{algorithm}%
\usepackage{algorithmicx}%
\usepackage{algpseudocode}%
\usepackage{listings}%
\usepackage{siunitx}
\usepackage{multirow}
\usepackage{adjustbox}
\usepackage{graphicx}
\usepackage{natbib}
\usepackage{lscape} 
\usepackage[table,xcdraw]{xcolor}
\usepackage{changepage}
\usepackage{color}
\usepackage{tabularray}
\usepackage{colortbl}
\usepackage{rotating}
\usepackage{pdflscape}
\usepackage{array}
\usepackage[normalem]{ulem}

\definecolor{Gallery}{rgb}{0.937,0.937,0.937}
\definecolor{Alto}{rgb}{0.85,0.85,0.85}
\definecolor{Pippin}{rgb}{1,0.87,0.87}
\definecolor{FairPink}{rgb}{1,0.909,0.905}
\definecolor{VividTangerine}{rgb}{1,0.549,0.552}
\definecolor{Cosmos}{rgb}{1,0.835,0.835}
\definecolor{Sundown}{rgb}{1,0.694,0.694}
\definecolor{Pink}{rgb}{1,0.8,0.803}
\definecolor{YourPink}{rgb}{1,0.764,0.764}
\definecolor{Chablis}{rgb}{1,0.941,0.941}
\definecolor{RoseWhite}{rgb}{1,0.976,0.976}
\definecolor{MonaLisa}{rgb}{1,0.623,0.623}
\definecolor{CornflowerLilac}{rgb}{1,0.662,0.654}
\definecolor{YourPink1}{rgb}{1,0.729,0.729}
\definecolor{Bittersweet}{rgb}{1,0.443,0.443}
\definecolor{MonaLisa1}{rgb}{1,0.584,0.588}
\definecolor{VividTangerine1}{rgb}{1,0.513,0.517}
\definecolor{Nobel}{rgb}{0.717,0.717,0.717}
\definecolor{VistaBlue}{rgb}{0.592,0.835,0.717}
\definecolor{SilverTree}{rgb}{0.341,0.733,0.541}
\definecolor{FringyFlower}{rgb}{0.67,0.866,0.772}
\definecolor{SilverTree1}{rgb}{0.4,0.756,0.58}
\definecolor{FringyFlower1}{rgb}{0.666,0.866,0.768}
\definecolor{FringyFlower2}{rgb}{0.65,0.858,0.756}

\theoremstyle{thmstyleone}%
\theoremstyle{thmstyletwo}%

\theoremstyle{thmstylethree}%

\begin{document}

\title[Article Title]{Analytical Study on the Exposedness of Potential Positions for External Human-Machine Interfaces}


\author[1,2]{\fnm{Jose} \sur{Gonzalez-Belmonte}}\email{josegonz@umich.com}

\author*[1]{\fnm{Jaerock} \sur{Kwon}}\email{jrkwon@umich.edu}

\affil*[1]{\orgdiv{Electrical and Computer Engineering}, \orgname{University of Michigan-Dearborn}, \orgaddress{\street{4901 Evergreen Rd}, \city{Dearborn}, \postcode{48128}, \state{Michigan}, \country{United States of America}}}

\affil[2]{\orgdiv{Math and Computer Science}, \orgname{Lawrence Technological University}, \orgaddress{\street{21000 W Ten Mile Rd}, \city{Southfield}, \postcode{48076}, \state{Michigan}, \country{United States of America}}}


\abstract{As we move towards a future of autonomous vehicles, questions regarding their method of communication have arisen. One of the common questions concerns the placement of the signaling used to communicate with pedestrians and road users, but few works have been fully dedicated to the matter. This paper uses a simulation made in the Unity game engine to record the fifteen different vehicles under fifty-seven different scenarios each for the first time, in order to find how often its forward-facing exterior surfaces can be seen by a pedestrian on the sidewalk. Variables include the vehicle type, position, number of vehicles on the road, camera position and direction, as well as its minimum and maximum distance from the recorded points. It was concluded that the areas of the vehicle most often seen by pedestrians on the sidewalk attempting to cross the road were the wheels, front fenders, and headlights. Based on these results, a suggestion is made to implement displays on at least two of the following regions: windshield, front fenders, side mirrors. These findings are valuable in the future design of signaling for autonomous vehicles in order to ensure pedestrians are able to see them on approaching vehicles. The software used provides a platform for similar works in the future to be conducted.}




\maketitle

\section{Introduction}\label{Introduction}

In the field of Autonomous Vehicles (AVs), external Human-Machine Interfaces (eHMIs) refer to methods of communication, such as screens or lights, that an AV can use to communicate with other road users. With traditional vehicles, drivers are able to use body language to communicate implicitly with pedestrians, but the adoption of AVs introduces a loss of implicit social cues from drivers to handle ambiguous road interactions (\cite{de_winter_external_2022}). eHMIs were introduced as a solution to this problem, aiming to eliminate ambiguity and the necessity of inter-personal social cues by letting the AV outwardly express its intent to pedestrians explicitly.

Previous studies have addressed the effectiveness of eHMIs (\cite{de_clercq_external_2019, rodriguez_palmeiro_interaction_2018, alhawiti_effectiveness_2024}) and the symbols they should use (\cite{de_winter_external_2022}), but few studies have been focused on determining the ideal position for these displays. Those available are concerned primarily with the behavior and perception of the pedestrian (\cite{guo_video-based_2022, de_winter_external_2022, dey_gaze_2019}), but few have attempted to optimize the position of the display to maximize when it is visible, and the ones that do are limited in the data collected and represented (\cite{troel-madec_ehmi_2019}). As a result, there exists a gap in the knowledge of eHMI placement, including information on where it will be most seen by pedestrians, how distance may affect its perception, and how the type of vehicle may limit these choices. As observers must be able to perceive an eHMI before making a decision based on its information, this gap is must be filled in order to set standards for the technology.

The present study aims to contribute to answering these questions with a data-capturing method that measures the exposedness of different vehicle parts, using the Merriam-Webster definition of exposed as "open to view" and "not shielded" \cite{noauthor_definition_2026}. The intent is to develop an open-source software capable of simulating the possible positions for a vehicle and produce a heatmap indicating what parts and areas of the vehicle were visible in the highest amount of scenarios, then utilize this tool with a variety of vehicle types and layout permutations in order to determine general recommendations for eHMI placement based on the results. This is not a study on pedestrian behavior or response, but one that aims to collect information on where an eHMI can be placed to communicate with pedestrians uninterruptedly. 

\section{Related Work}\label{Related Work}

The effectiveness of eHMIs has been thoroughly investigated through experiments using virtual reality (\cite{de_clercq_external_2019}), video-based eye tracking techniques (\cite{guo_video-based_2022, dey_gaze_2019}), the Wizard of Oz method (\cite{rodriguez_palmeiro_interaction_2018}), and outdoor experiments with real autonomous vehicles (\cite{alhawiti_effectiveness_2024}). In each of these examples the presence of an eHMI proved to increase response time and feelings of safety on pedestrians. However each these experiments feature vastly different proposals of eHMIs, often varying in their positioning within the same study. Research on the positioning of eHMI is less common than those focused on their effectiveness. 

Dey et al. \cite{dey_gaze_2019} had participants stand on a sidewalk wearing eye tracking glasses, and indicate their willingness to cross as four vehicles approached the pedestrian crossing in front of them. The experiment found that pedestrians lose their willingness to cross the road when the vehicle is at a distance of 40\si{m} from them, and that the part of the vehicle they focus on varies as a function of the vehicle's distance, moving from the vehicle bumper to the windshield as it approaches. This suggests that a distributed approach to eHMI design is required, as pedestrian attention shifts thorough their interaction with an autonomous vehicle.

de Winter et al. \cite{de_winter_how_2020} conducted an experiment where participants walk around a busy parking lot with an eye tracking device, and found that most participants focus on the general parts of a vehicle that are affected by their movement, such as the vehicle's front, back, and wheels. A significant portion of pedestrians also looked at the windshield of the vehicle, in accordance with \cite{dey_gaze_2019}. This study also  concluded that eHMIs must be visible from all angles due to the varied conditions in which pedestrians encounter them.  

Guo et al. \cite{guo_video-based_2022} used an eye tracking device as well. In this experiment participants watched a video of a vehicle approaching from the right as their gaze was being recorded, in order to determine what parts of the vehicle they paid attention to. The vehicles illustrated in this experiment either employed no eHMI, or showed one of six different eHMI designs (flashing text, sweeping pedestrian icon, sweeping arrow, flashing smiley, flashing light band, and sweeping light bar) located in one of three different positions (grill, windshield, and roof). The results determined the windshield to be the locations of the vehicle with the longest visual attention from participants, followed closely by the grill, but found little difference in the efficiency of an eHMI based on said location.

A similar experiment to the one done for the present paper is that of Troel-Madec et al. \cite{troel-madec_ehmi_2019}, where a 3D simulation was created to measure what parts of the vehicle were visible most of the time to pedestrians looking at an incoming vehicle, and hence which areas were best suited for the placement of eHMIs. For this, the team placed three 3D models of a sedan vehicle in a virtual space, forming an unmoving line on the side closest to the sidewalk and facing towards the camera. Each vehicle was then assigned a different primary color, with differences in the shading based on what part of the vehicle they were in: front, side, and back. A virtual camera then took a series photos from a predefined set of points on the sidewalk at five different heights, in order to simulate how children, teenagers, and adults may perceive the situation. The pixels on the resulting images were analyzed, using the solid color of the vehicles to find what parts of the each vehicle were visible the most. The results pointed to the side of the vehicle to be the most consistently visible part of the vehicle, and the team used these results to develop a prototype for an eHMI that was positioned on the front doors and windshield rails.

These studies overall conclude that the specific location of an eHMI matters less than the fact said eHMI is visible to the observer (\cite{de_clercq_external_2019}), and that pedestrians tend to fixate on parts of the vehicle such as the bumper, the windshield, the wheels, and the general back and front, while the sides of the vehicle seem to be most visible in most scenarios. While these results are promising, they lack granularity and are always done with a single type of vehicle, often coming from the same direction. 

Similarly to the work of \cite{troel-madec_ehmi_2019}, this study aims to find the forward-facing exterior element most visible to most pedestrians in most situations, while widening the scope to include a variety of vehicle types and scenarios not tested in the previous literature.

\section{Methods}\label{Methodology}
The following requirements must be met by the results in order to provide a specific visible area for the placement of eHMIs:
\begin{enumerate}
    \item The raw data must be granular, not just pointing at general parts of the car, but showing how visible specific points on the surface of the vehicle are. 
    \item Since pedestrians may find their view of a vehicle blocked by another (e.g. when vehicles are in a line), the software must be able to calculate all possible scenarios where this may happen and simulate them. 
    \item Multiple vehicles of different sizes must be simulated, as different vehicle types may present different challenges and limitations for eHMI design.
    \item Vehicles may come from four possible directions, respective to the sidewalk: two perpendicular to it, and two parallel to it. The simulation must present cases where this can be explored.
\end{enumerate}

\begin{figure}[h]
    \centering
    \includegraphics[width=3in,height=3in,clip,keepaspectratio]{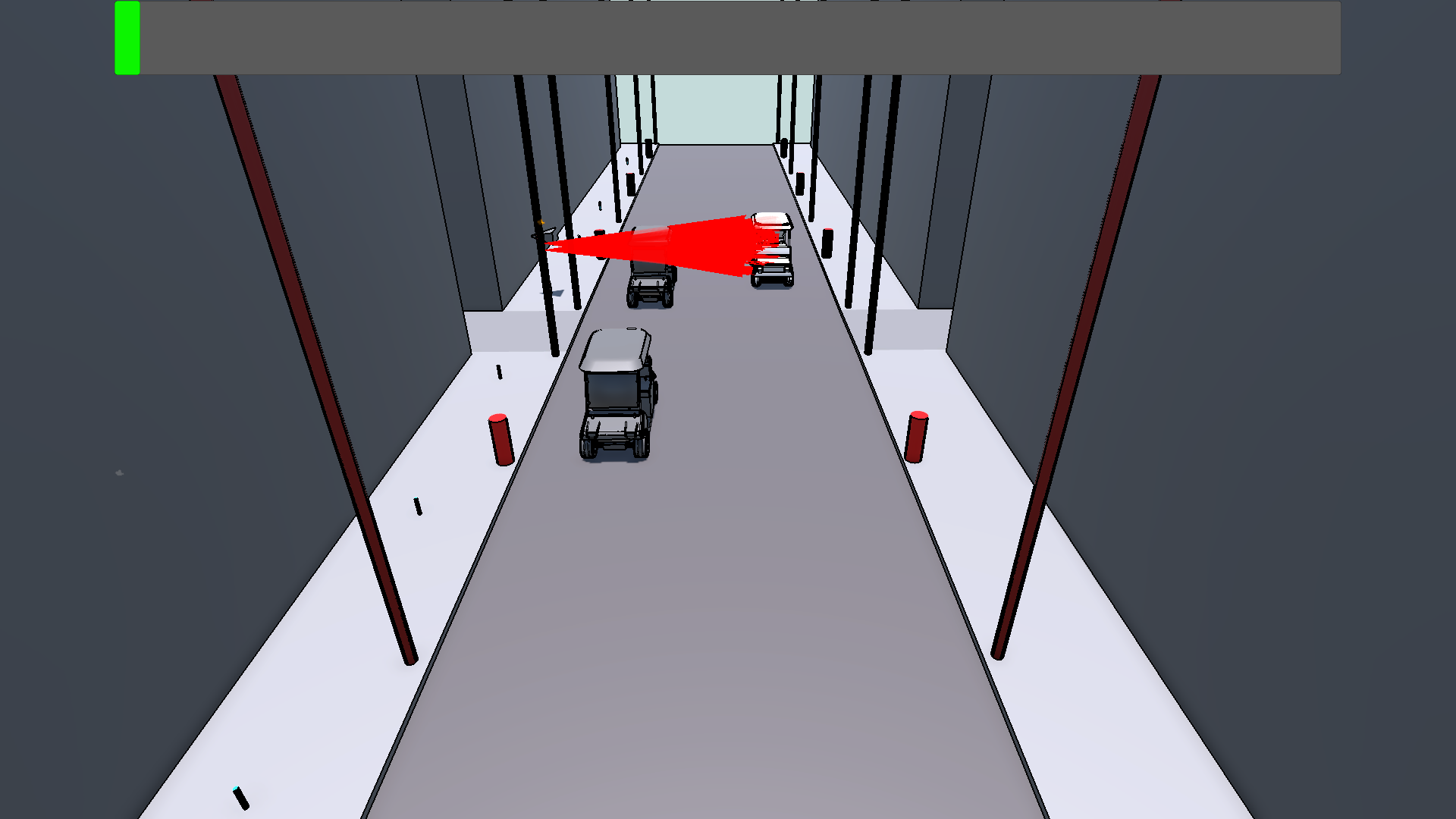}
    \caption{Screenshot of the simulation running. Red lines indicate the trajectory of the ray-casts that have hit the \textit{Target Vehicle}.}
    \label{fig:run_sample}
\end{figure}

A simulation software was created to achieve these goals using the Unity game engine. The execution of a simulation can be divided into a hierarchy of five levels, each made up of the ones after:
\begin{enumerate}
    \item A \textit{Simulation Run} is the highest level, where the user provides the program with the parameters that it will use during the simulation. These include:
        \begin{itemize}
            \item The set of positions on the virtual road where camera may be positioned (\textit{Camera Positions}) and the directions that the camera may face at each (\textit{Camera Directions}).
            \item The set of positions on the virtual road that the \textit{Target Vehicle} may occupy. This is the vehicle that is being measured, and the only one to record whenever it is visible to the camera.
            \item The type of vehicle used for the \textit{Target Vehicle}, each corresponding to a pre-selected 3D model implemented into the program.
            \item The set of positions that the \textit{Blocking Vehicles} may occupy. These are other vehicles placed on the scene that do not record data, and only exist to block the view of the \textit{Target Vehicle} from the camera as illustrated in Fig.\ref{fig:filler_vehicle_diagram}.
            \begin{figure}[h]
            \centering
                \includegraphics[width=4in,height=4in,clip,keepaspectratio]{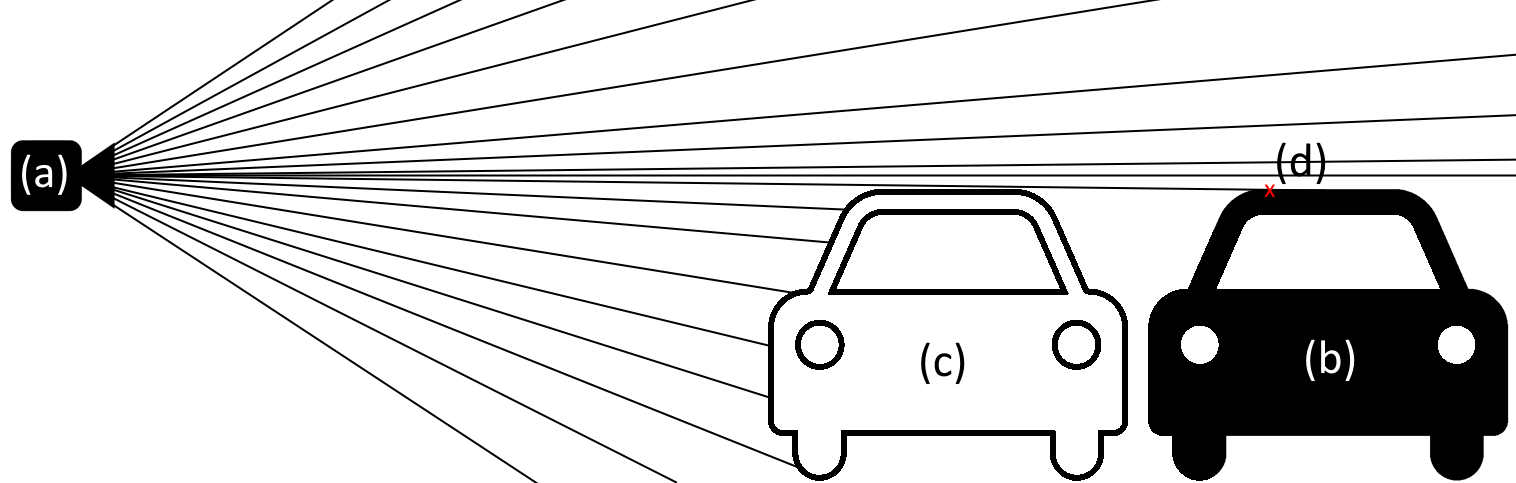}
                \caption{Diagram showing the camera (a) attempting to capture points on the surface of the \textit{Target Vehicle} (b) but being blocked by a \textit{Blocking Vehicle} (c). Only a single point (d, marked with an X) can be seen being recorded.}
                \label{fig:filler_vehicle_diagram}
            \end{figure}
            \item The type of vehicle used for the \textit{Blocking Vehicles}.
            \item The number of vehicles in the scenario, including the \textit{Target Vehicle} and \textit{Blocking Vehicles}. There will always be one and only one \textit{Target Vehicle} in any given simulation. There may never be more \textit{Blocking Vehicles} in the road than there are positions for it. Entering '1' will make it so no \textit{Blocking Vehicles} are present in the \textit{Simulation Run}.
            \item The minimum and maximum distance that a point on the \textit{Target Vehicle} must be from the camera to be recorded.
        \end{itemize}
        Based on these parameters, the software will instantiate a \textit{Target Vehicle} in the pre-made environment, calculate all possible permutations of the \textit{Target Vehicle} and the \textit{Blocking Vehicles} in their possible positions, then cycle through each of these \textit{Scenarios} using the specified settings.
    \item A \textit{Scenario} represents a combination of different placements on the road for the \textit{Target Vehicle} and other \textit{Blocking Vehicles} on the road. Since we want to find the points of the \textit{Target Vehicle} that are visible the most often, we want to cycle through all possible combinations of \textit{Blocking Vehicle} placements to simulate the ways in which other vehicles on the road may block the view of the pedestrian.
    \item For each \textit{Scenario}, the camera will cycle through each of the \textit{Camera Positions}. For this paper, these all correspond to positions where a pedestrian may be located on the left sidewalk, although the program has capability to select positions on the road as well.
    \item For each \textit{Camera Direction} available at each \textit{Camera Position}, the camera executes a series of \textit{Data Captures} at five different heights from the floor: 1\si{m}, 1.2\si{m}, 1.4\si{m}, 1.6\si{m}, and 1.8\si{m}. 
    \item To perform a \textit{Data Capture}, the camera simulates a series of ray-casts projected outwards from the camera in a grid pattern in order to replicate where it is observing the \textit{Target Vehicle}. Each ray-cast starts at the minimum distance away from the camera, and ends at its maximum distance. Ray-casting allows us to find the intersection point between an origin point and a destination point, as well as the details of the first collider hit this way, such as what object it belongs to or what components are attached to that game object (\cite{aversa_unity_2022}). An example of this process can be seen in Fig. \ref{fig:run_sample}. Ray-casting does not ignore transparent surfaces, it treats them as if they were solid matter.
\end{enumerate}

\begin{figure}[t]
        \centering
        \includegraphics[width=1.8in,keepaspectratio]{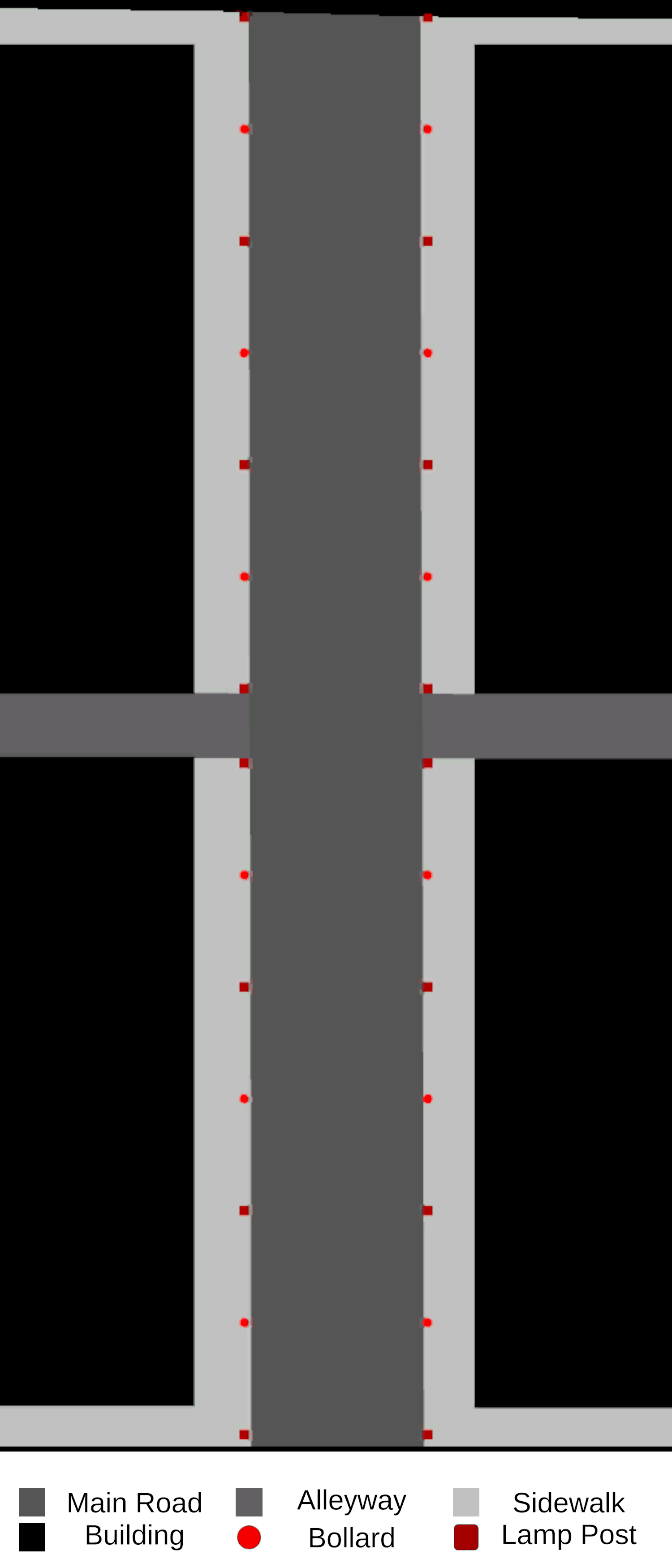}
        \caption{Top-Down View of the Virtual Street Layout.}
        \label{fig:empty_layout}
\end{figure}

The virtual environment for this simulation, as seen on Fig. \ref{fig:empty_layout}, consists of a two-way street with alleyways intersecting it from the left and the right. The main road in this virtual intersection is \num{6.94}\si{m} wide and \num{57.719}\si{m} long; the alleyways are \num{2.604}\si{m} wide; both sidewalks are \num{2.216}\si{m} wide and \num{0.102}\si{m} above the road; the four cuboids that stand for buildings in this model are each \num{21.5}\si{m} tall, and cover the space not occupied by the alleyways or the road. 

The virtual sidewalk was also populated by static objects that serve as obstacles in the observer's view. There are two kinds of obstacles, each placed in alternate order on the sidewalk \num{4.5}\si{m} from each other. Both sidewalks are mirrored.

\begin{itemize}
    \item \textbf{Bollards:} Each a cylinder \num{0.305}\si{m} in radius and \num{0.914}\si{m} in height. No cylinder is placed in the space taken by an alleyway intersecting a sidewalk.  All cylinders are positioned \num{0.063}\si{m} away from the edge of the curb. There are a total of 12 bollards on the scene. Visible as red circles in Fig. \ref{fig:empty_layout}.
    
    \item \textbf{Lamp Posts:} A series of tall cylinders representing lamp posts placed on the sidewalk, \num{0.161}\si{m} from the edge. Each tall cylinder is \num{0.152}\si{m} in radius and \num{9}\si{m} in height. Six are placed on each sidewalk, \num{9}\si{m} from each other. There are a total of 16 lamp posts on the scene. Visible as red squares in Fig. \ref{fig:empty_layout}.

\end{itemize}

The vehicles to be used in this simulation had to be representative of the variety of vehicles that can be found on the road. The fifteen vehicles seen in Table \ref{tab:vehicle_list} were selected using the list of vehicle types outlined by the \cite{federal_highway_administration_office_nodate} as a base. All models were found and downloaded for free from the online website Sketchfab.com under the Creative Commons license. Each model, the vehicle type it corresponds to, and its closest real-life equivalent can be seen in Table \ref{tab:vehicle_list}, with notes providing additional context if the specific model was not specified.

\begin{table*}[h]
\caption{List of vehicles used in the simulations.}
\centering
\begin{tabular}{|l|p{3cm}|l|p{4cm}|}
\hline
\textbf{Category}     & \textbf{Model}                    & \textbf{Source}                                                                & \textbf{Notes}             \\ \hline
Motorcycle        & Honda Shadow RS 2010              & \cite{alexka_honda_2022}                                    &                            \\ \hline
Smart Car         & Smart Fortwo                      & \cite{beztao01_smart_2021}                                  &                            \\ \hline
Carryall Car      & N/A                               & \cite{maxdragon_area_2021}                                  & Described as a "Golf Cart" \\ \hline
Sedan &
  N/A &
  \cite{daniel_zhabotinsky_danielzhabotinsky_generic_2020} &
  Described as "Based on asian design features of last decade {[}2010s{]}." \\ \hline
SUV               & Ford Edge 2006                    & \cite{fishermans_ford_2024}                                 &                            \\ \hline
Panel Van         & D3S MB Vito Panel Van (W447) '15  & \cite{danzzzeg_d3s_2017}                                    &                            \\ \hline
Topless Convertible &
  N/A &
  \cite{brian_n_chrysler_2007} &
  Chrysler Sebring Convertible. Specific model not specified. Page removed. \\ \hline
Station Wagon     & 80s Lincoln Zephyr Mercury        & \cite{daniel_zhabotinsky_danielzhabotinsky_american_2020} &                            \\ \hline
Four wheel Truck &
  N/A &
  \cite{daniel_zhabotinsky_danielzhabotinsky_light_2020} &
  Described as a generic Japanese-made small-cabin commerial truck from the 90s. \\ \hline
Minibus           & Peugeot J5                        & \cite{boitaloran_peugeot_2022}                              &                            \\ \hline
Pickup Truck      & 2015 Ford F150 King Ranch Edition & \cite{david_holiday_david_holiday_2015_2021}             &                            \\ \hline
Moving Truck &
  N/A &
  \cite{daniel_zhabotinsky_danielzhabotinsky_90_2021} &
  Described as a generic Japanese-made 2-door pickup-cabin box truck. \\ \hline
Motor home        & N/A                               & \cite{lion_sharp_lionsharp_free_2013}                     & Described as a "Modified GMC Motorhome"     \\ \hline
Double Decker Bus &
  N/A &
  \cite{businessyuen_bus_2020} &
  Described as a "Hong Kong Bus", corresponding to a Volvo B8L \\ \hline
Single Decker Bus & N/A                               & \cite{ownguest_generic_2021}                                &                            \\ \hline
\end{tabular}
\label{tab:vehicle_list}
\end{table*}

\begin{figure}
    \centering
    \includegraphics[width=3.5in,clip,keepaspectratio]{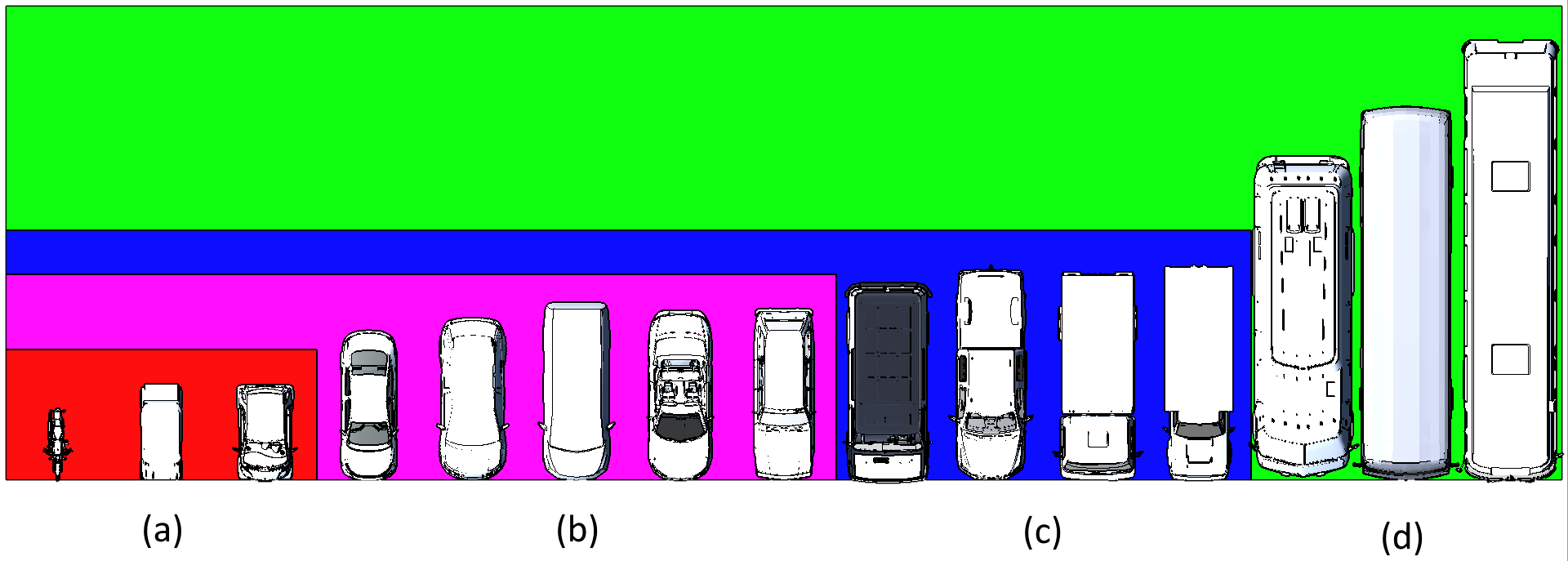}
    \caption{Top-Down View of the fifteen vehicles used in the simulations. Color denotes the category of each vehicle: (a) Small (b) Medium (c) Large (d) Extra Large}
    \label{fig:vehicle_size_categories}
\end{figure}

Based on their dimensions, the vehicles selected were broken into four categories, based on their length and classification in \cite{federal_highway_administration_office_nodate}:
\begin{itemize}
    \item \textbf{Small (S):} Vehicles of Class 1 and vehicles of Class 2 less than \num{4}\si{m} in length (Fig. \ref{fig:vehicle_size_categories}(a)).
    
    \item \textbf{Medium (M):} Vehicles of Class 2 and Class 3 between \num{4}\si{m} and \num{6}\si{m} in length (Fig. \ref{fig:vehicle_size_categories}(b)).
    
    \item \textbf{Large (L):} Vehicles of Class 3, Class 4, and Class 5 between \num{6}\si{m} and \num{8}\si{m} in length (Fig. \ref{fig:vehicle_size_categories}(c)).
    
    \item \textbf{Extra Large (XL):} Vehicles of Class 4, Class 5, and Class 6 longer than \num{8}\si{m} (Fig. \ref{fig:vehicle_size_categories}(d)).
\end{itemize}

Since all vehicle models were made by a third-party, it is important to compare the dimensions of the virtual vehicles with the real counterpart of those vehicles to ensure accuracy. Adjustments were made to the size of the vehicles to more closely resemble their real-life equivalents, if available. A comparison of both can be see in Table \ref{tab:dimensions_table}. 

\begin{table}[ht]
\caption{Dimensions of the virtual vehicles and their real-life equivalents, in meters.}
\centering
\begin{tabular}{|>{\hspace{0pt}}m{0.265\linewidth}|>{\hspace{0pt}}m{0.06\linewidth}|>{\hspace{0pt}}m{0.06\linewidth}|>{\hspace{0pt}}m{0.06\linewidth}|>{\hspace{0pt}}m{0.06\linewidth}|>{\hspace{0pt}}m{0.06\linewidth}|>{\hspace{0pt}}m{0.06\linewidth}|} 
\hline
\multirow{2}{0.9\linewidth}{\hspace{0pt}Vehicle Size Category} & \multicolumn{3}{>{\hspace{0pt}}m{0.23\linewidth}|}{Real Life \newline Dimensions (m)} & \multicolumn{3}{>{\hspace{0pt}}m{0.23\linewidth}|}{Dimensions in Simulation (m)~}  \\ 

                                                                 & H    & W    & L                                                               & H    & W    & L                                                                       \\ 
\hline
Motorcycle                                                       & 1.13 & 0.92 & 2.51                                                            & 1.15 & 0.86 & 2.14                                                                    \\ 
\hline
Smart Car                                                        & 1.54 & 1.56 & 2.70                                                            & 1.71 & 1.88 & 2.78                                                                    \\ 
\hline
Carryall Car                                                     & -    & -    & -                                                               & 1.78 & 1.22 & 2.79                                                                    \\ 
\hline
Sedan                                                            & -    & -    & -                                                               & 1.44 & 1.71 & 4.35                                                                    \\ 
\hline
SUV                                                              & 1.70 & 1.93 & 4.72                                                            & 1.63 & 2.08 & 4.71                                                                    \\ 
\hline
Panel Van                                                        & 1.90 & 1.93 & 4.90                                                            & 1.91 & 1.91 & 4.89                                                                    \\ 
\hline
Topless Convertible                                              & -    & -    & -                                                               & 1.31 & 1.93 & 4.94                                                                    \\ 
\hline
Station Wagon                                                    & 1.38 & 1.80 & 4.97                                                            & 1.36 & 1.83 & 4.98                                                                    \\ 
\hline
Four wheel Truck                                                 & 2.92 & 1.97 & 4.77                                                            & 2.61 & 2.84 & 6.02                                                                    \\ 
\hline
Minibus                                                          & 2.42 & 1.97 & 5.49                                                            & 2.46 & 2.62 & 5.81                                                                    \\ 
\hline
Pickup Truck                                                     & 2.00 & 2.03 & 6.18                                                            & 1.86 & 2.35 & 6.03                                                                    \\ 
\hline
Moving Truck                                                     & -    & -    & -                                                               & 2.46 & 2.09 & 6.25                                                                    \\ 
\hline
Motor home                                                       & -    & -    & -                                                               & 3.21 & 3.02 & 9.12                                                                    \\ 
\hline
Double Decker Bus                                                & -    & -    & -                                                               & 4.79 & 2.77 & 11.10                                                                   \\ 
\hline
Single Decker Bus                                                & -    & -    & -                                                               & 3.12 & 3.17 & 12.80                                                                   \\
\hline
\end{tabular}
\label{tab:dimensions_table}
\end{table}

\begin{figure}[h]
    \centering
    \includegraphics[width=3in,height=3in,clip,keepaspectratio]{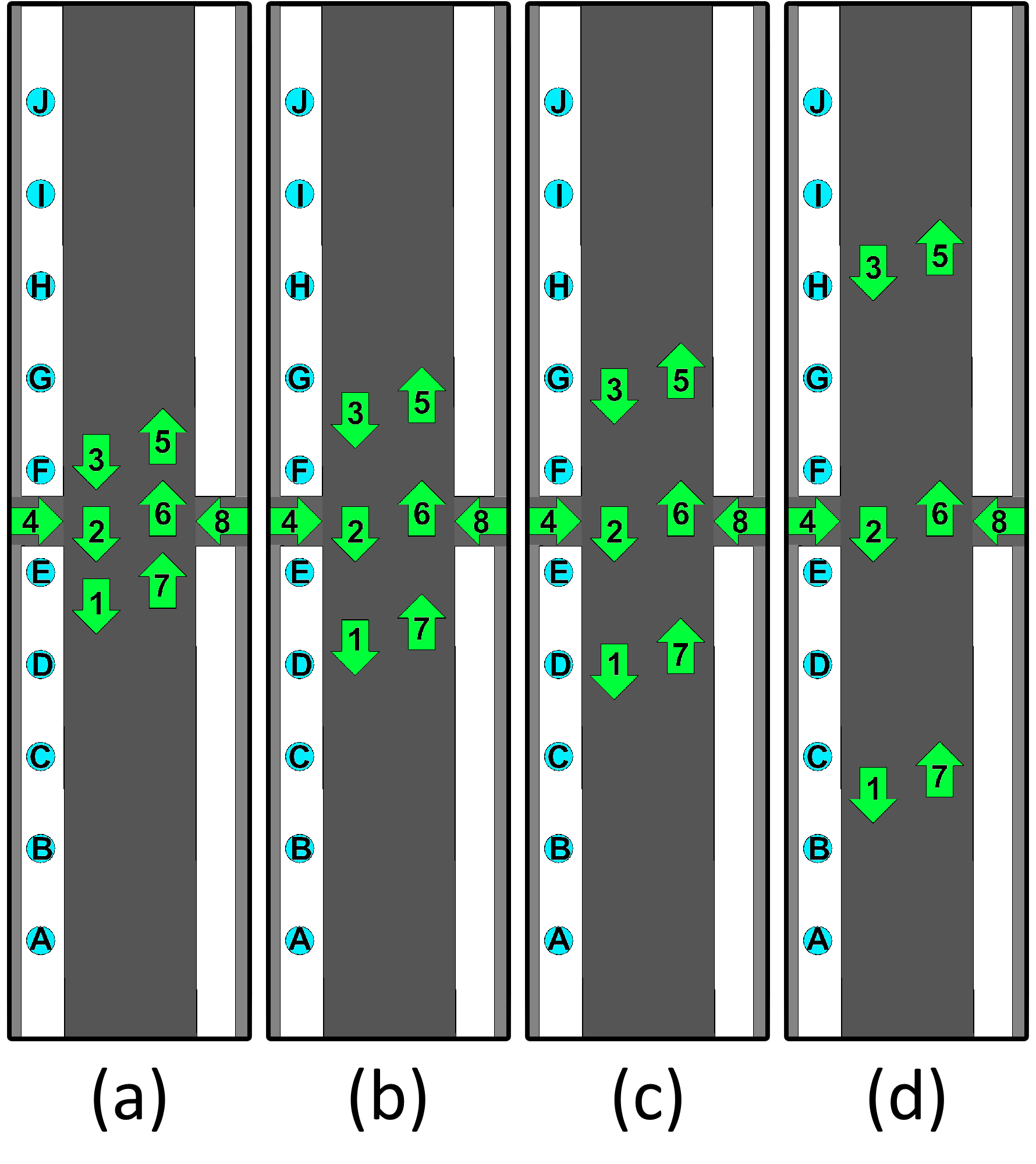}
    \caption{Diagram illustrating the ten possible sidewalk camera positions, and the eight possible vehicle positions, based on the size categories of the target vehicle. (a) Small vehicles (b) Medium vehicles (c) Large vehicles (d) Extra large vehicles.}
    \label{fig:simulation_diagram}
\end{figure}

The program has eight possible positions for vehicles.  
\begin{itemize}
    \item \textbf{On the Road:} Three of the positions are on the left side (Fig. \ref{fig:simulation_diagram}(a to d)(1 to 3)) and three on the right side (Fig. \ref{fig:simulation_diagram}(a to d)(5 to 7)). Vehicles in each set of positions face in the same direction, forming a line. For this work, the American driving direction was used (right side), so vehicles on the left side of the image drive towards the south, while those on the right side drive northward.
    \item \textbf{Exiting Alleyway:} Two of the positions place the vehicle intersecting with the sidewalk, facing east (Fig. \ref{fig:simulation_diagram}(a to d)(4)) or west (Fig. \ref{fig:simulation_diagram}(a to d)(8)) towards the road. This emulates vehicles turning into the road, as well as vehicles driving over the intersection.
\end{itemize}

The front of the spawned vehicles are aligned with the pivot point they are placed in. To ensure that vehicles wouldn't overlap, the specific position of each of the six pivots on the main road varies depending on the size of the vehicle being evaluated. Table \ref{tab:vehicle_positions} shows the distance between pivots for each size category, while Fig. \ref{fig:simulation_diagram} illustrates them.

\label{tab:VehiclePositionDistance}
\begin{table}[h]
\caption{Distance between each vehicle position on the road}
\centering
\begin{tabular}{|c|c|}
\hline
\textbf{Vehicle Category} & \textbf{Distance (m)} \\ \hline
Small                    & 3.792                 \\ \hline
Medium                   & 5.970                 \\ \hline
Large                    & 7.240                 \\ \hline
Extra Large              & 13.730                \\ \hline
\end{tabular}
\label{tab:vehicle_positions}
\end{table}

This means that the space between the front of a vehicle, and the back of the one its facing will vary between vehicles. Table \ref{tab:vehicle_distance} shows what this distance is for all fifteen vehicles used in the simulation.

\begin{table}
\caption{Distance between each vehicle when forming a line on the same lane.}
\centering
\begin{tabular}{|>{\centering\hspace{0pt}}m{0.137\linewidth}|>{\centering\hspace{0pt}}m{0.221\linewidth}|>{\centering\hspace{0pt}}m{0.2\linewidth}|>{\centering\arraybackslash\hspace{0pt}}m{0.2\linewidth}|} 
\hline
Category    & Vehicle Type        & Bumped-to-Bumper Distance (m) & Vehicle Length (m)  \\ 
\hline
Small       & Motorcycle          & 1.65                          & 2.14                \\ 
\hline
Small       & Smart Car           & 1.01                          & 2.78                \\ 
\hline
Small       & Carryall Car        & 1.00                          & 2.79                \\ 
\hline
Medium      & Sedan               & 1.62                          & 4.35                \\ 
\hline
Medium      & SUV                 & 1.26                          & 4.71                \\ 
\hline
Medium      & Panel Van           & 1.08                          & 4.89                \\ 
\hline
Medium      & Topless Convertible & 1.03                          & 4.94                \\ 
\hline
Medium      & Station Wagon       & 0.99                          & 4.98                \\ 
\hline
Large       & Four wheel Truck    & 1.22                          & 6.02                \\ 
\hline
Large       & Minibus             & 1.43                          & 5.81                \\ 
\hline
Large       & Pickup Truck        & 1.21                          & 6.03                \\ 
\hline
Large       & Moving Truck        & 0.99                          & 6.25                \\ 
\hline
Extra Large & Motor home          & 4.61                          & 9.12                \\ 
\hline
Extra Large & Double Decker Bus   & 2.63                          & 11.10               \\ 
\hline
Extra Large & Single Decker Bus   & 0.93                          & 12.80               \\
\hline
\end{tabular}
\label{tab:vehicle_distance}
\end{table}

At any given time, one and only one of the vehicles currently in the environment may record the points where it is being observed by the recording camera, this is considered the \textit{Target Vehicle}. In each \textit{Scenario}, the rest of the spaces that are occupied by vehicles are all filled with \textit{Blocking Vehicles} that use the same model as the \textit{Target Vehicle}, but do not record any data. 

Because the position of a vehicle on the road will vary regularly, and hence parts of said vehicle may be blocked by another vehicle, the simulation generates all possible combinations of \textit{Target Vehicle} and \textit{Blocking Vehicle} positions, based on whether a position is occupied by the target vehicle, a blocking vehicle, or no vehicle. 

The simulation has a single camera that, through a series of \textit{Data Captures}, records what parts of the \textit{Target Vehicle} are being observed. During the simulation, the camera cycles through a list of \textit{Camera Positions}, their possible \textit{Camera Directions}, and their possible heights. All \textit{Camera Positions} are placed on the sidewalk, 1.2\si{m} from the street, and are almost all spaced out from the next by 4.84\si{m}, except for the two adjacent to the alleyway which are  5.38\si{m} apart from each other. The \textit{Camera Positions} can be seen in Fig. \ref{fig:simulation_diagram} (A to J).

\begin{figure}
    \centering
    \includegraphics[width=2.5in,height=2.5in,clip,keepaspectratio]{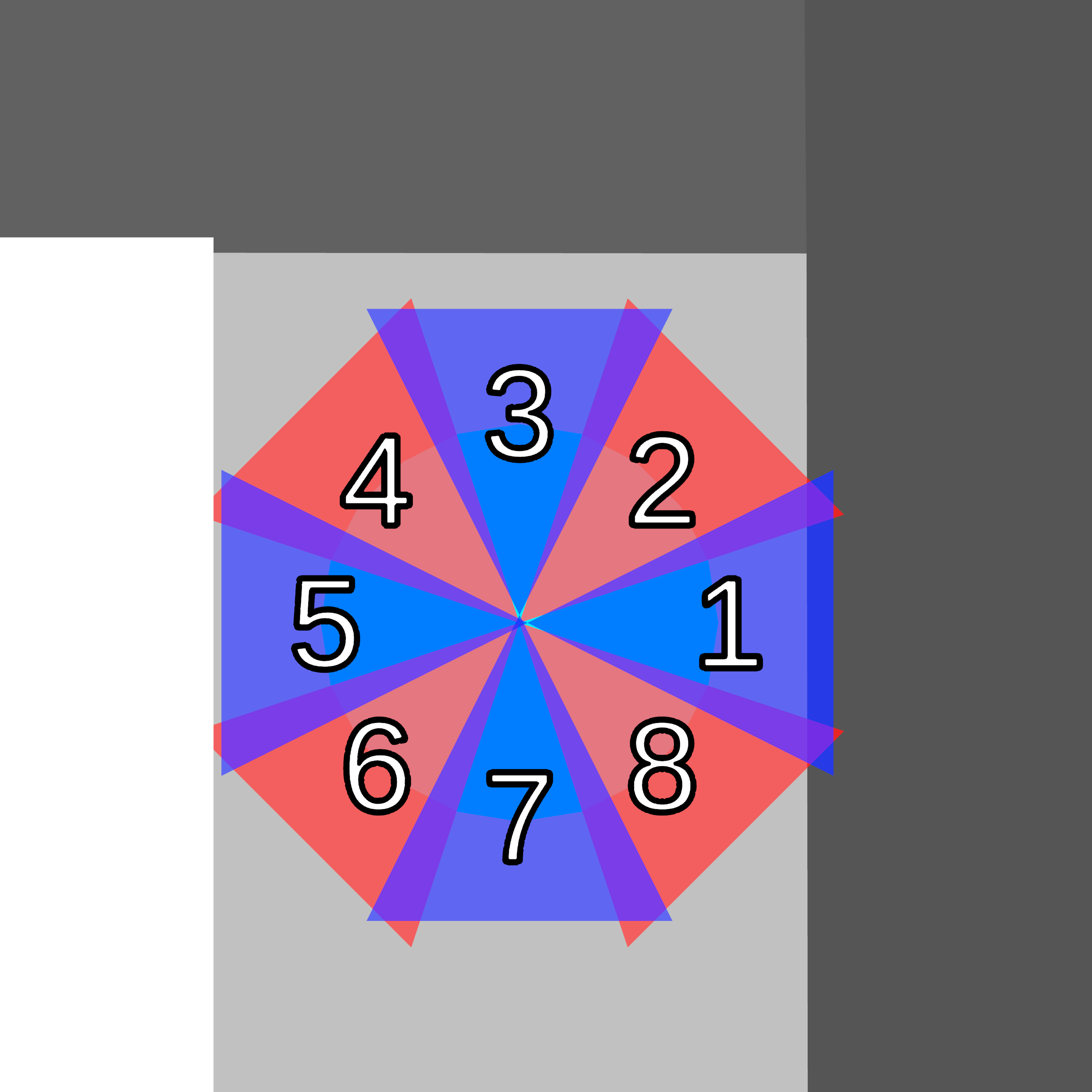}
    \caption{Diagram illustrating the eight \textit{Camera Directions} that a camera may take at any given \textit{Camera Position}, each identified by the cardinal or ordinal direction relative to the overhead camera: East (1), North-East (2), North (3), North-West (4), West (5), South-West (6), South (7), and South-East (8).}
    \label{fig:camera_directions}
\end{figure}

When the camera is at a specific \textit{Camera Position}, it must take a \textit{Data Capture} for each possible facing that a road user may have in that position. To replicate this, each point has a list of valid facing directions that the camera cycles through, and that the user selects during setup. Since the environment used for this simulation is made up entirely of right angles, the \textit{Camera Directions} are implemented in increments of 45 \si{degrees}, as seen in Fig. \ref{fig:camera_directions}

Since our roads are shared with people of varied heights, the camera needs to get \textit{Data Captures} from different heights. For each \textit{Camera Direction}, the camera cycles through five camera heights: 1.0\si{m}, 1.2\si{m}, 1.4\si{m}, 1.6\si{m}, and 1.8\si{m}. These correspond to the ones used by \cite{troel-madec_ehmi_2019} to simulate a variety of age groups. According to the United States of America's National Center for Health Statistics \cite{national_center_for_health_statistics_growth_2024}, these would approximately match the height of the median American at the ages of 3, 6, 10, 13, and 20 years old respectively. 

The system for recording data during a \textit{Data Capture} is mainly controlled by three scripts: The \textit{Heatmap Sender}, placed on the camera; the \textit{Heatmap Display} placed on the \textit{Target Vehicle}; and the the \textit{Heatmap Receiver}, placed also on the \textit{Target Vehicle}.

In each \textit{Data Capture}, the \textit{Heatmap Sender} casts a total of 14 400 straight rays in a $160\times90$ grid pattern. This resolution was chosen for two reasons: 
\begin{enumerate}
    \item Ray-casting is resource intensive, so a lower resolution was needed to ensure the simulation times remained achievable and in-scope.
    \item During development $1920\times1080$ was the most common screen resolution, according to W3Counter \cite{noauthor_w3counter_2024}. Since the Camera component in Unity makes use of the application resolution, and 1920 and 1080 can be divided evenly by 160 and 90 respectively, this reduces effect of rounding on the results.
\end{enumerate}

At the start of the simulation, the \textit{Heatmap Display} simplifies the mesh of the \textit{Target Vehicle} into a tridimensional array of \textit{Grid Points}, where each is a structure containing the internal local coordinates of that point on the grid. Each \textit{Grid Point} also holds a long integer value representing how many times that point has been hit by the \textit{Heatmap Sender}. All points in the grid are in contact with the mesh of the vehicle. This system is used to round the information received for later display. Note that, by using collisions and ray-casting as the methods of capturing data, the simulation is only capable of capturing the exposedness of points in the geometry of the vehicle, and does not account for factors such as reflectiveness and color.

When one of the rays casted by the \textit{Heatmap Sender} hits the the \textit{Target Vehicle}, the tridimensional point of contact is recorded internally in local coordinates by the \textit{Heatmap Receiver}. These rays are cast in the facing direction of the camera using Unity's \textit{ScreenPointToRay} function, which adjusts for the camera's default 60 degree field of view. At the end of each \textit{Data Capture}, once all rays have been cast by the \textit{Heatmap Sender}, the \textit{Heatmap Display} rounds each of the points recorded to the nearest equivalent \textit{Grid Point} (if applicable). Every time a point on the \textit{Heatmap Receiver} is rounded to a \textit{Grid Point}, the value associated with said \textit{Grid Point} is increased by one. 

Since the projection of the ray-casts is radial in nature, the distance between points captured is proportional to their radius from the camera. This introduces a proximity bias, making it so whenever the camera is closer to the \textit{Target Vehicle}, multiple local coordinates found by the \textit{Heatmap Sender} during a \textit{Data Capture} would be rounded to the same \textit{Grid Point}. To prevent this, it was made so \textit{Grid Points} may only increase their value by one during a given \textit{Data Capture}.

Once the simulation has run through all \textit{Scenarios}, the points stored and their associated values are serialized into a JavaScript Object Notation (JSON) file that can be read by the software, like the one in Listing 1. This represents the end of the simulation. 

\begin{lstlisting} [caption=Example of the contents of one of the JSON files]
    {"FileName":"Sample","maxValue":2,"points":[
    {"point":{"x":4,"y":38,"z":102},"value":2},
    {"point":{"x":4,"y":38,"z":103},"value":0},
    {"point":{"x":4,"y":38,"z":104},"value":1}]}
\end{lstlisting}

\begin{figure}[h]
    \centering
    \includegraphics[width=3in,height=3in,clip,keepaspectratio]{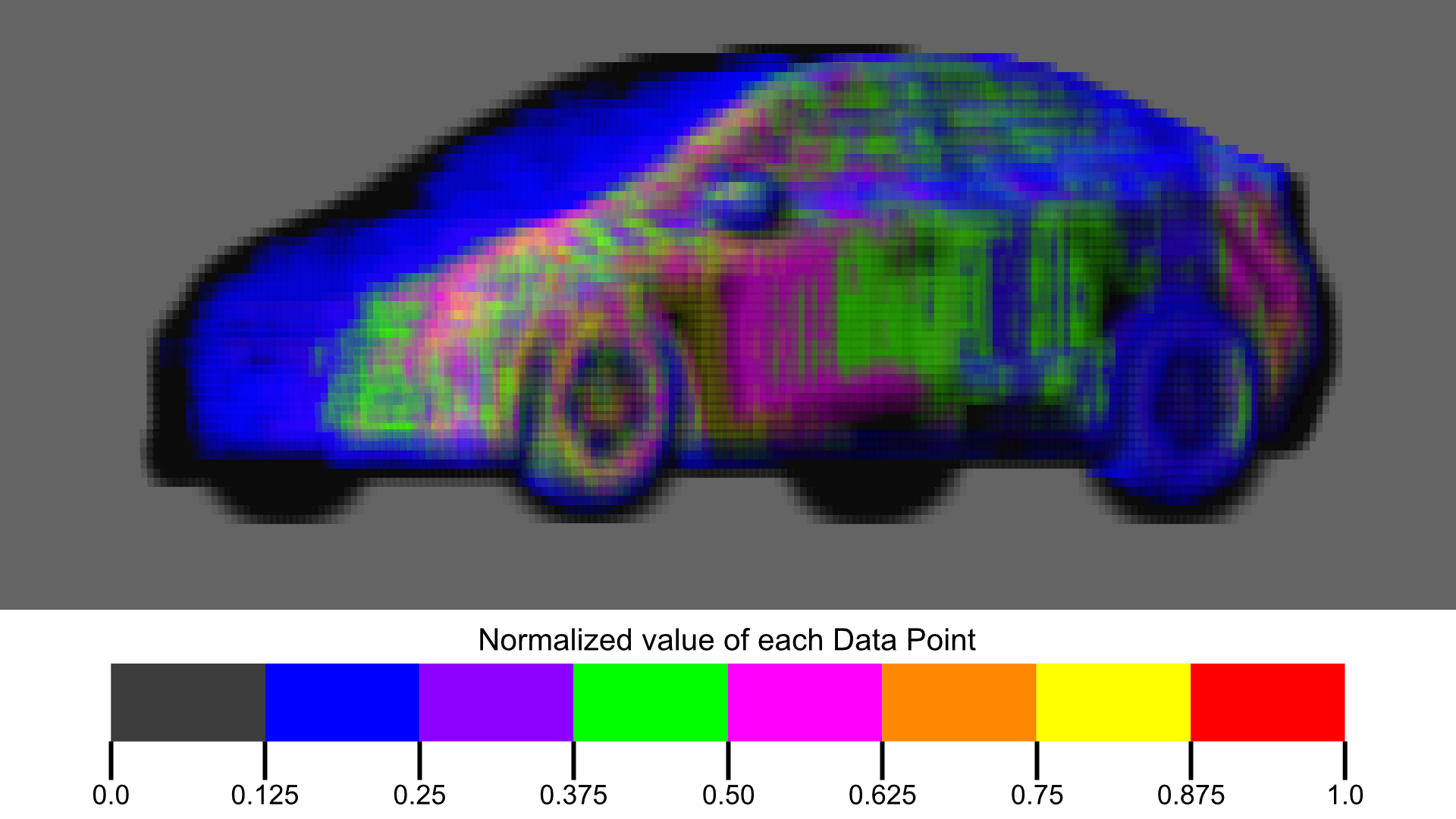}
    \caption{Diagram illustrating the coloring of \textit{Grid Points} based on their normalized value, with the highest value recorded as the upper limit.}
    \label{fig:normalized_values}
\end{figure}

In a separate Unity scene, the software is able to load said JSON files, then the user may select one or more of the JSON files to display, as well as one of the fifteen vehicles to display it on (Table \ref{tab:vehicle_list}). The software will then instantiate a copy of the selected vehicle and add the data from the JSON files to that vehicle's Heatmap Display through an additive process. The value of each \textit{Grid Points} is then normalized with the highest value in the file, and evaluated with a gradient selected by the user. By default, points are given one of eight colors corresponding to their normalized value, as seen on Fig. \ref{fig:normalized_values}.

The resulting 3D visualization is then presented to the user so they can inspect it and take virtual photographs of the vehicle facing nine of its possible angles: front, front-right corner, right flank, back-right corner, back, front-left corner, left flank, front-left corner, and directly from above.

\section{Experiment Setup}
\label{Experiment Setup}

The software developed for the simulation was made using the Unity game engine, version 2022.3.28f1, and C\#. Blender 4.2 was used to model the street, as well as adjust the pivot and size of the vehicles. Most of the data capture process was done using Unity's built-in physics and collision system for its ray-casting functionality. Physics were not used for anything else.

\begin{landscape}

    \begin{figure}
        \centering
        \includegraphics[width=7.7in,clip,keepaspectratio]{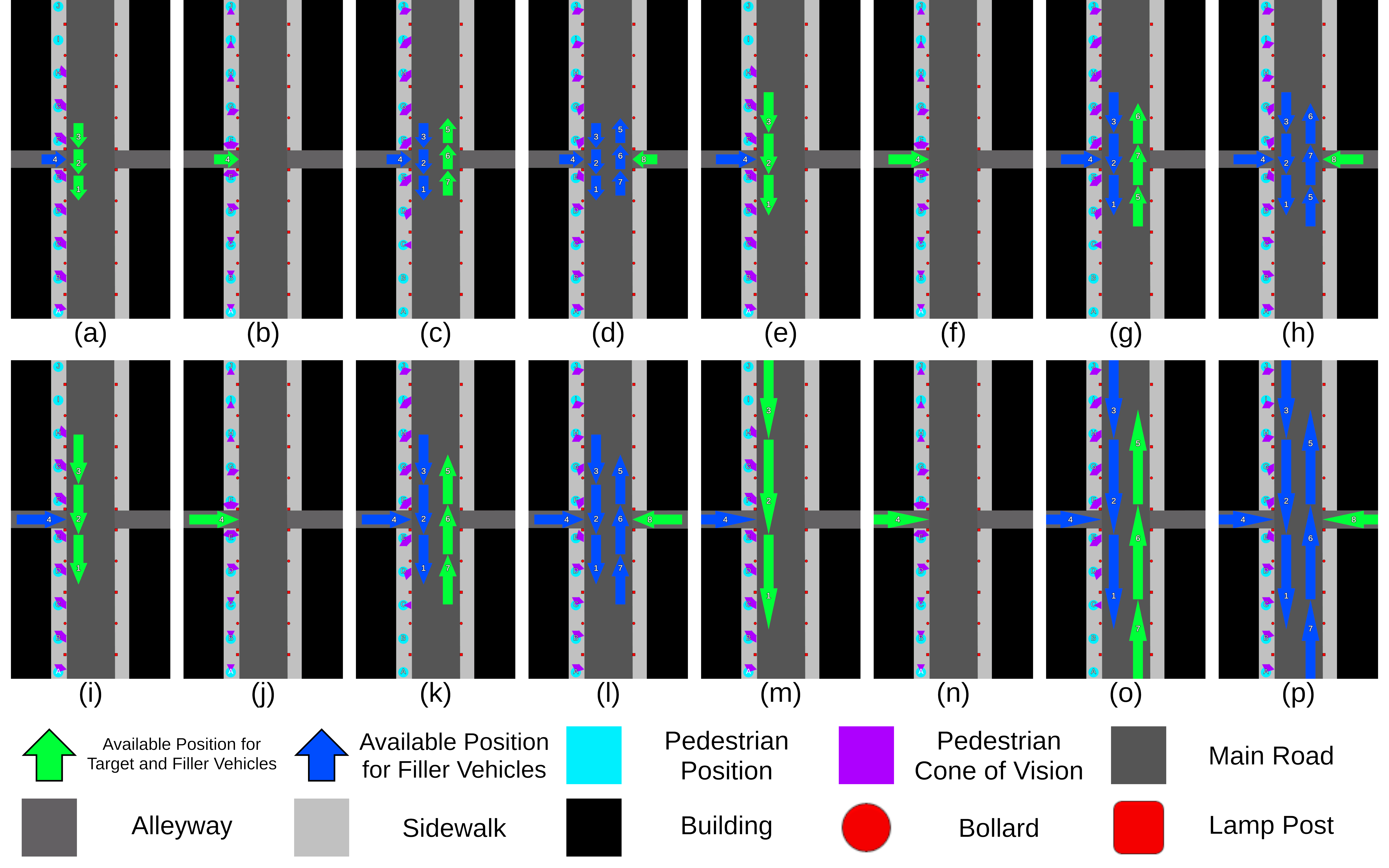}
        \caption{Diagram illustrating all sixteen setups based on the \textit{Target and Blocking Vehicle} available positions (green arrows in (a) to (p)), exclusive \textit{Blocking Vehicle} available positions (blue arrows in (a) to (o)), vehicle type ((a) to (d) for Small, (e) to (h) for Medium, (i) to (l) for Large, (m) to (p) for Extra Large) , \textit{Camera Positions} ((a to p)(A to J)), and \textit{Camera Directions} (Marked in purple).}
        \label{fig:all_vehicle_setups}
    \end{figure}
\end{landscape}

Since the amount of \textit{Blocking Vehicles} on the road will influence how much of the \textit{Target Vehicle} is visible, multiple \textit{Simulation Runs} were done varying the number of \textit{Blocking Vehicles} on the road. Fig. \ref{fig:all_vehicle_setups} illustrates the configuration of all positions that \textit{Target Vehicles} and \textit{Blocking Vehicles} occupied for this experiment, depending on their type.
\begin{itemize}
    \item The total amount of vehicles may be between 1 and 4 for scenarios where the \textit{Target Vehicle} is on the road closest to the sidewalk (Fig. \ref{fig:all_vehicle_setups} (a, e, j, n)(1,2,3)). This is because exposedness in that case can only be affected by \textit{Blocking Vehicles} placed on the sidewalk (Fig. \ref{fig:all_vehicle_setups} (a, e, j, n)(4)) or in front of the \textit{Target Vehicle}. 
    
    \item The total amount of vehicles is always 1 when the \textit{Target Vehicle} is located on the west alleyway (Fig. \ref{fig:all_vehicle_setups} (b, f, j, n)(4)), since there is no position where \textit{Blocking Vehicles} may block it.
    
    \item The total amount of vehicles may be between 1 and 7 when the \textit{Target Vehicle} is located on the right-hand side of the road  (Fig. \ref{fig:all_vehicle_setups}(c, g, k, o)(5,6,7)), since it may be blocked by other vehicles in those positions, as well as vehicles on the left side (Fig. \ref{fig:all_vehicle_setups} (c, g, k, o)(1 to 4)).
    
    \item The total amount of vehicles may be between 1 and 8 when the \textit{Target Vehicle} is positioned on the east alleyway (Fig. \ref{fig:all_vehicle_setups} (d, h, l, p)(8)), since it can be blocked by any other vehicle on the road.
    
\end{itemize}

Each \textit{Camera Position} has a total of ten possible positions on the sidewalk that it may cycle through (Fig. \ref{fig:all_vehicle_setups}(a to p)(A to J)). The user may chose which ones to use for a \textit{Simulation Run}, as well as the direction the camera may face at said location. For the simulations in this paper, camera positions were chosen based on four possible vehicle facings, minimizing cases where the pedestrian would not be able to see the front of the vehicle:
\begin{itemize}
    \item If the \textit{Target Vehicle} is going southwards, \textit{Data Captures} will only happen towards the east, north-east, and north directions (Fig. \ref{fig:all_vehicle_setups} (a, e, i, m), Fig. \ref{fig:camera_directions}(1,2,3)).
    \item If the \textit{Target Vehicle} is coming out of the left alleyway, going eastwards, \textit{Data Captures} will only be made in the north, north-east, south, and south-east directions (Fig. \ref{fig:all_vehicle_setups} (b, f, j, n), Fig. \ref{fig:camera_directions}(2,3,7,8)).
    \item If the \textit{Target Vehicle} is going northwards, \textit{Data Captures} will only be made in the east, south-east, and south directions (Fig. \ref{fig:all_vehicle_setups} (c, g, k, o), Fig. \ref{fig:camera_directions}(1,7,8)).
    \item If the \textit{Target Vehicle} is coming out of the right alleyway, going westwards, \textit{Data Captures} will only happen towards the north, north-east, south, and south-east (Fig. \ref{fig:all_vehicle_setups} (d, g, l, p), Fig. \ref{fig:camera_directions}(2,3,7,8)).
\end{itemize}

\begin{figure}[h]
    \centering
    \includegraphics[width=3in,height=3in,clip,keepaspectratio]{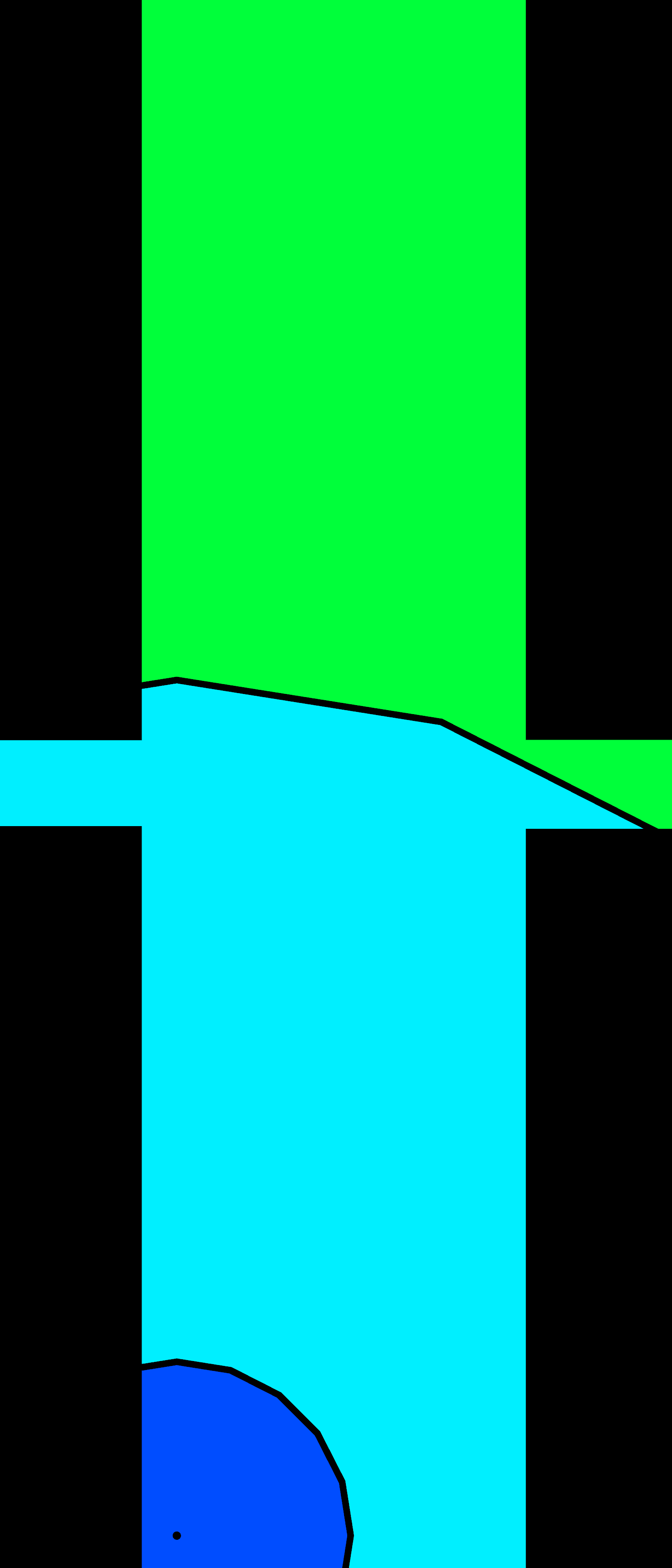}
    \caption{Visualization of the three ranges used for \textit{Data Capturing} for this experiment: Blue: \num{0}\si{m} to \num{5}\si{m}; Cyan: \num{5}\si{m} to \num{25}\si{m}; Green: \num{25}\si{m} to \num{75}\si{m}.}
    \label{fig:range_diagram}
\end{figure}

For this experiment, in order to produce data on how the distance between the \textit{Target Vehicle} and the camera may affect the exposedness of certain elements, each set of nineteen scenarios was run three times with different minimum and maximum distances for the \textit{Data Capture}: \num{0}\si{m} to \num{5}\si{m}, \num{5}\si{m} to \num{25}\si{m}, and \num{25}\si{m} to \num{75}\si{m}. These values were chosen to correspond to a short, medium, and long distance threshold. Fig. \ref{fig:range_diagram} puts these values into scale within the environment of the simulation.

The simulation was run a total of nineteen times for each vehicle, using the codes found in Appendix \ref{secA2} and producing 860 output files. \textit{Simulation Runs} were done in two computers at the same time, each running multiple parallel instances of the program. The main computer used for these simulations was a desktop personal computer equipped with an AMD Ryzen 7 2700 eight-core processor, an NVIDIA GeForce RTX 2060, and 16 GB of RAM running a 64-bit version of Windows 10. The second computer was an Acer Nitro V 15 laptop equipped with a 13th gen Intel Core i7-13620H, an NVIDIA GeForce RTX 4060 Laptop GPU, and 16 GB of RAM running a 64-bit version of Windows 11 Home.

The time taken by each \textit{Simulation Run} varied from 1 minute to 8 hours, depending on factors such as:
\begin{itemize}
    \item The size of the model, as larger models were hit by more ray-casts.
    \item The amount of polygons in the model, which takes up more resources to render.
    \item The number of vehicles and possible vehicle positions, as a low number of the former with a higher number of the later means more permutations to test.
    \item The total number of \textit{Camera Directions} visited by the \textit{Data Captures}.
\end{itemize}

\section{Results and Discussion} \label{Results}

A total of 860 separate simulations were run in order to produce data that could represent the exposedness of all fifteen selected vehicles under the three different ranges. Since the number of \textit{Scenarios} varies depending on the position of the vehicle, simply adding all files corresponding to a given vehicle type would produce results biased towards scenarios where the vehicle was on the right side of the street, as those have a higher amount of \textit{Blocking Vehicle} positions and permutations. Because of this, results were divided into four categories to be processed and normalized independently:
\begin{itemize}
    \item Vehicle is facing south (Fig. \ref{fig:all_vehicle_setups} (a,e,i,m)). 
    \item Vehicle is facing east (Fig. \ref{fig:all_vehicle_setups} (b,f,j,n)).
    \item Vehicle is facing north (Fig. \ref{fig:all_vehicle_setups} (c,g,k,o)).
    \item Vehicle is facing west (Fig. \ref{fig:all_vehicle_setups} (d,h,l,p)).
\end{itemize}
For each category, the recorded data was divided into three subsets based on their minimum and maximum distance parameters (0\si{m} to 5\si{m}, 5\si{m} to 25\si{m}, 25\si{m} to 75\si{m}), then added unweighted to produce a total of 12 heatmaps per vehicle, where each point on the vehicle shows how many times it was captured by the camera. This presents a total of 180 total different heatmaps to be evaluated ($15\times4\times3$). 

Since each \textit{Grid Point} can only be recorded once per \textit{Data Capture}, the color of a point also represents the amount of circumstances under which it was recorded. These heatmaps were normalized using 0 as the lower bound, and the maximum highest value across all points in the subset as the higher bound. We decided to apply this normalization instead of using absolute values to prevent biases introduced by larger vehicles or by conditions where the \textit{Target Vehicle} was closer to the camera. 

To make analyzing the data possible, we chose to use solid color delimitation in the figures instead of a gradient. We divided the evaluation of point values into eight separate colors, numbered 0 through 7, where each corresponds to a range of values one-eight of the total range of the normalization. These colors were chosen because of their contrast with each other: 0 for Black (\#000000), 1 for Blue (\#0000FF), 2 for Purple (\#7900FF), 3 for Lime Green (\#84FF0C), 4 for Magenta (\#FF00FF), 5 for Yellow (\#FFFF00), 6 for Cyan (\#00FFFF), and 7 for Red (\#FF0000). The number associated with a \textit{Data Point} color is hereon referred to as the \textit{Exposedness Score} of that \textit{Data Point}.

\begin{figure} [h]
    \centering
    \includegraphics[width=5in,clip,keepaspectratio]{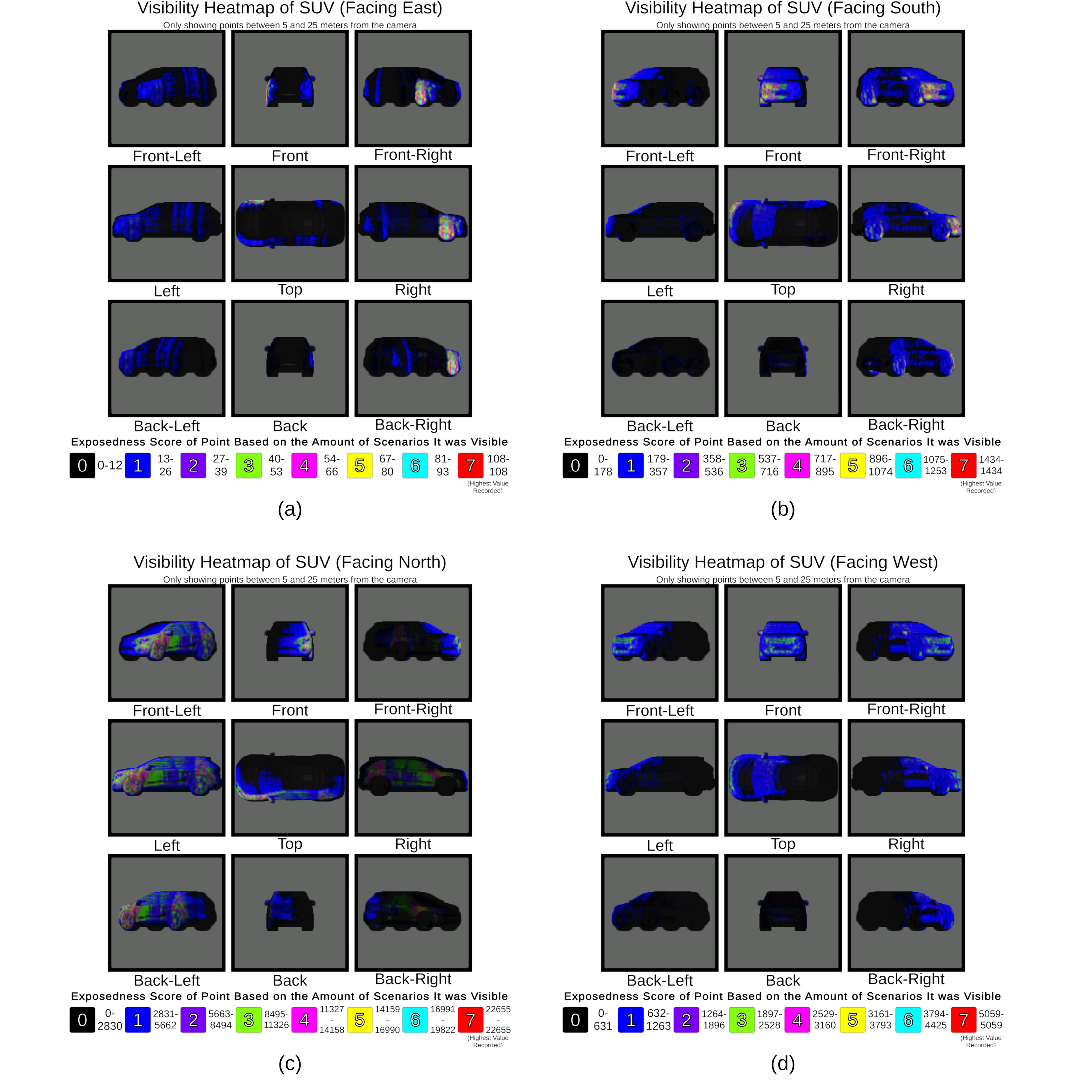}
    \caption{Results showing how many times the \textit{Grid Points} of the SUV vehicle was recorded, when these points are between 5\si{m} and 25\si{m} from the camera, and based on the direction the vehicle was driving: (a) Driving north, on the right side of the virtual environment; (b) driving east, out of the alleyway on the left side of the virtual environment; (c) driving south, on the left side of the virtual environment; (d) driving west, out of the alleyway on the right side of the virtual environment.}
    \label{fig:suv_facing_all_5_25}
\end{figure}

Afterwards, we took a series of screenshots for each resulting heatmap. Fig. \ref{fig:suv_facing_all_5_25} shows an example of the heatmap for the SUV vehicle at each of the four facings described above, and for \textit{Data Points} captured between 5\si{m} and 25\si{m} of the camera.


\subsection{Data Analysis} \label{Data Analysis}

\begin{figure}[!h]
    \centering
    \includegraphics[width=4in,clip,keepaspectratio]{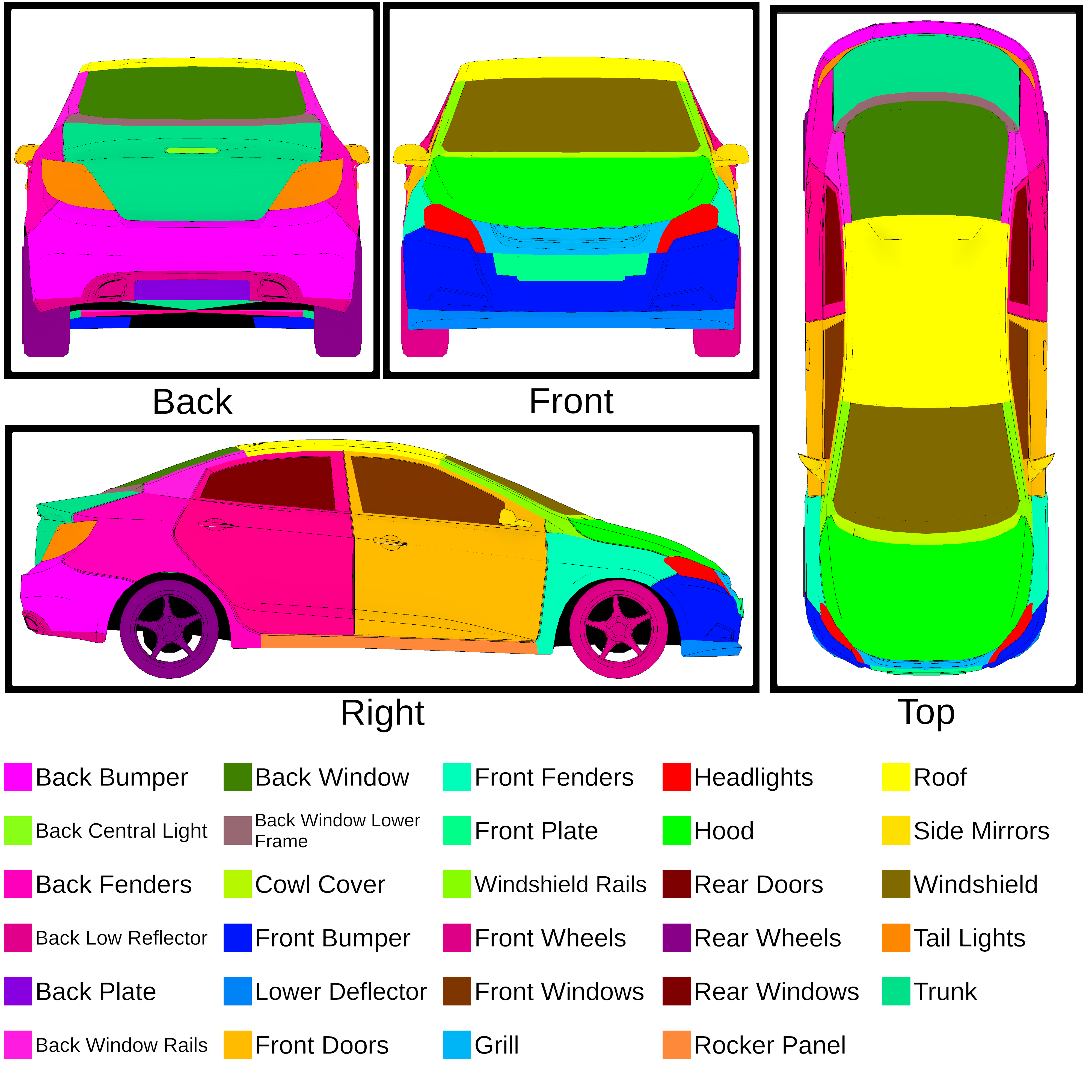}
    \caption{Illustration of the different exterior parts of a sedan, and the terminology used to refer to each.}
    \label{fig:vehicle_elements}
\end{figure}

During data analysis, we manually reviewed each of the 180 images to evaluate the exposedness of each vehicle part. This required a visual aid and consistent terminology that could be used to identify each vehicle part, Fig. \ref{fig:vehicle_elements} provides both for the sedan vehicle type. 

These parts were extrapolated for each of the 15 vehicle types, noting any time a part was not present in a vehicle type (e.g. doors on a motorcycle, roof on a convertible). Since the approach used for this study does not offer a way to automatically measure the average color of a vehicle part (and therefore its \textit{Exposedness Score}), we only annotated the highest one that can be found on each individual part, for each vehicle type and driving direction. 

Table \ref{tab:top5_exposedness} shows the five highest recorded \textit{Exposedness Scores} across each of the 15 vehicles, for each of the three distance thresholds. Each intersecting cell has four values separated by spaces, each value determining what the highest \textit{Exposedness Score} was for that part when that vehicle type was facing south, east, north, and west respectively. These cells are also shaded based on the sum of their contents, with those closer to 28 (the maximum possible value) being darker. The "Average Highest" row shows the average of each column across all vehicles in a distance threshold, and is shaded based on the smallest and largest value in that row. Whenever a vehicle does not feature the specified part, the intersecting cell is marked as "0 (NA)", which is valued the same as "0 0 0 0". A full breakdown of every part can be found on Appendix \ref{tab:0-5Results}, \ref{tab:5-25Results}, and  \ref{tab:25-75Results}.

\begin{table}[h]
\label{tab:top5_exposedness}
\resizebox{\columnwidth}{!}{%
\begin{tabular}{|
>{\columncolor[HTML]{EFEFEF}}l llllll|}
\hline
\multicolumn{7}{|c|}{\cellcolor[HTML]{B7B7B7}\textbf{Five Highest registered Exposedness Score in each facing (South East North West)}} \\ \hline
\multicolumn{7}{|c|}{\cellcolor[HTML]{CCCCCC}\textbf{For points between 0 and 5 meters from the camera.}} \\ \hline
\multicolumn{1}{|r|}{\cellcolor[HTML]{D9D9D9}\textbf{Vehicle Part \& Type}} &
  \multicolumn{1}{l|}{\cellcolor[HTML]{EFEFEF}\textbf{Front Doors}} &
  \multicolumn{1}{l|}{\cellcolor[HTML]{EFEFEF}\textbf{Front Fenders}} &
  \multicolumn{1}{l|}{\cellcolor[HTML]{EFEFEF}\textbf{Windshield Rails}} &
  \multicolumn{1}{l|}{\cellcolor[HTML]{EFEFEF}\textbf{Front Wheels}} &
  \multicolumn{1}{l|}{\cellcolor[HTML]{EFEFEF}\textbf{Front Windows}} &
  \cellcolor[HTML]{EFEFEF}\textbf{Side Mirrors} \\ \hline
\multicolumn{1}{|l|}{\cellcolor[HTML]{EFEFEF}\textbf{Carryall Car}} &
  \multicolumn{1}{l|}{\cellcolor[HTML]{FFFFFF}0 (NA)} &
  \multicolumn{1}{l|}{\cellcolor[HTML]{FFDEDE}2 2 0 0} &
  \multicolumn{1}{l|}{\cellcolor[HTML]{FFE8E7}2 1 0 0} &
  \multicolumn{1}{l|}{\cellcolor[HTML]{FFDEDE}2 2 0 0} &
  \multicolumn{1}{l|}{\cellcolor[HTML]{FFFFFF}0 (NA)} &
  \cellcolor[HTML]{FFFFFF}0 (NA) \\ \hline
\multicolumn{1}{|l|}{\cellcolor[HTML]{EFEFEF}\textbf{Double decker bus}} &
  \multicolumn{1}{l|}{\cellcolor[HTML]{FFFFFF}0 (NA)} &
  \multicolumn{1}{l|}{\cellcolor[HTML]{FF8C8D}5 7 0 1} &
  \multicolumn{1}{l|}{\cellcolor[HTML]{FFD5D5}4 0 0 1} &
  \multicolumn{1}{l|}{\cellcolor[HTML]{FFB1B1}7 1 0 1} &
  \multicolumn{1}{l|}{\cellcolor[HTML]{FFCCCD}3 3 0 0} &
  \cellcolor[HTML]{FFC3C3}0 0 0 7 \\
\hline
\end{tabular}%
}
\caption{The five parts with the highest total \textit{Exposedness Score} in each distance threshold, as well as what was said score at each facing direction.}
\end{table}

\subsection{Effects of Vehicle Position} \label{Effects of Vehicle Position}

Table \ref{tab:facing_summary} summarizes the average highest \textit{Exposedness Score} for each part across all fifteen vehicles, for each of the facing directions described above, and for each of the three distance thresholds.

\begin{table}[]
\caption{The average Highest Exposedness Score for each vehicle part when the vehicle is facing each of the four directions described.}
\begin{tabular}{|
>{\columncolor[HTML]{F3F3F3}}l cccc|}
\hline
\multicolumn{5}{|c|}{\cellcolor[HTML]{CCCCCC}\textbf{\begin{tabular}[c]{@{}c@{}}Average highest Exposedness Score found in part, for each facing direction.\end{tabular}}} \\ \hline
\multicolumn{5}{|c|}{\cellcolor[HTML]{D9D9D9}\textbf{For Data Points 0m to 5m from Camera}} \\ \hline
\multicolumn{1}{|l|}{\cellcolor[HTML]{EFEFEF}\textbf{Vehicle Part \& Facing Direction}} &
  \multicolumn{1}{c|}{\cellcolor[HTML]{F3F3F3}\textbf{South}} &
  \multicolumn{1}{c|}{\cellcolor[HTML]{F3F3F3}\textbf{East}} &
  \multicolumn{1}{c|}{\cellcolor[HTML]{F3F3F3}\textbf{North}} &
  \cellcolor[HTML]{F3F3F3}\textbf{West} \\ \hline
\multicolumn{1}{|l|}{\cellcolor[HTML]{F3F3F3}\textbf{Front Bumper}} &
  \multicolumn{1}{c|}{\cellcolor[HTML]{FAE3E1}1.53} &
  \multicolumn{1}{c|}{\cellcolor[HTML]{FFFDFD}0.13} &
  \multicolumn{1}{c|}{\cellcolor[HTML]{FFFFFF}0.00} &
  \cellcolor[HTML]{FFFFFF}0.00 \\ \hline
\multicolumn{1}{|l|}{\cellcolor[HTML]{F3F3F3}\textbf{Front Doors}} &
  \multicolumn{1}{c|}{\cellcolor[HTML]{F3BCB8}3.58} &
  \multicolumn{1}{c|}{\cellcolor[HTML]{F4C6C2}3.08} &
  \multicolumn{1}{c|}{\cellcolor[HTML]{FFFFFF}0.00} &
  \cellcolor[HTML]{FFFFFF}0.00 \\ \hline
\multicolumn{1}{|l|}{\cellcolor[HTML]{F3F3F3}\textbf{Front Fenders}} &
  \multicolumn{1}{c|}{\cellcolor[HTML]{F0B0AA}4.27} &
  \multicolumn{1}{c|}{\cellcolor[HTML]{F1B6B1}3.93} &
  \multicolumn{1}{c|}{\cellcolor[HTML]{FEF9F9}0.33} &
  \cellcolor[HTML]{FFFEFE}0.07 \\ \hline
\multicolumn{1}{|l|}{\cellcolor[HTML]{F3F3F3}\textbf{Front Windows}} &
  \multicolumn{1}{c|}{\cellcolor[HTML]{F6CCC8}2.75} &
  \multicolumn{1}{c|}{\cellcolor[HTML]{F6CCC8}2.75} &
  \multicolumn{1}{c|}{\cellcolor[HTML]{FFFFFF}0.00} &
  \cellcolor[HTML]{FFFFFF}0.00 \\ \hline
\multicolumn{1}{|l|}{\cellcolor[HTML]{F3F3F3}\textbf{Front Wheels}} &
  \multicolumn{1}{c|}{\cellcolor[HTML]{F0ACA6}4.47} &
  \multicolumn{1}{c|}{\cellcolor[HTML]{F4C1BD}3.33} &
  \multicolumn{1}{c|}{\cellcolor[HTML]{FFFFFF}0.00} &
  \cellcolor[HTML]{FFFEFE}0.07 \\ \hline
\multicolumn{1}{|l|}{\cellcolor[HTML]{F3F3F3}\textbf{Headlights}} &
  \multicolumn{1}{c|}{\cellcolor[HTML]{F4C5C1}3.13} &
  \multicolumn{1}{c|}{\cellcolor[HTML]{FEF7F6}0.47} &
  \multicolumn{1}{c|}{\cellcolor[HTML]{FFFFFF}0.00} &
  \cellcolor[HTML]{FFFEFE}0.07 \\ \hline
\multicolumn{1}{|l|}{\cellcolor[HTML]{F3F3F3}\textbf{Rear Wheels}} &
  \multicolumn{1}{c|}{\cellcolor[HTML]{F7D3CF}2.40} &
  \multicolumn{1}{c|}{\cellcolor[HTML]{FFFDFD}0.13} &
  \multicolumn{1}{c|}{\cellcolor[HTML]{FFFFFF}0.00} &
  \cellcolor[HTML]{FFFFFF}0.00 \\ \hline
\multicolumn{1}{|l|}{\cellcolor[HTML]{F3F3F3}\textbf{Side Mirrors}} &
  \multicolumn{1}{c|}{\cellcolor[HTML]{F7D3D0}2.36} &
  \multicolumn{1}{c|}{\cellcolor[HTML]{FCEBEA}1.07} &
  \multicolumn{1}{c|}{\cellcolor[HTML]{FCEDEB}1.00} &
  \cellcolor[HTML]{FEF6F5}0.50 \\ \hline
\multicolumn{1}{|l|}{\cellcolor[HTML]{F3F3F3}\textbf{Windshield}} &
  \multicolumn{1}{c|}{\cellcolor[HTML]{FAE2E0}1.57} &
  \multicolumn{1}{c|}{\cellcolor[HTML]{FDF1F0}0.79} &
  \multicolumn{1}{c|}{\cellcolor[HTML]{FFFFFF}0.00} &
  \cellcolor[HTML]{FFFEFE}0.07 \\ \hline
\multicolumn{1}{|l|}{\cellcolor[HTML]{F3F3F3}\textbf{Windshield Rails}} &
  \multicolumn{1}{c|}{\cellcolor[HTML]{F6CBC8}2.79} &
  \multicolumn{1}{c|}{\cellcolor[HTML]{F9DEDC}1.79} &
  \multicolumn{1}{c|}{\cellcolor[HTML]{FFFFFF}0.00} &
  \cellcolor[HTML]{FFFEFE}0.07 \\ \hline
\multicolumn{5}{|c|}{\cellcolor[HTML]{D9D9D9}\textbf{For Data Points 5m to 25m from Camera}} \\ \hline
\multicolumn{1}{|l|}{\cellcolor[HTML]{EFEFEF}\textbf{Vehicle Part \& Facing Direction}} &
  \multicolumn{1}{c|}{\cellcolor[HTML]{F3F3F3}\textbf{South}} &
  \multicolumn{1}{c|}{\cellcolor[HTML]{F3F3F3}\textbf{East}} &
  \multicolumn{1}{c|}{\cellcolor[HTML]{F3F3F3}\textbf{North}} &
  \cellcolor[HTML]{F3F3F3}\textbf{West} \\ \hline
\multicolumn{1}{|l|}{\cellcolor[HTML]{F3F3F3}\textbf{Front Bumper}} &
  \multicolumn{1}{c|}{\cellcolor[HTML]{FCEDEB}1.00} &
  \multicolumn{1}{c|}{\cellcolor[HTML]{FEF9F9}0.33} &
  \multicolumn{1}{c|}{\cellcolor[HTML]{FCECEA}1.07} &
  \cellcolor[HTML]{FFFFFF}0.00 \\ \hline
\multicolumn{1}{|l|}{\cellcolor[HTML]{F3F3F3}\textbf{Front Doors}} &
  \multicolumn{1}{c|}{\cellcolor[HTML]{FBE7E5}1.33} &
  \multicolumn{1}{c|}{\cellcolor[HTML]{F8DAD7}2.00} &
  \multicolumn{1}{c|}{\cellcolor[HTML]{F4C1BD}3.33} &
  \cellcolor[HTML]{F7D1CD}2.50 \\ \hline
\multicolumn{1}{|l|}{\cellcolor[HTML]{F3F3F3}\textbf{Front Fenders}} &
  \multicolumn{1}{c|}{\cellcolor[HTML]{F4C5C1}3.13} &
  \multicolumn{1}{c|}{\cellcolor[HTML]{EFABA5}4.53} &
  \multicolumn{1}{c|}{\cellcolor[HTML]{EB9790}5.60} &
  \cellcolor[HTML]{F5C7C3}3.00 \\ \hline
\multicolumn{1}{|l|}{\cellcolor[HTML]{F3F3F3}\textbf{Front Windows}} &
  \multicolumn{1}{c|}{\cellcolor[HTML]{FCEDEB}1.00} &
  \multicolumn{1}{c|}{\cellcolor[HTML]{F8DAD7}2.00} &
  \multicolumn{1}{c|}{\cellcolor[HTML]{F3BCB8}3.58} &
  \cellcolor[HTML]{F5C7C3}3.00 \\ \hline
\multicolumn{1}{|l|}{\cellcolor[HTML]{F3F3F3}\textbf{Front Wheels}} &
  \multicolumn{1}{c|}{\cellcolor[HTML]{EFA8A2}4.67} &
  \multicolumn{1}{c|}{\cellcolor[HTML]{EFA8A2}4.67} &
  \multicolumn{1}{c|}{\cellcolor[HTML]{EFA7A1}4.73} &
  \cellcolor[HTML]{F4C1BD}3.33 \\ \hline
\multicolumn{1}{|l|}{\cellcolor[HTML]{F3F3F3}\textbf{Headlights}} &
  \multicolumn{1}{c|}{\cellcolor[HTML]{EEA69F}4.80} &
  \multicolumn{1}{c|}{\cellcolor[HTML]{FBEAE9}1.13} &
  \multicolumn{1}{c|}{\cellcolor[HTML]{EEA69F}4.80} &
  \cellcolor[HTML]{F0B1AB}4.20 \\ \hline
\multicolumn{1}{|l|}{\cellcolor[HTML]{F3F3F3}\textbf{Rear Wheels}} &
  \multicolumn{1}{c|}{\cellcolor[HTML]{EEA49E}4.87} &
  \multicolumn{1}{c|}{\cellcolor[HTML]{F9DBD9}1.93} &
  \multicolumn{1}{c|}{\cellcolor[HTML]{F3BDB9}3.53} &
  \cellcolor[HTML]{FBE9E7}1.20 \\ \hline
\multicolumn{1}{|l|}{\cellcolor[HTML]{F3F3F3}\textbf{Side Mirrors}} &
  \multicolumn{1}{c|}{\cellcolor[HTML]{EC9A93}5.43} &
  \multicolumn{1}{c|}{\cellcolor[HTML]{FCEFEE}0.86} &
  \multicolumn{1}{c|}{\cellcolor[HTML]{F0B1AB}4.21} &
  \cellcolor[HTML]{EC9A93}5.43 \\ \hline
\multicolumn{1}{|l|}{\cellcolor[HTML]{F3F3F3}\textbf{Windshield}} &
  \multicolumn{1}{c|}{\cellcolor[HTML]{F4C3BF}3.21} &
  \multicolumn{1}{c|}{\cellcolor[HTML]{FDF5F4}0.57} &
  \multicolumn{1}{c|}{\cellcolor[HTML]{F5C6C2}3.07} &
  \cellcolor[HTML]{F2B9B4}3.79 \\ \hline
\multicolumn{1}{|l|}{\cellcolor[HTML]{F3F3F3}\textbf{Windshield Rails}} &
  \multicolumn{1}{c|}{\cellcolor[HTML]{F8D7D5}2.14} &
  \multicolumn{1}{c|}{\cellcolor[HTML]{FAE2E0}1.57} &
  \multicolumn{1}{c|}{\cellcolor[HTML]{EFAAA4}4.57} &
  \cellcolor[HTML]{F4C2BE}3.29 \\ \hline
\multicolumn{5}{|c|}{\cellcolor[HTML]{D9D9D9}\textbf{For Data Points 25m to 75m from Camera}} \\ \hline
\multicolumn{1}{|l|}{\cellcolor[HTML]{EFEFEF}\textbf{Vehicle Part \& Facing Direction}} &
  \multicolumn{1}{c|}{\cellcolor[HTML]{F3F3F3}\textbf{South}} &
  \multicolumn{1}{c|}{\cellcolor[HTML]{F3F3F3}\textbf{East}} &
  \multicolumn{1}{c|}{\cellcolor[HTML]{F3F3F3}\textbf{North}} &
  \cellcolor[HTML]{F3F3F3}\textbf{West} \\ \hline
\multicolumn{1}{|l|}{\cellcolor[HTML]{F3F3F3}\textbf{Front Bumper}} &
  \multicolumn{1}{c|}{\cellcolor[HTML]{FCECEA}1.07} &
  \multicolumn{1}{c|}{\cellcolor[HTML]{FEF9F9}0.33} &
  \multicolumn{1}{c|}{\cellcolor[HTML]{FEF7F6}0.47} &
  \cellcolor[HTML]{F8D9D6}2.07 \\ \hline
\multicolumn{1}{|l|}{\cellcolor[HTML]{F3F3F3}\textbf{Front Doors}} &
  \multicolumn{1}{c|}{\cellcolor[HTML]{F7D2CF}2.42} &
  \multicolumn{1}{c|}{\cellcolor[HTML]{FFFFFF}0.00} &
  \multicolumn{1}{c|}{\cellcolor[HTML]{FCEDEB}1.00} &
  \cellcolor[HTML]{FDF1F0}0.75 \\ \hline
\multicolumn{1}{|l|}{\cellcolor[HTML]{F3F3F3}\textbf{Front Fenders}} &
  \multicolumn{1}{c|}{\cellcolor[HTML]{F6D0CD}2.53} &
  \multicolumn{1}{c|}{\cellcolor[HTML]{FFFFFF}0.00} &
  \multicolumn{1}{c|}{\cellcolor[HTML]{FBE8E6}1.27} &
  \cellcolor[HTML]{FFFFFF}0.00 \\ \hline
\multicolumn{1}{|l|}{\cellcolor[HTML]{F3F3F3}\textbf{Front Windows}} &
  \multicolumn{1}{c|}{\cellcolor[HTML]{F8D9D6}2.08} &
  \multicolumn{1}{c|}{\cellcolor[HTML]{FFFFFF}0.00} &
  \multicolumn{1}{c|}{\cellcolor[HTML]{FBEAE8}1.17} &
  \cellcolor[HTML]{FCEBEA}1.08 \\ \hline
\multicolumn{1}{|l|}{\cellcolor[HTML]{F3F3F3}\textbf{Front Wheels}} &
  \multicolumn{1}{c|}{\cellcolor[HTML]{F3C0BB}3.40} &
  \multicolumn{1}{c|}{\cellcolor[HTML]{FFFFFF}0.00} &
  \multicolumn{1}{c|}{\cellcolor[HTML]{F6D0CD}2.53} &
  \cellcolor[HTML]{FDF3F2}0.67 \\ \hline
\multicolumn{1}{|l|}{\cellcolor[HTML]{F3F3F3}\textbf{Headlights}} &
  \multicolumn{1}{c|}{\cellcolor[HTML]{F3C0BB}3.40} &
  \multicolumn{1}{c|}{\cellcolor[HTML]{FFFFFF}0.00} &
  \multicolumn{1}{c|}{\cellcolor[HTML]{F7D4D1}2.33} &
  \cellcolor[HTML]{FFFFFF}0.00 \\ \hline
\multicolumn{1}{|l|}{\cellcolor[HTML]{F3F3F3}\textbf{Rear Wheels}} &
  \multicolumn{1}{c|}{\cellcolor[HTML]{ED9F99}5.13} &
  \multicolumn{1}{c|}{\cellcolor[HTML]{FFFFFF}0.00} &
  \multicolumn{1}{c|}{\cellcolor[HTML]{F0B0AA}4.27} &
  \cellcolor[HTML]{EEA49E}4.87 \\ \hline
\multicolumn{1}{|l|}{\cellcolor[HTML]{F3F3F3}\textbf{Side Mirrors}} &
  \multicolumn{1}{c|}{\cellcolor[HTML]{F5C6C2}3.07} &
  \multicolumn{1}{c|}{\cellcolor[HTML]{FFFFFF}0.00} &
  \multicolumn{1}{c|}{\cellcolor[HTML]{F8D6D3}2.21} &
  \cellcolor[HTML]{FDF5F4}0.57 \\ \hline
\multicolumn{1}{|l|}{\cellcolor[HTML]{F3F3F3}\textbf{Windshield}} &
  \multicolumn{1}{c|}{\cellcolor[HTML]{F5CAC6}2.86} &
  \multicolumn{1}{c|}{\cellcolor[HTML]{FFFFFF}0.00} &
  \multicolumn{1}{c|}{\cellcolor[HTML]{F6CECB}2.64} &
  \cellcolor[HTML]{FFFFFF}0.00 \\ \hline
\multicolumn{1}{|l|}{\cellcolor[HTML]{F3F3F3}\textbf{Windshield Rails}} &
  \multicolumn{1}{c|}{\cellcolor[HTML]{F9DBD9}1.93} &
  \multicolumn{1}{c|}{\cellcolor[HTML]{FFFFFF}0.00} &
  \multicolumn{1}{c|}{\cellcolor[HTML]{FBE9E7}1.21} &
  \cellcolor[HTML]{FFFEFE}0.07 \\ \hline
\end{tabular}
\label{tab:facing_summary}
\end{table}

Based on this data, we can determine position on the road plays a significant role in the general exposedness of a vehicle's elements. Across the three distance thresholds, vehicle parts were most consistently captured when the vehicle was on the lane adjacent to the pedestrian. This is explained by a lower amount of permutations and \textit{Blocking Vehicles} that could occlude the \textit{Target Vehicle}. These conditions are also the most most similar to those in \cite{troel-madec_ehmi_2019}, which is reflected on higher \textit{Exposedness Score} for side elements such as the wheels, doors, side mirrors, and fenders.

\subsection{Effects of Distance} \label{Effects of Distance}

\begin{figure}[!h]
    \centering
    \includegraphics[width=4in,clip,keepaspectratio]{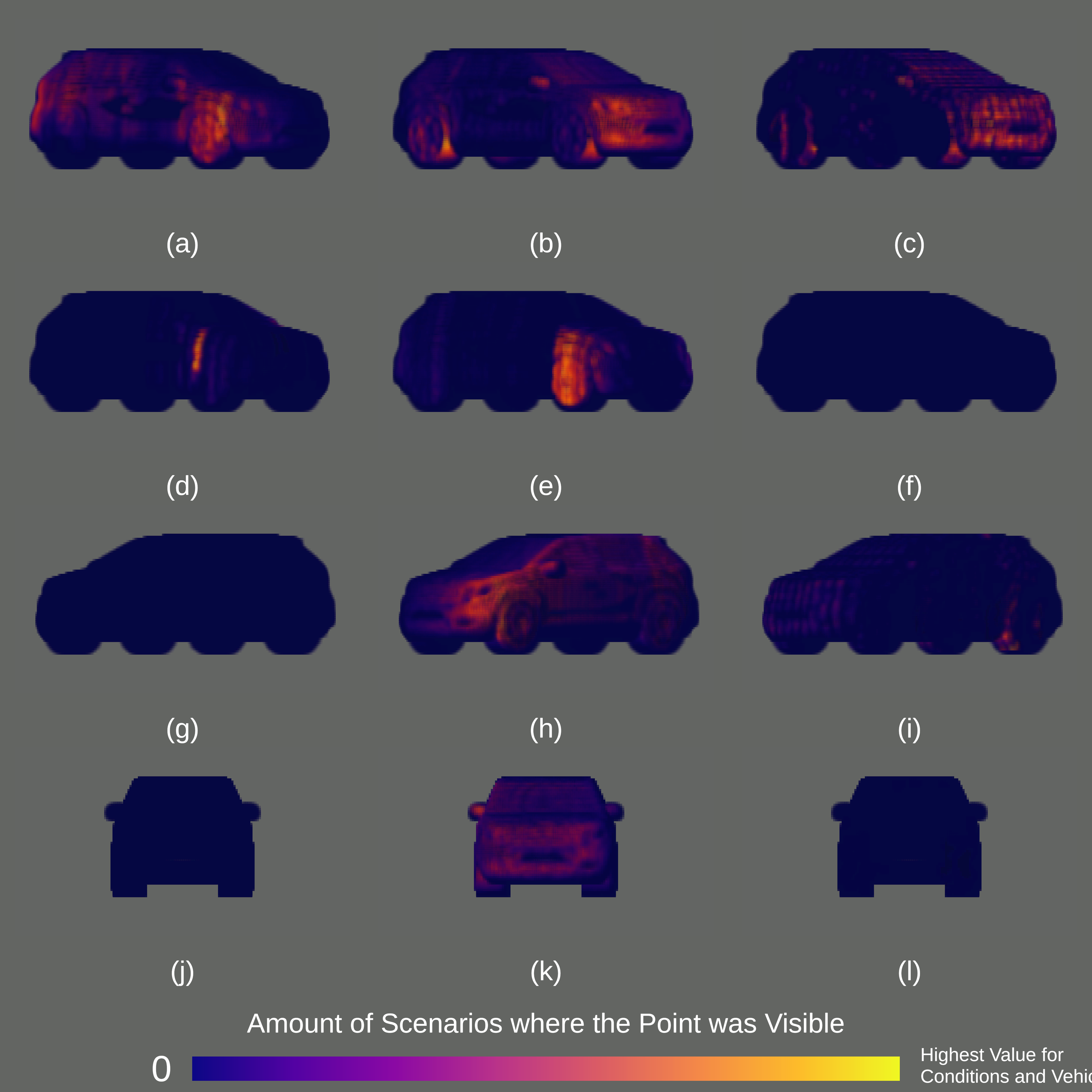}
    \caption{Example showing the heatmap of the SUV based on facing orientation and distance: south (a,b,c), east (d,e,f), south (g,h,i), and west (j,k,l). It also showcases differences based on the distance between the camera and the \textit{Data Points}: 0\si{m} to 5\si{m} (a,d,g,j), 5\si{m} to 25\si{m} (b,e,h,k), and 25\si{m} to 75\si{m} (c,f,i,l).}
    \label{fig:distance_comparison}
\end{figure}

By separating the results based on the distance thresholds we can analyze how this distance may affect the exposedness of certain elements. Fig. \ref{fig:distance_comparison} illustrates a plasma heatmap of the SUV vehicle type accross the three distance ranges: 0\si{m} to 5\si{m} (Fig. \ref{fig:distance_comparison}(a,d,g,j)), 5\si{m} to 25\si{m} (Fig. \ref{fig:distance_comparison}(b,e,h,k)), 25\si{m} to 75\si{m} (Fig. \ref{fig:distance_comparison}(c,f,i,l)).

By looking at Fig. \ref{fig:distance_comparison} and Table \ref{tab:top5_exposedness}, we can make the following observations:

\begin{itemize}
    \item \textbf{In the 0\si{m} to 5\si{m} range: }
        \begin{itemize}
            \item The five parts with the highest average registered \textit{Exposedness Score} are the front fenders (2.15), front wheels (1.97), side mirrors (1.14), front windows (1.10), and front doors (1.33).
            \item Whenever the vehicle is on one of the positions facing north or west, it is also further than 5\si{m} away from the camera, so they are rarely captured in the 0\si{m} to 5\si{m} range. 
        \end{itemize}
    \item \textbf{In the 5\si{m} to 25\si{m} range: }
        \begin{itemize}
            \item The five parts with the highest average registered \textit{Exposedness Score} are the front wheels (4.35), front fenders (4.07), headlights (3.73), side mirrors (3.72), and rear wheels (2.88).
            \item At this distance we see higher \textit{Exposedness Scores} on average and a wider distribution of higher values across different parts.
        \end{itemize}
    \item \textbf{In the 25\si{m} to 75\si{m} range: }
        \begin{itemize}
            \item The five parts with the highest average registered \textit{Exposedness Score} are the rear wheels (2.35), front wheels (1.48), headlights (1.43), front bumper (1.32), and windshield (1.28).
            \item On Fig. \ref{fig:distance_comparison} we see a noticeable drop in resolution in this range due to the space between points increasing proportionally with their distance from the camera.
            \item Since pedestrians tend to lose their willingness to cross the road when the vehicle is within 40\si{m} of them \cite{dey_gaze_2019}, this is a particularly significant range to examine.
        \end{itemize}
    \end{itemize}

\subsection{Effects of Vehicle Size} \label{Effects of Vehicle Size}

In order to determine how the different vehicles affect recommendations for eHMI placement, Table \ref{tab:vehicle_type_comparison} notes the average highest \textit{Exposedness Score} for the top five vehicle parts in each vehicle size category, across all facings. 

\begin{table}[]
\caption{The average Highest \textit{Exposedness Score} for the top five vehicle parts across each facing direction, grouped by vehicle size category.}
\label{tab:vehicle_type_comparison}
\begin{tabular}{|
>{\columncolor[HTML]{F3F3F3}}l cccc|}
\hline
\multicolumn{5}{|c|}{\cellcolor[HTML]{CCCCCC}\textbf{\begin{tabular}[c]{@{}c@{}}Average highest Exposedness Score found in part, \\ for each vehicle size category\end{tabular}}} \\ \hline
\multicolumn{5}{|c|}{\cellcolor[HTML]{D9D9D9}\textbf{For Data Points 0m to 5m from Camera}} \\ \hline
\multicolumn{1}{|l|}{\cellcolor[HTML]{EFEFEF}\textbf{Vehicle Part \& Size}} &
  \multicolumn{1}{c|}{\cellcolor[HTML]{F3F3F3}\textbf{S}} &
  \multicolumn{1}{c|}{\cellcolor[HTML]{F3F3F3}\textbf{M}} &
  \multicolumn{1}{c|}{\cellcolor[HTML]{F3F3F3}\textbf{L}} &
  \cellcolor[HTML]{F3F3F3}\textbf{XL} \\ \hline
\multicolumn{1}{|l|}{\cellcolor[HTML]{F3F3F3}\textbf{Front Bumper}} &
  \multicolumn{1}{c|}{\cellcolor[HTML]{FEF7F7}0.42} &
  \multicolumn{1}{c|}{\cellcolor[HTML]{FDF1F0}0.65} &
  \multicolumn{1}{c|}{\cellcolor[HTML]{FEFAF9}0.31} &
  \cellcolor[HTML]{FFFDFD}0.17 \\ \hline
\multicolumn{1}{|l|}{\cellcolor[HTML]{F3F3F3}\textbf{Front Doors}} &
  \multicolumn{1}{c|}{\cellcolor[HTML]{FFFBFB}0.25} &
  \multicolumn{1}{c|}{\cellcolor[HTML]{F9DDDB}1.45} &
  \multicolumn{1}{c|}{\cellcolor[HTML]{F8DAD8}1.56} &
  \cellcolor[HTML]{F9DCD9}1.50 \\ \hline
\multicolumn{1}{|l|}{\cellcolor[HTML]{F3F3F3}\textbf{Front Fenders}} &
  \multicolumn{1}{c|}{\cellcolor[HTML]{FBE6E5}1.08} &
  \multicolumn{1}{c|}{\cellcolor[HTML]{F3BEB9}2.70} &
  \multicolumn{1}{c|}{\cellcolor[HTML]{F8DAD8}1.56} &
  \cellcolor[HTML]{F1B4AF}3.08 \\ \hline
\multicolumn{1}{|l|}{\cellcolor[HTML]{F3F3F3}\textbf{Front Windows}} &
  \multicolumn{1}{c|}{\cellcolor[HTML]{F2B6B1}3.00} &
  \multicolumn{1}{c|}{\cellcolor[HTML]{FBEAE9}0.94} &
  \multicolumn{1}{c|}{\cellcolor[HTML]{FBE8E7}1.00} &
  \cellcolor[HTML]{F7D1CE}1.92 \\ \hline
\multicolumn{1}{|l|}{\cellcolor[HTML]{F3F3F3}\textbf{Front Wheels}} &
  \multicolumn{1}{c|}{\cellcolor[HTML]{FEF8F8}0.38} &
  \multicolumn{1}{c|}{\cellcolor[HTML]{FDF3F3}0.56} &
  \multicolumn{1}{c|}{\cellcolor[HTML]{FBE8E7}1.00} &
  \cellcolor[HTML]{FEF9F9}0.33 \\ \hline
\multicolumn{1}{|l|}{\cellcolor[HTML]{F3F3F3}\textbf{Headlights}} &
  \multicolumn{1}{c|}{\cellcolor[HTML]{FEF7F7}0.42} &
  \multicolumn{1}{c|}{\cellcolor[HTML]{FCEBEA}0.90} &
  \multicolumn{1}{c|}{\cellcolor[HTML]{FBEAE9}0.94} &
  \cellcolor[HTML]{F9DEDC}1.42 \\ \hline
\multicolumn{1}{|l|}{\cellcolor[HTML]{F3F3F3}\textbf{Rear Wheels}} &
  \multicolumn{1}{c|}{\cellcolor[HTML]{FDF3F2}0.58} &
  \multicolumn{1}{c|}{\cellcolor[HTML]{FDF1F0}0.65} &
  \multicolumn{1}{c|}{\cellcolor[HTML]{FDF2F1}0.63} &
  \cellcolor[HTML]{FDF1F0}0.67 \\ \hline
\multicolumn{1}{|l|}{\cellcolor[HTML]{F3F3F3}\textbf{Side Mirrors}} &
  \multicolumn{1}{c|}{\cellcolor[HTML]{FCECEA}0.88} &
  \multicolumn{1}{c|}{\cellcolor[HTML]{FBE6E4}1.10} &
  \multicolumn{1}{c|}{\cellcolor[HTML]{FBE8E7}1.00} &
  \cellcolor[HTML]{F6CFCC}2.00 \\ \hline
\multicolumn{1}{|l|}{\cellcolor[HTML]{F3F3F3}\textbf{Windshield}} &
  \multicolumn{1}{c|}{\cellcolor[HTML]{FFFBFB}0.25} &
  \multicolumn{1}{c|}{\cellcolor[HTML]{FDF1F0}0.65} &
  \multicolumn{1}{c|}{\cellcolor[HTML]{FDF2F1}0.63} &
  \cellcolor[HTML]{FCEFEE}0.75 \\ \hline
\multicolumn{1}{|l|}{\cellcolor[HTML]{F3F3F3}\textbf{Windshield Rails}} &
  \multicolumn{1}{c|}{\cellcolor[HTML]{FAE2E0}1.25} &
  \multicolumn{1}{c|}{\cellcolor[HTML]{F8D8D5}1.65} &
  \multicolumn{1}{c|}{\cellcolor[HTML]{FEF5F4}0.50} &
  \cellcolor[HTML]{FAE4E2}1.17 \\ \hline
\multicolumn{1}{|l|}{\cellcolor[HTML]{B7B7B7}\textbf{Average}} &
  \multicolumn{1}{c|}{\cellcolor[HTML]{FFFFFF}0.85} &
  \multicolumn{1}{c|}{\cellcolor[HTML]{99D6B8}1.13} &
  \multicolumn{1}{c|}{\cellcolor[HTML]{E8F6EF}0.91} &
  \cellcolor[HTML]{57BB8A}1.30 \\ \hline
\multicolumn{5}{|c|}{\cellcolor[HTML]{D9D9D9}\textbf{For Data Points 5m to 25m from Camera}} \\ \hline
\multicolumn{1}{|l|}{\cellcolor[HTML]{EFEFEF}\textbf{Vehicle Part \& Size}} &
  \multicolumn{1}{c|}{\cellcolor[HTML]{F3F3F3}\textbf{S}} &
  \multicolumn{1}{c|}{\cellcolor[HTML]{F3F3F3}\textbf{M}} &
  \multicolumn{1}{c|}{\cellcolor[HTML]{F3F3F3}\textbf{L}} &
  \cellcolor[HTML]{F3F3F3}\textbf{XL} \\ \hline
\multicolumn{1}{|l|}{\cellcolor[HTML]{F3F3F3}\textbf{Front Bumper}} &
  \multicolumn{1}{c|}{\cellcolor[HTML]{FFFFFF}0.08} &
  \multicolumn{1}{c|}{\cellcolor[HTML]{FDF1F0}0.65} &
  \multicolumn{1}{c|}{\cellcolor[HTML]{FAE2E0}1.25} &
  \cellcolor[HTML]{FFFDFD}0.17 \\ \hline
\multicolumn{1}{|l|}{\cellcolor[HTML]{F3F3F3}\textbf{Front Doors}} &
  \multicolumn{1}{c|}{\cellcolor[HTML]{F2B6B1}3.00} &
  \multicolumn{1}{c|}{\cellcolor[HTML]{F5C7C3}2.35} &
  \multicolumn{1}{c|}{\cellcolor[HTML]{F4C4C0}2.44} &
  \cellcolor[HTML]{FBE8E7}1.00 \\ \hline
\multicolumn{1}{|l|}{\cellcolor[HTML]{F3F3F3}\textbf{Front Fenders}} &
  \multicolumn{1}{c|}{\cellcolor[HTML]{EC9992}4.17} &
  \multicolumn{1}{c|}{\cellcolor[HTML]{E67C73}5.30} &
  \multicolumn{1}{c|}{\cellcolor[HTML]{F2BBB6}2.81} &
  \cellcolor[HTML]{EFA8A2}3.58 \\ \hline
\multicolumn{1}{|l|}{\cellcolor[HTML]{F3F3F3}\textbf{Front Windows}} &
  \multicolumn{1}{c|}{\cellcolor[HTML]{EFAAA4}3.50} &
  \multicolumn{1}{c|}{\cellcolor[HTML]{F3C0BB}2.63} &
  \multicolumn{1}{c|}{\cellcolor[HTML]{F8D9D6}1.63} &
  \cellcolor[HTML]{F3BDB8}2.75 \\ \hline
\multicolumn{1}{|l|}{\cellcolor[HTML]{F3F3F3}\textbf{Front Wheels}} &
  \multicolumn{1}{c|}{\cellcolor[HTML]{F8D6D3}1.75} &
  \multicolumn{1}{c|}{\cellcolor[HTML]{F8DAD8}1.56} &
  \multicolumn{1}{c|}{\cellcolor[HTML]{FCECEA}0.88} &
  \cellcolor[HTML]{FEF9F9}0.33 \\ \hline
\multicolumn{1}{|l|}{\cellcolor[HTML]{F3F3F3}\textbf{Headlights}} &
  \multicolumn{1}{c|}{\cellcolor[HTML]{EFA8A2}3.58} &
  \multicolumn{1}{c|}{\cellcolor[HTML]{EC9790}4.25} &
  \multicolumn{1}{c|}{\cellcolor[HTML]{F2BBB6}2.81} &
  \cellcolor[HTML]{EC9790}4.25 \\ \hline
\multicolumn{1}{|l|}{\cellcolor[HTML]{F3F3F3}\textbf{Rear Wheels}} &
  \multicolumn{1}{c|}{\cellcolor[HTML]{F6CBC8}2.17} &
  \multicolumn{1}{c|}{\cellcolor[HTML]{F3BDB8}2.75} &
  \multicolumn{1}{c|}{\cellcolor[HTML]{EC9891}4.19} &
  \cellcolor[HTML]{F6CDCA}2.08 \\ \hline
\multicolumn{1}{|l|}{\cellcolor[HTML]{F3F3F3}\textbf{Side Mirrors}} &
  \multicolumn{1}{c|}{\cellcolor[HTML]{EFA7A0}3.63} &
  \multicolumn{1}{c|}{\cellcolor[HTML]{ED9E98}3.95} &
  \multicolumn{1}{c|}{\cellcolor[HTML]{EEA59F}3.69} &
  \cellcolor[HTML]{EA8C84}4.67 \\ \hline
\multicolumn{1}{|l|}{\cellcolor[HTML]{F3F3F3}\textbf{Windshield}} &
  \multicolumn{1}{c|}{\cellcolor[HTML]{EFAAA4}3.50} &
  \multicolumn{1}{c|}{\cellcolor[HTML]{F6CECB}2.05} &
  \multicolumn{1}{c|}{\cellcolor[HTML]{F8DAD8}1.56} &
  \cellcolor[HTML]{EA8E87}4.58 \\ \hline
\multicolumn{1}{|l|}{\cellcolor[HTML]{F3F3F3}\textbf{Windshield Rails}} &
  \multicolumn{1}{c|}{\cellcolor[HTML]{F0B0AB}3.25} &
  \multicolumn{1}{c|}{\cellcolor[HTML]{F2B8B3}2.95} &
  \multicolumn{1}{c|}{\cellcolor[HTML]{F5C9C5}2.25} &
  \cellcolor[HTML]{F0ACA6}3.42 \\ \hline
\multicolumn{1}{|l|}{\cellcolor[HTML]{B7B7B7}\textbf{Average}} &
  \multicolumn{1}{c|}{\cellcolor[HTML]{57BB8A}2.86} &
  \multicolumn{1}{c|}{\cellcolor[HTML]{5EBE8F}2.84} &
  \multicolumn{1}{c|}{\cellcolor[HTML]{FFFFFF}2.35} &
  \cellcolor[HTML]{92D3B3}2.68 \\ \hline
\multicolumn{5}{|c|}{\cellcolor[HTML]{D9D9D9}\textbf{For Data Points 25m to 75m from Camera}} \\ \hline
\multicolumn{1}{|l|}{\cellcolor[HTML]{EFEFEF}\textbf{Vehicle Part \& Size}} &
  \multicolumn{1}{c|}{\cellcolor[HTML]{F3F3F3}\textbf{S}} &
  \multicolumn{1}{c|}{\cellcolor[HTML]{F3F3F3}\textbf{M}} &
  \multicolumn{1}{c|}{\cellcolor[HTML]{F3F3F3}\textbf{L}} &
  \cellcolor[HTML]{F3F3F3}\textbf{XL} \\ \hline
\multicolumn{1}{|l|}{\cellcolor[HTML]{F3F3F3}\textbf{Front Bumper}} &
  \multicolumn{1}{c|}{\cellcolor[HTML]{FFFDFD}0.17} &
  \multicolumn{1}{c|}{\cellcolor[HTML]{FDF0EF}0.70} &
  \multicolumn{1}{c|}{\cellcolor[HTML]{F7D1CE}1.94} &
  \cellcolor[HTML]{FBE8E7}1.00 \\ \hline
\multicolumn{1}{|l|}{\cellcolor[HTML]{F3F3F3}\textbf{Front Doors}} &
  \multicolumn{1}{c|}{\cellcolor[HTML]{FCEFEE}0.75} &
  \multicolumn{1}{c|}{\cellcolor[HTML]{FBEAE8}0.95} &
  \multicolumn{1}{c|}{\cellcolor[HTML]{FAE4E2}1.19} &
  \cellcolor[HTML]{FBE5E4}1.13 \\ \hline
\multicolumn{1}{|l|}{\cellcolor[HTML]{F3F3F3}\textbf{Front Fenders}} &
  \multicolumn{1}{c|}{\cellcolor[HTML]{FEF5F4}0.50} &
  \multicolumn{1}{c|}{\cellcolor[HTML]{F9DEDC}1.40} &
  \multicolumn{1}{c|}{\cellcolor[HTML]{FCEDEC}0.81} &
  \cellcolor[HTML]{FCEDEB}0.83 \\ \hline
\multicolumn{1}{|l|}{\cellcolor[HTML]{F3F3F3}\textbf{Front Windows}} &
  \multicolumn{1}{c|}{\cellcolor[HTML]{FEF5F4}0.50} &
  \multicolumn{1}{c|}{\cellcolor[HTML]{FDF0EF}0.69} &
  \multicolumn{1}{c|}{\cellcolor[HTML]{F8DAD8}1.56} &
  \cellcolor[HTML]{FAE4E2}1.17 \\ \hline
\multicolumn{1}{|l|}{\cellcolor[HTML]{F3F3F3}\textbf{Front Wheels}} &
  \multicolumn{1}{c|}{\cellcolor[HTML]{FBE5E4}1.13} &
  \multicolumn{1}{c|}{\cellcolor[HTML]{F9DFDD}1.38} &
  \multicolumn{1}{c|}{\cellcolor[HTML]{F9DCD9}1.50} &
  \cellcolor[HTML]{F6CDCA}2.08 \\ \hline
\multicolumn{1}{|l|}{\cellcolor[HTML]{F3F3F3}\textbf{Headlights}} &
  \multicolumn{1}{c|}{\cellcolor[HTML]{FBE8E7}1.00} &
  \multicolumn{1}{c|}{\cellcolor[HTML]{F8D7D4}1.70} &
  \multicolumn{1}{c|}{\cellcolor[HTML]{FBE7E5}1.06} &
  \cellcolor[HTML]{F7D1CE}1.92 \\ \hline
\multicolumn{1}{|l|}{\cellcolor[HTML]{F3F3F3}\textbf{Rear Wheels}} &
  \multicolumn{1}{c|}{\cellcolor[HTML]{F6CFCC}2.00} &
  \multicolumn{1}{c|}{\cellcolor[HTML]{EC9992}4.15} &
  \multicolumn{1}{c|}{\cellcolor[HTML]{EA8F87}4.56} &
  \cellcolor[HTML]{F2BAB6}2.83 \\ \hline
\multicolumn{1}{|l|}{\cellcolor[HTML]{F3F3F3}\textbf{Side Mirrors}} &
  \multicolumn{1}{c|}{\cellcolor[HTML]{FCECEA}0.88} &
  \multicolumn{1}{c|}{\cellcolor[HTML]{FAE5E3}1.15} &
  \multicolumn{1}{c|}{\cellcolor[HTML]{F8D9D6}1.63} &
  \cellcolor[HTML]{F6CBC8}2.17 \\ \hline
\multicolumn{1}{|l|}{\cellcolor[HTML]{F3F3F3}\textbf{Windshield}} &
  \multicolumn{1}{c|}{\cellcolor[HTML]{FAE2E0}1.25} &
  \multicolumn{1}{c|}{\cellcolor[HTML]{FAE3E2}1.20} &
  \multicolumn{1}{c|}{\cellcolor[HTML]{FBE8E7}1.00} &
  \cellcolor[HTML]{F5C9C5}2.25 \\ \hline
\multicolumn{1}{|l|}{\cellcolor[HTML]{F3F3F3}\textbf{Windshield Rails}} &
  \multicolumn{1}{c|}{\cellcolor[HTML]{FBE8E7}1.00} &
  \multicolumn{1}{c|}{\cellcolor[HTML]{FBEAE8}0.95} &
  \multicolumn{1}{c|}{\cellcolor[HTML]{FCEFEE}0.75} &
  \cellcolor[HTML]{FEF5F4}0.50 \\ \hline
\multicolumn{1}{|l|}{\cellcolor[HTML]{B7B7B7}\textbf{Average}} &
  \multicolumn{1}{c|}{\cellcolor[HTML]{FFFFFF}0.92} &
  \multicolumn{1}{c|}{\cellcolor[HTML]{82CDA8}1.43} &
  \multicolumn{1}{c|}{\cellcolor[HTML]{57BB8A}1.60} &
  \cellcolor[HTML]{5BBD8D}1.59 \\ \hline
\end{tabular}
\end{table}

Based on this information we can make the following statements of note:
\begin{itemize}
    \item Small vehicles capture less \textit{Data Points} in the 25\si{m} to 75\si{m} distance range, as well as the 0\si{m} to 5\si{m} range.
    \item Large vehicles have lower \textit{Exposedness Scores} in the 5\si{m} to 25\si{m} distance range.
    \item In the 25\si{m} to 75\si{m} distance range, the front fenders have the highest score in small and medium vehicles, the rear wheels in the large vehicles, and the side mirrors in extra large vehicles.
    \item Across all four vehicle size categories, the rear wheels have the highest \textit{Exposedness Score} in the 25\si{m} to 75\si{m}. This is followed in second by the windshield for small and extra large vehicles, the headlights for medium vehicles, and the front bumper for large vehicles. 
\end{itemize}

\subsection{Effects of Vehicle Type} \label{Effects of Vehicle type}

\begin{figure}[!h]
    \centering
    \includegraphics[width=4in,clip,keepaspectratio]{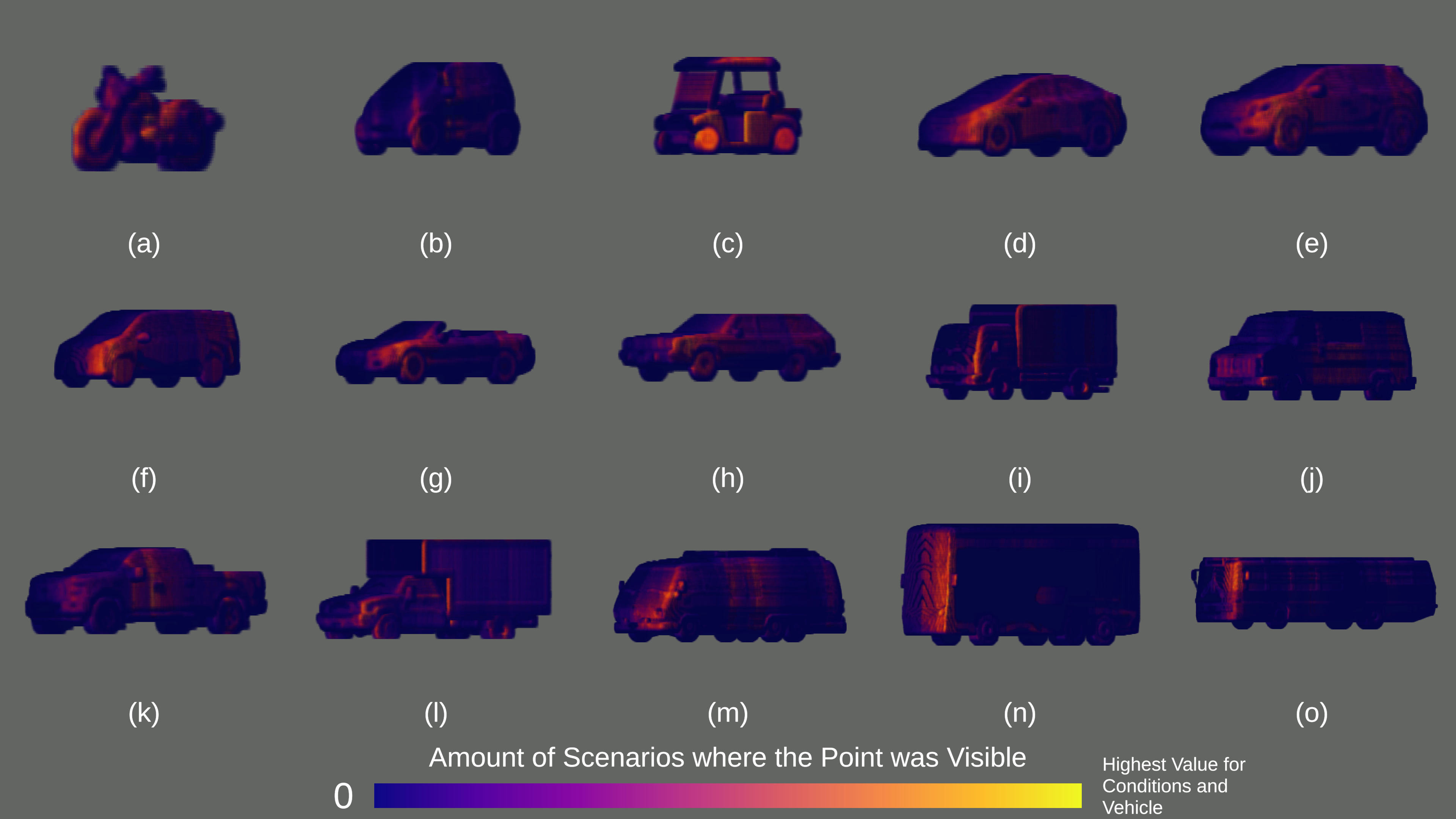}
    \caption{Comparison of the heatmaps produced by each of the fifteen vehicle types when in a north-facing position, in the 5\si{m} to 25\si{m} distance range. (a) Motorcycle, (b) Smart Car, (c) Carryall Car, (d) Sedan, (e) SUV, (f) Panel Van, (g) Topless Convertible, (h) Station Wagon, (i) Four Wheel Truck, (j) Minibus, (k) Pickup Truck, (l) Moving Truck, (m) Motor Home, (n) Double Decker Bus, (o) Single Decker bus.}
    \label{fig:vehicle_type_comparison}
\end{figure}

Fig. \ref{fig:vehicle_type_comparison} shows a comparison of the fifteen heatmaps produced by the vehicles when facing north, and using only \textit{Data Points} between 5\si{m} to 25\si{m} away from the camera. Note the following:
\begin{itemize}
    \item The frontal-right corner of the vehicle is consistently the most visible across all types in this configuration.
    \item The windshield only shows relevance for the motor home (Fig. \ref{fig:vehicle_type_comparison}(m)), single decker bus (Fig. \ref{fig:vehicle_type_comparison}(n)), and double decker bus (Fig. \ref{fig:vehicle_type_comparison}(o)). 
\end{itemize}

\subsection{Parts Visible Most Often}

As seen on Appendix \ref{tab:0-5Results}, \ref{tab:5-25Results}, and \ref{tab:25-75Results}, among the twenty-nice individual vehicle parts illustrated in Fig. \ref{fig:vehicle_elements}, only seven are present in all vehicle types. Of those, the front wheels, rear wheels, front fenders, and headlights were among the highest-scored elements in Table \ref{tab:top5_exposedness}, pointing to these parts as the ones most visible most often.

\subsection{Limitations of Study} \label{Limitations of Study}
Since this is a study that makes use of ray-casting to determine the exposedness of objects, it does not account for non-geometrical factors that may affect how well a pedestrian may perceive an element of a vehicle. This includes reflections, such as the one from glass or chrome details, as well as the color and finish of the vehicle. This study also does not account for other visual difficulties such as poor weather, low light conditions, or visual imparity.

The scope of this work also limits the complexity of its environment. We did not feature other obstacles that may interfere with the visibility of certain elements, such as parked vehicles, kiosks, or parked bikes. Each side of the road is composed of a single lane as well, so there is no data for how vehicles may occlude other vehicles driving in the same direction.  

Additionally, since the program makes use of models freely available on Sketchfab, the level of detail between them is not consistent, and some are of much lower poly-count than others. This is relevant, as the simulation relies entirely on the geometry of the objects, and a lower level of detail may signify much less clear results for some of the vehicle types. 

Finally, this study was done entirely with static vehicles. While the number of \textit{Camera Positions} and \textit{Camera Directions} was chosen to simulate the different possible angles that a pedestrian may encounter a vehicle, it does not fully replicate the subtle changes in vehicle rotation or motion that may alter the perspective of the viewer. 

\subsection{Discussion} \label{Discussion}

The conclusion by Troel-Madec et al. \cite{troel-madec_ehmi_2019} that the sides of vehicles are a good placement for eHMIs is backed by the present research, as the wheels and fenders are lateral elements with high \textit{Exposedness Scores}. These results also align with the findings that moving elements of the vehicle are points of interest for pedestrians \cite{de_winter_how_2020}. Wheels, however, can be obstructed by high sidewalks and puddles, not to mention the difficulties and costs of placing a display on the part of such a part, so this is not a recommended placement in most situations.

Another noteworthy exterior part are the headlights, which already feature a signaling device in the form of turning lights. This is consistent with findings that pedestrians focus on the front of the vehicle during crossing events \cite{de_winter_how_2020, dey_gaze_2019, guo_video-based_2022}, but raises concerns about visual distinction and overload.

Noticeably, the \textit{Exposedness Score} of the windshield only stands out once when the vehicle is in the 25\si{m} to 75\si{m} range, which does bring attention to the observation that pedestrians lose the confidence to cross a road as a vehicle approaches within 40\si{m}\cite{dey_gaze_2019} of them. This is noteworthy, as previous studies have consistently found that the gaze of pedestrians naturally gravitate towards the windshield of vehicles, while our study reveals that it is also an area easily occluded by other elements on the road.

Another factor to consider is how the effects of distance as observed in Fig. \ref{fig:distance_comparison}, inform us of a need for large displays on autonomous vehicles. This means that we must prioritize placement of eHMIs that allow us to implement larger displays, particularly for smaller vehicles.

Our results suggest that a distributed approach to eHMI placement is required to fulfill both the behavioral response of pedestrians and ensure that signaling is as visible as often as possible. Additionally, while it is preferable to standardize the placement of eHMIs to prevent pedestrian confusion \cite{de_winter_how_2020}, the diversity of multi-purpose vehicles on the road requires that we make specific recommendations for cases where this isn't possible. Based on the existing research and the results presented, the recommended placement for eHMIs would be at least into one of the following pairs, listed in no particular order:

\begin{itemize}
    \item \textbf{Windshield and front fenders:} Provides three large surfaces to place displays. Works for vehicle types such as sedans, SUVs, panel vans, topless convertibles, station wagons, minibuses, pickup trucks, and motor homes.
    \item \textbf{Windshield and side mirrors:} Provides one large surface to place a display and two secondary smaller ones. Works for vehicle types such as smart cars, sedans, SUVs, panel vans, topless convertibles, station wagons, four wheel trucks, minibuses, pickup trucks, moving trucks, motor homes, and single or double decker buses. 
    \item \textbf{Front fender(s) and side mirrors:} Provides one large surface to place a display and two secondary smaller ones, none of which are usually gazed by pedestrians. Works for vehicle types such as motorcycles, sedans, SUVs, panel vans, topless convertibles, station wagons, minibuses, pickup trucks, motor homes, and single or double decker buses. 
\end{itemize}

\section{Future Work}

This paper aims to generate knowledge on the general exposedness of different parts of the vehicles tested. The results presented were obtained under specific circumstances, and more specific patterns may arise if changes are made to it. 

Some future iterations of this experiment may find data for left-hand driving, find the exposedness heat map of vehicles in respect to other drivers, evaluate the effect of disability on the results (e.g. wheelchair height, visual impairment), include autonomous vehicle models and evaluate parts associated with them (e.g. sensor pods), evaluate the back of the vehicles, or add different obstacles such as traffic signs and parked vehicles. Different layouts could also be tested, making changes to the currently used one such as widening the sidewalk, adding biking lanes or parked vehicles between the vehicles and the sidewalk, increasing the amount of lanes, adding a traffic light intersection, raising the sidewalk et cetera.

This study is entirely limited to measuring the exposedness of the geometry of objects, and does not test other factors that might compromise visibility such as color and reflections. The software used is also entirely static, and does not feature an animation system that might affect the results, or an in-depth fine-tuning system to adjust angles in other than 45 \si{degrees} intervals. Additional work may include testing different exposedness conditions, testing the recommended external human-machine interface placements in an virtual reality experiment, or introducing an animation of a vehicle in motion to collect data on how its exposedness changes. 

Future work can also be done to design an experiment capable of measuring the value of points on a given part and produce a total for each, instead of requiring manual analysis which introduces the possibility of human error and uses only the maximum values found. This would also allow us to get more granular data to find the specific areas of a vehicle element where an eHMI can be placed. 

Finally, the resolution of the capture methodology used for this software limits the accuracy of the results, particularly at long distances. A future study may increase the resolution of the data capture, or attempt a different method altogether that is capable of processing more data faster and with more detail.

\section{Conclusion} \label{Conclusion}
In this paper, we used a virtual simulation to study fifteen different vehicles of drastically different shapes and sizes, and find which of their parts are most often unobstructed to pedestrians in a two-way two-lane right-driving street with sidewalks. The results led us to propose a distributed approach to external Human-Machine Interface (eHMI) implementation, with three recommended pairings of forward-facing exterior surfaces: windshield and front fenders, windshield and side mirrors, and front fenders and side mirrors.

This correlates with the results and observations of previous works. Additional testing might prove useful to better understand and contextualize these findings, but the data revealed in this experiment should be valuable to design better forms of signaling in our modern roads that can be seen by their intended audience, especially as we steadily keep progressing to a future where autonomous vehicles roam our cities.

\section{Declarations}
\subsection{Funding}
Jaerock Kwon is an associate professor for University of Michigan Dearborn. Jose Gonzalez-Belmonte is an assistant professor for Lawrence Technological University (LTU), and his ongoing doctoral studies at University of Michigan Dearborn are being partially funded by LTU. The authors did not receive support from any other organization for the submitted work. 
\subsection{Conflict of Interest}
The author of this paper declares that they have no known competing financial interests or personal relationships that could have appeared to influence the work reported in this paper.
\subsection{Ethics approval and consent to participate}
Not applicable.
\subsection{Consent for publication}
Not applicable.
\subsection{Data availability }
The data produced and used for this paper can be found at https://jgonzalez-uom.github.io/exposedness/
\subsection{Materials availability}
The software used for this paper, along with a manual to use it, can be found at https://jgonzalez-uom.github.io/exposedness/
\subsection{Code availability }
The source code of the software used for this paper can be found at https://github.com/bimilab/paper-exposedness-study-for-ehmi
The latest commit for the repository as of the time of writing was made on February 12 at 1:12AM. The code is fully functional.
\subsection{Author contribution}
All authors contributed to the study conception and design. Material preparation, software development, data collection, and analysis were performed by Jose Gonzalez-Belmonte. The first draft of the manuscript was written by Jose Gonzalez-Belmonte, and Jaerock Kwon commented and suggested edits on previous versions of the manuscript. All authors have read and approved the final manuscript.  
\subsection{Acknowledgements}
Not applicable.






\begin{appendices}

\section{Highest Exposedness Score of each Vehicle Part, for each facing, for each Vehicle Type, for each distance thresholds}\label{secA1}

\begin{table}[h]
\centering
\resizebox{\columnwidth}{!}{%
\begin{tabular}{|llllllllllllllllr|}
\hline
\rowcolor[HTML]{FFFFFF} 
\multicolumn{17}{|c|}{\cellcolor[HTML]{FFFFFF}\textbf{\begin{tabular}[c]{@{}c@{}}Highest registered exposedness index on any point between 0 and 5 meters from the camera that is found on the part when the vehicle is in each of the\\ four facings (South East North West).\end{tabular}}} \\ \hline
\rowcolor[HTML]{FFFFFF} 
\multicolumn{1}{|r|}{\cellcolor[HTML]{FFFFFF}\textbf{\begin{tabular}[c]{@{}r@{}}Vehicle \\ Type \&\\  Part\end{tabular}}} &
  \multicolumn{1}{c|}{\cellcolor[HTML]{FFFFFF}\textbf{\begin{tabular}[c]{@{}c@{}}Carryall\\ Car\end{tabular}}} &
  \multicolumn{1}{c|}{\cellcolor[HTML]{FFFFFF}\textbf{\begin{tabular}[c]{@{}c@{}}Double\\ Decker\\ Bus\end{tabular}}} &
  \multicolumn{1}{c|}{\cellcolor[HTML]{FFFFFF}\textbf{\begin{tabular}[c]{@{}c@{}}Four\\ Wheel\\ Truck\end{tabular}}} &
  \multicolumn{1}{c|}{\cellcolor[HTML]{FFFFFF}{\ul \textbf{Minibus}}} &
  \multicolumn{1}{c|}{\cellcolor[HTML]{FFFFFF}\textbf{\begin{tabular}[c]{@{}c@{}}Motor-\\ cycle\end{tabular}}} &
  \multicolumn{1}{c|}{\cellcolor[HTML]{FFFFFF}\textbf{\begin{tabular}[c]{@{}c@{}}Motor\\ Home\end{tabular}}} &
  \multicolumn{1}{c|}{\cellcolor[HTML]{FFFFFF}\textbf{\begin{tabular}[c]{@{}c@{}}Moving\\ Truck\end{tabular}}} &
  \multicolumn{1}{c|}{\cellcolor[HTML]{FFFFFF}\textbf{\begin{tabular}[c]{@{}c@{}}Panel\\ Van\end{tabular}}} &
  \multicolumn{1}{c|}{\cellcolor[HTML]{FFFFFF}\textbf{\begin{tabular}[c]{@{}c@{}}Pickup\\ Truck\end{tabular}}} &
  \multicolumn{1}{c|}{\cellcolor[HTML]{FFFFFF}{\ul \textbf{Sedan}}} &
  \multicolumn{1}{c|}{\cellcolor[HTML]{FFFFFF}\textbf{\begin{tabular}[c]{@{}c@{}}Single\\ Decker\\ Bus\end{tabular}}} &
  \multicolumn{1}{c|}{\cellcolor[HTML]{FFFFFF}\textbf{\begin{tabular}[c]{@{}c@{}}Smart\\ Car\end{tabular}}} &
  \multicolumn{1}{c|}{\cellcolor[HTML]{FFFFFF}\textbf{\begin{tabular}[c]{@{}c@{}}Station\\ Wagon\end{tabular}}} &
  \multicolumn{1}{c|}{\cellcolor[HTML]{FFFFFF}{\ul \textbf{SUV}}} &
  \multicolumn{1}{c|}{\cellcolor[HTML]{FFFFFF}\textbf{\begin{tabular}[c]{@{}c@{}}Topless \\ Conver-\\ tible\end{tabular}}} &
  \multicolumn{1}{c|}{\cellcolor[HTML]{FFFFFF}\textbf{\begin{tabular}[c]{@{}c@{}}Avg. \\ Highest\end{tabular}}} \\ \hline
\multicolumn{1}{|l|}{\cellcolor[HTML]{FFFFFF}\textbf{\begin{tabular}[c]{@{}l@{}}Back \\ Bumper\end{tabular}}} &
  \multicolumn{1}{l|}{\cellcolor[HTML]{FFE8E7}3 0 0 0} &
  \multicolumn{1}{l|}{\cellcolor[HTML]{FFF9F9}0 1 0 0} &
  \multicolumn{1}{l|}{\cellcolor[HTML]{FFFFFF}0 0 0 0} &
  \multicolumn{1}{l|}{\cellcolor[HTML]{FFDEDE}4 0 0 0} &
  \multicolumn{1}{l|}{\cellcolor[HTML]{FFFFFF}0 0 0 0} &
  \multicolumn{1}{l|}{\cellcolor[HTML]{FFF9F9}1 0 0 0} &
  \multicolumn{1}{l|}{\cellcolor[HTML]{FFFFFF}0 0 0 0} &
  \multicolumn{1}{l|}{\cellcolor[HTML]{FFFFFF}0 0 0 0} &
  \multicolumn{1}{l|}{\cellcolor[HTML]{FFF9F9}1 0 0 0} &
  \multicolumn{1}{l|}{\cellcolor[HTML]{FFE8E7}3 0 0 0} &
  \multicolumn{1}{l|}{\cellcolor[HTML]{FFFFFF}0 0 0 0} &
  \multicolumn{1}{l|}{\cellcolor[HTML]{FFF0F0}1 1 0 0} &
  \multicolumn{1}{l|}{\cellcolor[HTML]{FFD5D5}5 0 0 0} &
  \multicolumn{1}{l|}{\cellcolor[HTML]{FFF9F9}1 0 0 0} &
  \multicolumn{1}{l|}{\cellcolor[HTML]{FFDEDE}4 0 0 0} &
  \cellcolor[HTML]{DFF2E9}{\ul \textbf{0.42}} \\ \hline
\rowcolor[HTML]{FFFFFF} 
\multicolumn{1}{|l|}{\cellcolor[HTML]{FFFFFF}\textbf{\begin{tabular}[c]{@{}l@{}}Back \\ Central\\ Light\end{tabular}}} &
  \multicolumn{1}{l|}{\cellcolor[HTML]{FFFFFF}NA} &
  \multicolumn{1}{l|}{\cellcolor[HTML]{FFFFFF}0 0 0 0} &
  \multicolumn{1}{l|}{\cellcolor[HTML]{FFFFFF}NA} &
  \multicolumn{1}{l|}{\cellcolor[HTML]{FFFFFF}0 0 0 0} &
  \multicolumn{1}{l|}{\cellcolor[HTML]{FFFFFF}NA} &
  \multicolumn{1}{l|}{\cellcolor[HTML]{FFFFFF}0 0 0 0} &
  \multicolumn{1}{l|}{\cellcolor[HTML]{FFFFFF}NA} &
  \multicolumn{1}{l|}{\cellcolor[HTML]{FFFFFF}0 0 0 0} &
  \multicolumn{1}{l|}{\cellcolor[HTML]{FFFFFF}0 0 0 0} &
  \multicolumn{1}{l|}{\cellcolor[HTML]{FFFFFF}0 0 0 0} &
  \multicolumn{1}{l|}{\cellcolor[HTML]{FFFFFF}0 0 0 0} &
  \multicolumn{1}{l|}{\cellcolor[HTML]{FFFFFF}0 0 0 0} &
  \multicolumn{1}{l|}{\cellcolor[HTML]{FFFFFF}0 0 0 0} &
  \multicolumn{1}{l|}{\cellcolor[HTML]{FFFFFF}0 0 0 0} &
  \multicolumn{1}{l|}{\cellcolor[HTML]{FFFFFF}NA} &
  {\ul \textbf{0}} \\ \hline
\multicolumn{1}{|l|}{\cellcolor[HTML]{FFFFFF}\textbf{\begin{tabular}[c]{@{}l@{}}Back \\ Fenders\end{tabular}}} &
  \multicolumn{1}{l|}{\cellcolor[HTML]{FFDEDE}4 0 0 0} &
  \multicolumn{1}{l|}{\cellcolor[HTML]{FFC3C3}7 0 0 0} &
  \multicolumn{1}{l|}{\cellcolor[HTML]{FFFFFF}NA} &
  \multicolumn{1}{l|}{\cellcolor[HTML]{FFC3C3}7 0 0 0} &
  \multicolumn{1}{l|}{\cellcolor[HTML]{FFDEDE}3 1 0 0} &
  \multicolumn{1}{l|}{\cellcolor[HTML]{FFF9F9}1 0 0 0} &
  \multicolumn{1}{l|}{\cellcolor[HTML]{FFE8E7}3 0 0 0} &
  \multicolumn{1}{l|}{\cellcolor[HTML]{FFC3C3}7 0 0 0} &
  \multicolumn{1}{l|}{\cellcolor[HTML]{FFDEDE}4 0 0 0} &
  \multicolumn{1}{l|}{\cellcolor[HTML]{FFE8E7}3 0 0 0} &
  \multicolumn{1}{l|}{\cellcolor[HTML]{FFF9F9}1 0 0 0} &
  \multicolumn{1}{l|}{\cellcolor[HTML]{FFDEDE}2 2 0 0} &
  \multicolumn{1}{l|}{\cellcolor[HTML]{FFC3C3}7 0 0 0} &
  \multicolumn{1}{l|}{\cellcolor[HTML]{FFDEDE}4 0 0 0} &
  \multicolumn{1}{l|}{\cellcolor[HTML]{FFE8E7}3 0 0 0} &
  \cellcolor[HTML]{B3E0CA}{\ul \textbf{0.98}} \\ \hline
\rowcolor[HTML]{FFFFFF} 
\multicolumn{1}{|l|}{\cellcolor[HTML]{FFFFFF}\textbf{\begin{tabular}[c]{@{}l@{}}Back Low \\ Reflector\end{tabular}}} &
  \multicolumn{1}{l|}{\cellcolor[HTML]{FFFFFF}NA} &
  \multicolumn{1}{l|}{\cellcolor[HTML]{FFFFFF}0 0 0 0} &
  \multicolumn{1}{l|}{\cellcolor[HTML]{FFF9F9}1 0 0 0} &
  \multicolumn{1}{l|}{\cellcolor[HTML]{FFFFFF}0 0 0 0} &
  \multicolumn{1}{l|}{\cellcolor[HTML]{FFFFFF}NA} &
  \multicolumn{1}{l|}{\cellcolor[HTML]{FFFFFF}0 0 0 0} &
  \multicolumn{1}{l|}{\cellcolor[HTML]{FFFFFF}NA} &
  \multicolumn{1}{l|}{\cellcolor[HTML]{FFFFFF}0 0 0 0} &
  \multicolumn{1}{l|}{\cellcolor[HTML]{FFF9F9}1 0 0 0} &
  \multicolumn{1}{l|}{\cellcolor[HTML]{FFFFFF}0 0 0 0} &
  \multicolumn{1}{l|}{\cellcolor[HTML]{FFFFFF}0 0 0 0} &
  \multicolumn{1}{l|}{\cellcolor[HTML]{FFFFFF}0 0 0 0} &
  \multicolumn{1}{l|}{\cellcolor[HTML]{FFFFFF}0 0 0 0} &
  \multicolumn{1}{l|}{\cellcolor[HTML]{FFFFFF}0 0 0 0} &
  \multicolumn{1}{l|}{\cellcolor[HTML]{FFFFFF}0 0 0 0} &
  \cellcolor[HTML]{FDFEFE}{\ul \textbf{0.03}} \\ \hline
\rowcolor[HTML]{FFFFFF} 
\multicolumn{1}{|l|}{\cellcolor[HTML]{FFFFFF}{\ul \textbf{Back Plate}}} &
  \multicolumn{1}{l|}{\cellcolor[HTML]{FFFFFF}0 0 0 0} &
  \multicolumn{1}{l|}{\cellcolor[HTML]{FFFFFF}0 0 0 0} &
  \multicolumn{1}{l|}{\cellcolor[HTML]{FFFFFF}0 0 0 0} &
  \multicolumn{1}{l|}{\cellcolor[HTML]{FFFFFF}0 0 0 0} &
  \multicolumn{1}{l|}{\cellcolor[HTML]{FFFFFF}0 0 0 0} &
  \multicolumn{1}{l|}{\cellcolor[HTML]{FFFFFF}0 0 0 0} &
  \multicolumn{1}{l|}{\cellcolor[HTML]{FFFFFF}0 0 0 0} &
  \multicolumn{1}{l|}{\cellcolor[HTML]{FFFFFF}0 0 0 0} &
  \multicolumn{1}{l|}{\cellcolor[HTML]{FFFFFF}0 0 0 0} &
  \multicolumn{1}{l|}{\cellcolor[HTML]{FFFFFF}0 0 0 0} &
  \multicolumn{1}{l|}{\cellcolor[HTML]{FFFFFF}0 0 0 0} &
  \multicolumn{1}{l|}{\cellcolor[HTML]{FFFFFF}0 0 0 0} &
  \multicolumn{1}{l|}{\cellcolor[HTML]{FFFFFF}0 0 0 0} &
  \multicolumn{1}{l|}{\cellcolor[HTML]{FFFFFF}0 0 0 0} &
  \multicolumn{1}{l|}{\cellcolor[HTML]{FFFFFF}0 0 0 0} &
  {\ul \textbf{0}} \\ \hline
\multicolumn{1}{|l|}{\cellcolor[HTML]{FFFFFF}\textbf{\begin{tabular}[c]{@{}l@{}}Back\\ Window\\ Rails\end{tabular}}} &
  \multicolumn{1}{l|}{\cellcolor[HTML]{FFF9F9}1 0 0 0} &
  \multicolumn{1}{l|}{\cellcolor[HTML]{FFE8E7}3 0 0 0} &
  \multicolumn{1}{l|}{\cellcolor[HTML]{FFFFFF}NA} &
  \multicolumn{1}{l|}{\cellcolor[HTML]{FFC3C3}7 0 0 0} &
  \multicolumn{1}{l|}{\cellcolor[HTML]{FFFFFF}NA} &
  \multicolumn{1}{l|}{\cellcolor[HTML]{FFF9F9}1 0 0 0} &
  \multicolumn{1}{l|}{\cellcolor[HTML]{FFFFFF}NA} &
  \multicolumn{1}{l|}{\cellcolor[HTML]{FFE8E7}3 0 0 0} &
  \multicolumn{1}{l|}{\cellcolor[HTML]{FFF9F9}1 0 0 0} &
  \multicolumn{1}{l|}{\cellcolor[HTML]{FFF0F0}2 0 0 0} &
  \multicolumn{1}{l|}{\cellcolor[HTML]{FFFFFF}0 0 0 0} &
  \multicolumn{1}{l|}{\cellcolor[HTML]{FFF0F0}1 1 0 0} &
  \multicolumn{1}{l|}{\cellcolor[HTML]{FFE8E7}3 0 0 0} &
  \multicolumn{1}{l|}{\cellcolor[HTML]{FFF9F9}1 0 0 0} &
  \multicolumn{1}{l|}{\cellcolor[HTML]{FFFFFF}NA} &
  \cellcolor[HTML]{E0F3EA}{\ul \textbf{0.4}} \\ \hline
\rowcolor[HTML]{FFFFFF} 
\multicolumn{1}{|l|}{\cellcolor[HTML]{FFFFFF}\textbf{\begin{tabular}[c]{@{}l@{}}Back\\ Window\end{tabular}}} &
  \multicolumn{1}{l|}{\cellcolor[HTML]{FFFFFF}NA} &
  \multicolumn{1}{l|}{\cellcolor[HTML]{FFFFFF}0 0 0 0} &
  \multicolumn{1}{l|}{\cellcolor[HTML]{FFFFFF}NA} &
  \multicolumn{1}{l|}{\cellcolor[HTML]{FFFFFF}0 0 0 0} &
  \multicolumn{1}{l|}{\cellcolor[HTML]{FFFFFF}NA} &
  \multicolumn{1}{l|}{\cellcolor[HTML]{FFFFFF}0 0 0 0} &
  \multicolumn{1}{l|}{\cellcolor[HTML]{FFFFFF}NA} &
  \multicolumn{1}{l|}{\cellcolor[HTML]{FFFFFF}NA} &
  \multicolumn{1}{l|}{\cellcolor[HTML]{FFE8E7}3 0 0 0} &
  \multicolumn{1}{l|}{\cellcolor[HTML]{FFDEDE}4 0 0 0} &
  \multicolumn{1}{l|}{\cellcolor[HTML]{FFFFFF}0 0 0 0} &
  \multicolumn{1}{l|}{\cellcolor[HTML]{FFFFFF}NA} &
  \multicolumn{1}{l|}{\cellcolor[HTML]{FFFFFF}0 0 0 0} &
  \multicolumn{1}{l|}{\cellcolor[HTML]{FFFFFF}0 0 0 0} &
  \multicolumn{1}{l|}{\cellcolor[HTML]{FFFFFF}NA} &
  \cellcolor[HTML]{F6FCF9}{\ul \textbf{0.12}} \\ \hline
\rowcolor[HTML]{FFFFFF} 
\multicolumn{1}{|l|}{\cellcolor[HTML]{FFFFFF}\textbf{\begin{tabular}[c]{@{}l@{}}Back\\ Window\\ Lower\\ Frame\end{tabular}}} &
  \multicolumn{1}{l|}{\cellcolor[HTML]{FFFFFF}NA} &
  \multicolumn{1}{l|}{\cellcolor[HTML]{FFFFFF}0 0 0 0} &
  \multicolumn{1}{l|}{\cellcolor[HTML]{FFFFFF}NA} &
  \multicolumn{1}{l|}{\cellcolor[HTML]{FFFFFF}0 0 0 0} &
  \multicolumn{1}{l|}{\cellcolor[HTML]{FFFFFF}NA} &
  \multicolumn{1}{l|}{\cellcolor[HTML]{FFFFFF}0 0 0 0} &
  \multicolumn{1}{l|}{\cellcolor[HTML]{FFFFFF}NA} &
  \multicolumn{1}{l|}{\cellcolor[HTML]{FFFFFF}0 0 0 0} &
  \multicolumn{1}{l|}{\cellcolor[HTML]{FFFFFF}0 0 0 0} &
  \multicolumn{1}{l|}{\cellcolor[HTML]{FFF9F9}1 0 0 0} &
  \multicolumn{1}{l|}{\cellcolor[HTML]{FFFFFF}0 0 0 0} &
  \multicolumn{1}{l|}{\cellcolor[HTML]{FFFFFF}NA} &
  \multicolumn{1}{l|}{\cellcolor[HTML]{FFFFFF}0 0 0 0} &
  \multicolumn{1}{l|}{\cellcolor[HTML]{FFF9F9}1 0 0 0} &
  \multicolumn{1}{l|}{\cellcolor[HTML]{FFFFFF}NA} &
  \cellcolor[HTML]{FDFEFE}{\ul \textbf{0.03}} \\ \hline
\multicolumn{1}{|l|}{\cellcolor[HTML]{FFFFFF}\textbf{\begin{tabular}[c]{@{}l@{}}Cowl\\ Cover\end{tabular}}} &
  \multicolumn{1}{l|}{\cellcolor[HTML]{FFFFFF}NA} &
  \multicolumn{1}{l|}{\cellcolor[HTML]{FFE8E7}2 0 0 1} &
  \multicolumn{1}{l|}{\cellcolor[HTML]{FFF9F9}1 0 0 0} &
  \multicolumn{1}{l|}{\cellcolor[HTML]{FFF9F9}1 0 0 0} &
  \multicolumn{1}{l|}{\cellcolor[HTML]{FFFFFF}NA} &
  \multicolumn{1}{l|}{\cellcolor[HTML]{FFFFFF}0 0 0 0} &
  \multicolumn{1}{l|}{\cellcolor[HTML]{FFF9F9}1 0 0 0} &
  \multicolumn{1}{l|}{\cellcolor[HTML]{FFFFFF}0 0 0 0} &
  \multicolumn{1}{l|}{\cellcolor[HTML]{FFFFFF}0 0 0 0} &
  \multicolumn{1}{l|}{\cellcolor[HTML]{FFF9F9}1 0 0 0} &
  \multicolumn{1}{l|}{\cellcolor[HTML]{FFFFFF}NA} &
  \multicolumn{1}{l|}{\cellcolor[HTML]{FFF9F9}1 0 0 0} &
  \multicolumn{1}{l|}{\cellcolor[HTML]{FFE8E7}1 2 0 0} &
  \multicolumn{1}{l|}{\cellcolor[HTML]{FFE8E7}0 3 0 0} &
  \multicolumn{1}{l|}{\cellcolor[HTML]{FFE8E7}0 3 0 0} &
  \cellcolor[HTML]{E9F7F0}{\ul \textbf{0.28}} \\ \hline
\multicolumn{1}{|l|}{\cellcolor[HTML]{FFFFFF}\textbf{\begin{tabular}[c]{@{}l@{}}Front \\ Bumper\end{tabular}}} &
  \multicolumn{1}{l|}{\cellcolor[HTML]{FFFFFF}NA} &
  \multicolumn{1}{l|}{\cellcolor[HTML]{FFCCCD}5 0 0 1} &
  \multicolumn{1}{l|}{\cellcolor[HTML]{FFB1B1}4 5 0 0} &
  \multicolumn{1}{l|}{\cellcolor[HTML]{FFC3C3}4 3 0 0} &
  \multicolumn{1}{l|}{\cellcolor[HTML]{FFFFFF}NA} &
  \multicolumn{1}{l|}{\cellcolor[HTML]{FFC3C3}6 1 0 0} &
  \multicolumn{1}{l|}{\cellcolor[HTML]{FFC3C3}7 0 0 0} &
  \multicolumn{1}{l|}{\cellcolor[HTML]{FFF0F0}2 0 0 0} &
  \multicolumn{1}{l|}{\cellcolor[HTML]{FFC3C3}7 0 0 0} &
  \multicolumn{1}{l|}{\cellcolor[HTML]{FFD5D5}5 0 0 0} &
  \multicolumn{1}{l|}{\cellcolor[HTML]{FFDEDE}4 0 0 0} &
  \multicolumn{1}{l|}{\cellcolor[HTML]{FFF9F9}1 0 0 0} &
  \multicolumn{1}{l|}{\cellcolor[HTML]{FFF0F0}2 0 0 0} &
  \multicolumn{1}{l|}{\cellcolor[HTML]{FFF0F0}2 0 0 0} &
  \multicolumn{1}{l|}{\cellcolor[HTML]{FFDEDE}4 0 0 0} &
  \cellcolor[HTML]{ADDEC6}{\ul \textbf{1.05}} \\ \hline
\rowcolor[HTML]{FFFFFF} 
\multicolumn{1}{|l|}{\cellcolor[HTML]{FFFFFF}\textbf{\begin{tabular}[c]{@{}l@{}}Lower \\ Deflector\end{tabular}}} &
  \multicolumn{1}{l|}{\cellcolor[HTML]{FFFFFF}NA} &
  \multicolumn{1}{l|}{\cellcolor[HTML]{FFF0F0}1 0 0 1} &
  \multicolumn{1}{l|}{\cellcolor[HTML]{FFE8E7}3 0 0 0} &
  \multicolumn{1}{l|}{\cellcolor[HTML]{FFFFFF}NA} &
  \multicolumn{1}{l|}{\cellcolor[HTML]{FFFFFF}NA} &
  \multicolumn{1}{l|}{\cellcolor[HTML]{FFFFFF}0 0 0 0} &
  \multicolumn{1}{l|}{\cellcolor[HTML]{FFFFFF}0 0 0 0} &
  \multicolumn{1}{l|}{\cellcolor[HTML]{FFF9F9}1 0 0 0} &
  \multicolumn{1}{l|}{\cellcolor[HTML]{FFDEDE}4 0 0 0} &
  \multicolumn{1}{l|}{\cellcolor[HTML]{FFDEDE}4 0 0 0} &
  \multicolumn{1}{l|}{\cellcolor[HTML]{FFFFFF}NA} &
  \multicolumn{1}{l|}{\cellcolor[HTML]{FFFFFF}0 0 0 0} &
  \multicolumn{1}{l|}{\cellcolor[HTML]{FFFFFF}NA} &
  \multicolumn{1}{l|}{\cellcolor[HTML]{FFFFFF}0 0 0 0} &
  \multicolumn{1}{l|}{\cellcolor[HTML]{FFFFFF}0 0 0 0} &
  \cellcolor[HTML]{E5F5ED}{\ul \textbf{0.33}} \\ \hline
\multicolumn{1}{|l|}{\cellcolor[HTML]{FFFFFF}\textbf{\begin{tabular}[c]{@{}l@{}}Front\\ Doors\end{tabular}}} &
  \multicolumn{1}{l|}{\cellcolor[HTML]{FFFFFF}NA} &
  \multicolumn{1}{l|}{\cellcolor[HTML]{FFFFFF}NA} &
  \multicolumn{1}{l|}{\cellcolor[HTML]{FFB1B1}3 6 0 0} &
  \multicolumn{1}{l|}{\cellcolor[HTML]{FF9F9F}4 7 0 0} &
  \multicolumn{1}{l|}{\cellcolor[HTML]{FFFFFF}NA} &
  \multicolumn{1}{l|}{\cellcolor[HTML]{FFF9F9}1 0 0 0} &
  \multicolumn{1}{l|}{\cellcolor[HTML]{FFF0F0}1 1 0 0} &
  \multicolumn{1}{l|}{\cellcolor[HTML]{FFD5D5}4 1 0 0} &
  \multicolumn{1}{l|}{\cellcolor[HTML]{FFE8E7}2 1 0 0} &
  \multicolumn{1}{l|}{\cellcolor[HTML]{FFD5D5}4 1 0 0} &
  \multicolumn{1}{l|}{\cellcolor[HTML]{FF9F9F}4 7 0 0} &
  \multicolumn{1}{l|}{\cellcolor[HTML]{FF8384}7 7 0 0} &
  \multicolumn{1}{l|}{\cellcolor[HTML]{FFBABA}7 1 0 0} &
  \multicolumn{1}{l|}{\cellcolor[HTML]{FFC3C3}4 3 0 0} &
  \multicolumn{1}{l|}{\cellcolor[HTML]{FFDEDE}2 2 0 0} &
  \cellcolor[HTML]{97D5B7}{\ul \textbf{1.33}} \\ \hline
\multicolumn{1}{|l|}{\cellcolor[HTML]{FFFFFF}\textbf{\begin{tabular}[c]{@{}l@{}}Front \\ Fenders\end{tabular}}} &
  \multicolumn{1}{l|}{\cellcolor[HTML]{FFDEDE}2 2 0 0} &
  \multicolumn{1}{l|}{\cellcolor[HTML]{FF8C8D}5 7 0 1} &
  \multicolumn{1}{l|}{\cellcolor[HTML]{FFE8E7}2 1 0 0} &
  \multicolumn{1}{l|}{\cellcolor[HTML]{FFDEDE}2 2 0 0} &
  \multicolumn{1}{l|}{\cellcolor[HTML]{FFCCCD}3 3 0 0} &
  \multicolumn{1}{l|}{\cellcolor[HTML]{FF7171}4 7 5 0} &
  \multicolumn{1}{l|}{\cellcolor[HTML]{FFBABA}3 5 0 0} &
  \multicolumn{1}{l|}{\cellcolor[HTML]{FFC3C3}7 0 0 0} &
  \multicolumn{1}{l|}{\cellcolor[HTML]{FFA9A7}3 7 0 0} &
  \multicolumn{1}{l|}{\cellcolor[HTML]{FF9596}5 7 0 0} &
  \multicolumn{1}{l|}{\cellcolor[HTML]{FFBABA}7 1 0 0} &
  \multicolumn{1}{l|}{\cellcolor[HTML]{FFE8E7}3 0 0 0} &
  \multicolumn{1}{l|}{\cellcolor[HTML]{FF8384}7 7 0 0} &
  \multicolumn{1}{l|}{\cellcolor[HTML]{FF8C8D}6 7 0 0} &
  \multicolumn{1}{l|}{\cellcolor[HTML]{FFBABA}5 3 0 0} &
  \cellcolor[HTML]{57BB8A}{\ul \textbf{2.15}} \\ \hline
\rowcolor[HTML]{FFFFFF} 
\multicolumn{1}{|l|}{\cellcolor[HTML]{FFFFFF}\textbf{\begin{tabular}[c]{@{}l@{}}Front\\ Plate\end{tabular}}} &
  \multicolumn{1}{l|}{\cellcolor[HTML]{FFFFFF}0 0 0 0} &
  \multicolumn{1}{l|}{\cellcolor[HTML]{FFF9F9}1 0 0 0} &
  \multicolumn{1}{l|}{\cellcolor[HTML]{FFFFFF}0 0 0 0} &
  \multicolumn{1}{l|}{\cellcolor[HTML]{FFE8E7}3 0 0 0} &
  \multicolumn{1}{l|}{\cellcolor[HTML]{FFFFFF}NA} &
  \multicolumn{1}{l|}{\cellcolor[HTML]{FFFFFF}0 0 0 0} &
  \multicolumn{1}{l|}{\cellcolor[HTML]{FFD5D5}5 0 0 0} &
  \multicolumn{1}{l|}{\cellcolor[HTML]{FFFFFF}0 0 0 0} &
  \multicolumn{1}{l|}{\cellcolor[HTML]{FFFFFF}0 0 0 0} &
  \multicolumn{1}{l|}{\cellcolor[HTML]{FFFFFF}0 0 0 0} &
  \multicolumn{1}{l|}{\cellcolor[HTML]{FFFFFF}0 0 0 0} &
  \multicolumn{1}{l|}{\cellcolor[HTML]{FFFFFF}0 0 0 0} &
  \multicolumn{1}{l|}{\cellcolor[HTML]{FFF9F9}1 0 0 0} &
  \multicolumn{1}{l|}{\cellcolor[HTML]{FFFFFF}0 0 0 0} &
  \multicolumn{1}{l|}{\cellcolor[HTML]{FFFFFF}0 0 0 0} &
  \cellcolor[HTML]{F2FAF6}{\ul \textbf{0.17}} \\ \hline
\multicolumn{1}{|l|}{\cellcolor[HTML]{FFFFFF}\textbf{\begin{tabular}[c]{@{}l@{}}Front \\ Wheels\end{tabular}}} &
  \multicolumn{1}{l|}{\cellcolor[HTML]{FFDEDE}2 2 0 0} &
  \multicolumn{1}{l|}{\cellcolor[HTML]{FFB1B1}7 1 0 1} &
  \multicolumn{1}{l|}{\cellcolor[HTML]{FFCCCD}4 2 0 0} &
  \multicolumn{1}{l|}{\cellcolor[HTML]{FFA9A7}3 7 0 0} &
  \multicolumn{1}{l|}{\cellcolor[HTML]{FFC3C3}3 4 0 0} &
  \multicolumn{1}{l|}{\cellcolor[HTML]{FFDEDE}3 1 0 0} &
  \multicolumn{1}{l|}{\cellcolor[HTML]{FFBABA}4 4 0 0} &
  \multicolumn{1}{l|}{\cellcolor[HTML]{FFA9A7}7 3 0 0} &
  \multicolumn{1}{l|}{\cellcolor[HTML]{FFB1B1}3 6 0 0} &
  \multicolumn{1}{l|}{\cellcolor[HTML]{FF9596}7 5 0 0} &
  \multicolumn{1}{l|}{\cellcolor[HTML]{FFF9F9}1 0 0 0} &
  \multicolumn{1}{l|}{\cellcolor[HTML]{FFE8E7}2 1 0 0} &
  \multicolumn{1}{l|}{\cellcolor[HTML]{FFA9A7}7 3 0 0} &
  \multicolumn{1}{l|}{\cellcolor[HTML]{FF8384}7 7 0 0} &
  \multicolumn{1}{l|}{\cellcolor[HTML]{FF9F9F}7 4 0 0} &
  \cellcolor[HTML]{66C194}{\ul \textbf{1.97}} \\ \hline
\multicolumn{1}{|l|}{\cellcolor[HTML]{FFFFFF}\textbf{\begin{tabular}[c]{@{}l@{}}Front \\ Windows\end{tabular}}} &
  \multicolumn{1}{l|}{\cellcolor[HTML]{FFFFFF}NA} &
  \multicolumn{1}{l|}{\cellcolor[HTML]{FFCCCD}3 3 0 0} &
  \multicolumn{1}{l|}{\cellcolor[HTML]{FFF0F0}1 1 0 0} &
  \multicolumn{1}{l|}{\cellcolor[HTML]{FFA9A7}7 3 0 0} &
  \multicolumn{1}{l|}{\cellcolor[HTML]{FFFFFF}NA} &
  \multicolumn{1}{l|}{\cellcolor[HTML]{FFC3C3}2 5 0 0} &
  \multicolumn{1}{l|}{\cellcolor[HTML]{FFF0F0}1 1 0 0} &
  \multicolumn{1}{l|}{\cellcolor[HTML]{FFDEDE}3 1 0 0} &
  \multicolumn{1}{l|}{\cellcolor[HTML]{FFF0F0}1 1 0 0} &
  \multicolumn{1}{l|}{\cellcolor[HTML]{FFC3C3}3 4 0 0} &
  \multicolumn{1}{l|}{\cellcolor[HTML]{FFA9A7}3 7 0 0} &
  \multicolumn{1}{l|}{\cellcolor[HTML]{FF9596}7 5 0 0} &
  \multicolumn{1}{l|}{\cellcolor[HTML]{FFF0F0}1 1 0 0} &
  \multicolumn{1}{l|}{\cellcolor[HTML]{FFF0F0}1 1 0 0} &
  \multicolumn{1}{l|}{\cellcolor[HTML]{FFFFFF}NA} &
  \cellcolor[HTML]{AADDC4}{\ul \textbf{1.1}} \\ \hline
\multicolumn{1}{|l|}{\cellcolor[HTML]{FFFFFF}{\ul \textbf{Grill}}} &
  \multicolumn{1}{l|}{\cellcolor[HTML]{FFFFFF}NA} &
  \multicolumn{1}{l|}{\cellcolor[HTML]{FFF0F0}2 0 0 0} &
  \multicolumn{1}{l|}{\cellcolor[HTML]{FFE8E7}3 0 0 0} &
  \multicolumn{1}{l|}{\cellcolor[HTML]{FFDEDE}3 1 0 0} &
  \multicolumn{1}{l|}{\cellcolor[HTML]{FFFFFF}NA} &
  \multicolumn{1}{l|}{\cellcolor[HTML]{FFDEDE}4 0 0 0} &
  \multicolumn{1}{l|}{\cellcolor[HTML]{FFC3C3}7 0 0 0} &
  \multicolumn{1}{l|}{\cellcolor[HTML]{FFF9F9}1 0 0 0} &
  \multicolumn{1}{l|}{\cellcolor[HTML]{FFD5D5}5 0 0 0} &
  \multicolumn{1}{l|}{\cellcolor[HTML]{FFFFFF}0 0 0 0} &
  \multicolumn{1}{l|}{\cellcolor[HTML]{FFE8E7}3 0 0 0} &
  \multicolumn{1}{l|}{\cellcolor[HTML]{FFF9F9}1 0 0 0} &
  \multicolumn{1}{l|}{\cellcolor[HTML]{FFF9F9}1 0 0 0} &
  \multicolumn{1}{l|}{\cellcolor[HTML]{FFF9F9}1 0 0 0} &
  \multicolumn{1}{l|}{\cellcolor[HTML]{FFF9F9}1 0 0 0} &
  \cellcolor[HTML]{D5EEE2}{\ul \textbf{0.55}} \\ \hline
\multicolumn{1}{|l|}{\cellcolor[HTML]{FFFFFF}{\ul \textbf{Headlights}}} &
  \multicolumn{1}{l|}{\cellcolor[HTML]{FFFFFF}0 0 0 0} &
  \multicolumn{1}{l|}{\cellcolor[HTML]{FFC3C3}4 2 0 1} &
  \multicolumn{1}{l|}{\cellcolor[HTML]{FFDEDE}3 1 0 0} &
  \multicolumn{1}{l|}{\cellcolor[HTML]{FFE8E7}0 3 0 0} &
  \multicolumn{1}{l|}{\cellcolor[HTML]{FFDEDE}3 1 0 0} &
  \multicolumn{1}{l|}{\cellcolor[HTML]{FFC3C3}7 0 0 0} &
  \multicolumn{1}{l|}{\cellcolor[HTML]{FFDEDE}4 0 0 0} &
  \multicolumn{1}{l|}{\cellcolor[HTML]{FFE8E7}3 0 0 0} &
  \multicolumn{1}{l|}{\cellcolor[HTML]{FFDEDE}4 0 0 0} &
  \multicolumn{1}{l|}{\cellcolor[HTML]{FFD5D5}5 0 0 0} &
  \multicolumn{1}{l|}{\cellcolor[HTML]{FFE8E7}3 0 0 0} &
  \multicolumn{1}{l|}{\cellcolor[HTML]{FFF9F9}1 0 0 0} &
  \multicolumn{1}{l|}{\cellcolor[HTML]{FFC3C3}7 0 0 0} &
  \multicolumn{1}{l|}{\cellcolor[HTML]{FFE8E7}3 0 0 0} &
  \multicolumn{1}{l|}{\cellcolor[HTML]{FFFFFF}0 0 0 0} &
  \cellcolor[HTML]{B8E3CE}{\ul \textbf{0.92}} \\ \hline
\multicolumn{1}{|l|}{\cellcolor[HTML]{FFFFFF}{\ul \textbf{Hood}}} &
  \multicolumn{1}{l|}{\cellcolor[HTML]{FFF0F0}1 1 0 0} &
  \multicolumn{1}{l|}{\cellcolor[HTML]{FFFFFF}NA} &
  \multicolumn{1}{l|}{\cellcolor[HTML]{FFF0F0}2 0 0 0} &
  \multicolumn{1}{l|}{\cellcolor[HTML]{FFF9F9}1 0 0 0} &
  \multicolumn{1}{l|}{\cellcolor[HTML]{FFFFFF}NA} &
  \multicolumn{1}{l|}{\cellcolor[HTML]{FFE8E7}3 0 0 0} &
  \multicolumn{1}{l|}{\cellcolor[HTML]{FFF0F0}1 1 0 0} &
  \multicolumn{1}{l|}{\cellcolor[HTML]{FFF9F9}1 0 0 0} &
  \multicolumn{1}{l|}{\cellcolor[HTML]{FFF9F9}1 0 0 0} &
  \multicolumn{1}{l|}{\cellcolor[HTML]{FFF9F9}1 0 0 0} &
  \multicolumn{1}{l|}{\cellcolor[HTML]{FFFFFF}NA} &
  \multicolumn{1}{l|}{\cellcolor[HTML]{FFF9F9}1 0 0 0} &
  \multicolumn{1}{l|}{\cellcolor[HTML]{FFF0F0}1 1 0 0} &
  \multicolumn{1}{l|}{\cellcolor[HTML]{FFF9F9}1 0 0 0} &
  \multicolumn{1}{l|}{\cellcolor[HTML]{FFF9F9}1 0 0 0} &
  \cellcolor[HTML]{E8F6EF}{\ul \textbf{0.3}} \\ \hline
\rowcolor[HTML]{FFFFFF} 
\multicolumn{1}{|l|}{\cellcolor[HTML]{FFFFFF}{\ul \textbf{Rear Doors}}} &
  \multicolumn{1}{l|}{\cellcolor[HTML]{FFFFFF}NA} &
  \multicolumn{1}{l|}{\cellcolor[HTML]{FFFFFF}NA} &
  \multicolumn{1}{l|}{\cellcolor[HTML]{FFFFFF}NA} &
  \multicolumn{1}{l|}{\cellcolor[HTML]{FFC3C3}7 0 0 0} &
  \multicolumn{1}{l|}{\cellcolor[HTML]{FFFFFF}NA} &
  \multicolumn{1}{l|}{\cellcolor[HTML]{FFFFFF}NA} &
  \multicolumn{1}{l|}{\cellcolor[HTML]{FFFFFF}NA} &
  \multicolumn{1}{l|}{\cellcolor[HTML]{FFDEDE}4 0 0 0} &
  \multicolumn{1}{l|}{\cellcolor[HTML]{FFCCCD}6 0 0 0} &
  \multicolumn{1}{l|}{\cellcolor[HTML]{FFDEDE}4 0 0 0} &
  \multicolumn{1}{l|}{\cellcolor[HTML]{FFFFFF}NA} &
  \multicolumn{1}{l|}{\cellcolor[HTML]{FFFFFF}NA} &
  \multicolumn{1}{l|}{\cellcolor[HTML]{FFE8E7}3 0 0 0} &
  \multicolumn{1}{l|}{\cellcolor[HTML]{FFE8E7}3 0 0 0} &
  \multicolumn{1}{l|}{\cellcolor[HTML]{FFFFFF}NA} &
  \cellcolor[HTML]{D8F0E4}{\ul \textbf{0.5}} \\ \hline
\multicolumn{1}{|l|}{\cellcolor[HTML]{FFFFFF}\textbf{\begin{tabular}[c]{@{}l@{}}Rear\\ Wheels\end{tabular}}} &
  \multicolumn{1}{l|}{\cellcolor[HTML]{FFF9F9}1 0 0 0} &
  \multicolumn{1}{l|}{\cellcolor[HTML]{FFE8E7}3 0 0 0} &
  \multicolumn{1}{l|}{\cellcolor[HTML]{FFD5D5}5 0 0 0} &
  \multicolumn{1}{l|}{\cellcolor[HTML]{FFE8E7}3 0 0 0} &
  \multicolumn{1}{l|}{\cellcolor[HTML]{FFDEDE}3 1 0 0} &
  \multicolumn{1}{l|}{\cellcolor[HTML]{FFDEDE}4 0 0 0} &
  \multicolumn{1}{l|}{\cellcolor[HTML]{FFF9F9}1 0 0 0} &
  \multicolumn{1}{l|}{\cellcolor[HTML]{FFE8E7}3 0 0 0} &
  \multicolumn{1}{l|}{\cellcolor[HTML]{FFF9F9}1 0 0 0} &
  \multicolumn{1}{l|}{\cellcolor[HTML]{FFF0F0}2 0 0 0} &
  \multicolumn{1}{l|}{\cellcolor[HTML]{FFF9F9}1 0 0 0} &
  \multicolumn{1}{l|}{\cellcolor[HTML]{FFF0F0}1 1 0 0} &
  \multicolumn{1}{l|}{\cellcolor[HTML]{FFE8E7}3 0 0 0} &
  \multicolumn{1}{l|}{\cellcolor[HTML]{FFE8E7}3 0 0 0} &
  \multicolumn{1}{l|}{\cellcolor[HTML]{FFF0F0}2 0 0 0} &
  \cellcolor[HTML]{CEEBDD}{\ul \textbf{0.63}} \\ \hline
\multicolumn{1}{|l|}{\cellcolor[HTML]{FFFFFF}\textbf{\begin{tabular}[c]{@{}l@{}}Rear \\ Windows\end{tabular}}} &
  \multicolumn{1}{l|}{\cellcolor[HTML]{FFFFFF}NA} &
  \multicolumn{1}{l|}{\cellcolor[HTML]{FFC3C3}7 0 0 0} &
  \multicolumn{1}{l|}{\cellcolor[HTML]{FFFFFF}NA} &
  \multicolumn{1}{l|}{\cellcolor[HTML]{FFE8E7}3 0 0 0} &
  \multicolumn{1}{l|}{\cellcolor[HTML]{FFFFFF}NA} &
  \multicolumn{1}{l|}{\cellcolor[HTML]{FFF9F9}1 0 0 0} &
  \multicolumn{1}{l|}{\cellcolor[HTML]{FFFFFF}NA} &
  \multicolumn{1}{l|}{\cellcolor[HTML]{FFFFFF}NA} &
  \multicolumn{1}{l|}{\cellcolor[HTML]{FFE8E7}3 0 0 0} &
  \multicolumn{1}{l|}{\cellcolor[HTML]{FFDEDE}4 0 0 0} &
  \multicolumn{1}{l|}{\cellcolor[HTML]{FFDEDE}4 0 0 0} &
  \multicolumn{1}{l|}{\cellcolor[HTML]{FFFFFF}NA} &
  \multicolumn{1}{l|}{\cellcolor[HTML]{FFE8E7}3 0 0 0} &
  \multicolumn{1}{l|}{\cellcolor[HTML]{FFE8E7}3 0 0 0} &
  \multicolumn{1}{l|}{\cellcolor[HTML]{FFFFFF}NA} &
  \cellcolor[HTML]{DBF1E6}{\ul \textbf{0.47}} \\ \hline
\rowcolor[HTML]{FFFFFF} 
\multicolumn{1}{|l|}{\cellcolor[HTML]{FFFFFF}\textbf{\begin{tabular}[c]{@{}l@{}}Rocker\\ Panel\end{tabular}}} &
  \multicolumn{1}{l|}{\cellcolor[HTML]{FFFFFF}0 0 0 0} &
  \multicolumn{1}{l|}{\cellcolor[HTML]{FFFFFF}NA} &
  \multicolumn{1}{l|}{\cellcolor[HTML]{FFF9F9}1 0 0 0} &
  \multicolumn{1}{l|}{\cellcolor[HTML]{FFFFFF}0 0 0 0} &
  \multicolumn{1}{l|}{\cellcolor[HTML]{FFFFFF}NA} &
  \multicolumn{1}{l|}{\cellcolor[HTML]{FFF9F9}1 0 0 0} &
  \multicolumn{1}{l|}{\cellcolor[HTML]{FFFFFF}0 0 0 0} &
  \multicolumn{1}{l|}{\cellcolor[HTML]{FFFFFF}0 0 0 0} &
  \multicolumn{1}{l|}{\cellcolor[HTML]{FFF0F0}2 0 0 0} &
  \multicolumn{1}{l|}{\cellcolor[HTML]{FFFFFF}0 0 0 0} &
  \multicolumn{1}{l|}{\cellcolor[HTML]{FFFFFF}NA} &
  \multicolumn{1}{l|}{\cellcolor[HTML]{FFF9F9}1 0 0 0} &
  \multicolumn{1}{l|}{\cellcolor[HTML]{FFF9F9}1 0 0 0} &
  \multicolumn{1}{l|}{\cellcolor[HTML]{FFFFFF}0 0 0 0} &
  \multicolumn{1}{l|}{\cellcolor[HTML]{FFFFFF}0 0 0 0} &
  \cellcolor[HTML]{F4FBF7}{\ul \textbf{0.15}} \\ \hline
\multicolumn{1}{|l|}{\cellcolor[HTML]{FFFFFF}{\ul \textbf{Roof}}} &
  \multicolumn{1}{l|}{\cellcolor[HTML]{FF9F9F}4 7 0 0} &
  \multicolumn{1}{l|}{\cellcolor[HTML]{FFF9F9}0 0 0 1} &
  \multicolumn{1}{l|}{\cellcolor[HTML]{FFE8E7}2 1 0 0} &
  \multicolumn{1}{l|}{\cellcolor[HTML]{FFD5D5}4 1 0 0} &
  \multicolumn{1}{l|}{\cellcolor[HTML]{FFFFFF}NA} &
  \multicolumn{1}{l|}{\cellcolor[HTML]{FFF9F9}1 0 0 0} &
  \multicolumn{1}{l|}{\cellcolor[HTML]{FFE8E7}2 1 0 0} &
  \multicolumn{1}{l|}{\cellcolor[HTML]{FFE8E7}3 0 0 0} &
  \multicolumn{1}{l|}{\cellcolor[HTML]{FFF9F9}1 0 0 0} &
  \multicolumn{1}{l|}{\cellcolor[HTML]{FFE8E7}2 1 0 0} &
  \multicolumn{1}{l|}{\cellcolor[HTML]{FFFFFF}0 0 0 0} &
  \multicolumn{1}{l|}{\cellcolor[HTML]{FFCCCD}3 3 0 0} &
  \multicolumn{1}{l|}{\cellcolor[HTML]{FFDEDE}3 1 0 0} &
  \multicolumn{1}{l|}{\cellcolor[HTML]{FFF9F9}1 0 0 0} &
  \multicolumn{1}{l|}{\cellcolor[HTML]{FFFFFF}NA} &
  \cellcolor[HTML]{C9E9D9}{\ul \textbf{0.7}} \\ \hline
\multicolumn{1}{|l|}{\cellcolor[HTML]{FFFFFF}\textbf{\begin{tabular}[c]{@{}l@{}}Side\\ Mirrors\end{tabular}}} &
  \multicolumn{1}{l|}{\cellcolor[HTML]{FFFFFF}NA} &
  \multicolumn{1}{l|}{\cellcolor[HTML]{FFC3C3}0 0 0 7} &
  \multicolumn{1}{l|}{\cellcolor[HTML]{FFF9F9}1 0 0 0} &
  \multicolumn{1}{l|}{\cellcolor[HTML]{FFBABA}7 1 0 0} &
  \multicolumn{1}{l|}{\cellcolor[HTML]{FFD5D5}2 3 0 0} &
  \multicolumn{1}{l|}{\cellcolor[HTML]{FFA9A7}3 0 7 0} &
  \multicolumn{1}{l|}{\cellcolor[HTML]{FFDEDE}3 1 0 0} &
  \multicolumn{1}{l|}{\cellcolor[HTML]{FFD5D5}4 1 0 0} &
  \multicolumn{1}{l|}{\cellcolor[HTML]{FFE8E7}2 1 0 0} &
  \multicolumn{1}{l|}{\cellcolor[HTML]{FFA9A7}4 6 0 0} &
  \multicolumn{1}{l|}{\cellcolor[HTML]{FFC3C3}0 0 7 0} &
  \multicolumn{1}{l|}{\cellcolor[HTML]{FFF0F0}1 1 0 0} &
  \multicolumn{1}{l|}{\cellcolor[HTML]{FFF9F9}1 0 0 0} &
  \multicolumn{1}{l|}{\cellcolor[HTML]{FFD5D5}4 1 0 0} &
  \multicolumn{1}{l|}{\cellcolor[HTML]{FFF9F9}1 0 0 0} &
  \cellcolor[HTML]{A6DBC1}{\ul \textbf{1.15}} \\ \hline
\multicolumn{1}{|l|}{\cellcolor[HTML]{FFFFFF}{\ul \textbf{Tail Lights}}} &
  \multicolumn{1}{l|}{\cellcolor[HTML]{FFFFFF}0 0 0 0} &
  \multicolumn{1}{l|}{\cellcolor[HTML]{FFFFFF}0 0 0 0} &
  \multicolumn{1}{l|}{\cellcolor[HTML]{FFFFFF}0 0 0 0} &
  \multicolumn{1}{l|}{\cellcolor[HTML]{FFF0F0}2 0 0 0} &
  \multicolumn{1}{l|}{\cellcolor[HTML]{FFFFFF}0 0 0 0} &
  \multicolumn{1}{l|}{\cellcolor[HTML]{FFC3C3}7 0 0 0} &
  \multicolumn{1}{l|}{\cellcolor[HTML]{FFFFFF}0 0 0 0} &
  \multicolumn{1}{l|}{\cellcolor[HTML]{FFCCCD}6 0 0 0} &
  \multicolumn{1}{l|}{\cellcolor[HTML]{FFF9F9}1 0 0 0} &
  \multicolumn{1}{l|}{\cellcolor[HTML]{FFF9F9}1 0 0 0} &
  \multicolumn{1}{l|}{\cellcolor[HTML]{FFFFFF}0 0 0 0} &
  \multicolumn{1}{l|}{\cellcolor[HTML]{FFFFFF}0 0 0 0} &
  \multicolumn{1}{l|}{\cellcolor[HTML]{FFDEDE}4 0 0 0} &
  \multicolumn{1}{l|}{\cellcolor[HTML]{FFF9F9}1 0 0 0} &
  \multicolumn{1}{l|}{\cellcolor[HTML]{FFE8E7}3 0 0 0} &
  \cellcolor[HTML]{DFF2E9}{\ul \textbf{0.42}} \\ \hline
\rowcolor[HTML]{FFFFFF} 
\multicolumn{1}{|l|}{\cellcolor[HTML]{FFFFFF}{\ul \textbf{Trunk}}} &
  \multicolumn{1}{l|}{\cellcolor[HTML]{FFFFFF}NA} &
  \multicolumn{1}{l|}{\cellcolor[HTML]{FFFFFF}NA} &
  \multicolumn{1}{l|}{\cellcolor[HTML]{FFFFFF}0 0 0 0} &
  \multicolumn{1}{l|}{\cellcolor[HTML]{FFF9F9}1 0 0 0} &
  \multicolumn{1}{l|}{\cellcolor[HTML]{FFFFFF}NA} &
  \multicolumn{1}{l|}{\cellcolor[HTML]{FFFFFF}0 0 0 0} &
  \multicolumn{1}{l|}{\cellcolor[HTML]{FFF9F9}1 0 0 0} &
  \multicolumn{1}{l|}{\cellcolor[HTML]{FFFFFF}0 0 0 0} &
  \multicolumn{1}{l|}{\cellcolor[HTML]{FFFFFF}NA} &
  \multicolumn{1}{l|}{\cellcolor[HTML]{FFF9F9}1 0 0 0} &
  \multicolumn{1}{l|}{\cellcolor[HTML]{FFFFFF}NA} &
  \multicolumn{1}{l|}{\cellcolor[HTML]{FFFFFF}0 0 0 0} &
  \multicolumn{1}{l|}{\cellcolor[HTML]{FFF9F9}1 0 0 0} &
  \multicolumn{1}{l|}{\cellcolor[HTML]{FFF9F9}1 0 0 0} &
  \multicolumn{1}{l|}{\cellcolor[HTML]{FFFFFF}0 0 0 0} &
  \cellcolor[HTML]{F9FDFB}{\ul \textbf{0.08}} \\ \hline
\multicolumn{1}{|l|}{\cellcolor[HTML]{FFFFFF}{\ul \textbf{Windshield}}} &
  \multicolumn{1}{l|}{\cellcolor[HTML]{FFF9F9}1 0 0 0} &
  \multicolumn{1}{l|}{\cellcolor[HTML]{FFDEDE}3 0 0 1} &
  \multicolumn{1}{l|}{\cellcolor[HTML]{FFDEDE}3 1 0 0} &
  \multicolumn{1}{l|}{\cellcolor[HTML]{FFF9F9}1 0 0 0} &
  \multicolumn{1}{l|}{\cellcolor[HTML]{FFFFFF}NA} &
  \multicolumn{1}{l|}{\cellcolor[HTML]{FFF9F9}1 0 0 0} &
  \multicolumn{1}{l|}{\cellcolor[HTML]{FFDEDE}1 3 0 0} &
  \multicolumn{1}{l|}{\cellcolor[HTML]{FFF9F9}1 0 0 0} &
  \multicolumn{1}{l|}{\cellcolor[HTML]{FFF9F9}0 1 0 0} &
  \multicolumn{1}{l|}{\cellcolor[HTML]{FFD5D5}1 4 0 0} &
  \multicolumn{1}{l|}{\cellcolor[HTML]{FFDEDE}4 0 0 0} &
  \multicolumn{1}{l|}{\cellcolor[HTML]{FFF9F9}1 0 0 0} &
  \multicolumn{1}{l|}{\cellcolor[HTML]{FFDEDE}3 1 0 0} &
  \multicolumn{1}{l|}{\cellcolor[HTML]{FFF9F9}1 0 0 0} &
  \multicolumn{1}{l|}{\cellcolor[HTML]{FFF0F0}1 1 0 0} &
  \cellcolor[HTML]{D3EEE1}{\ul \textbf{0.57}} \\ \hline
\multicolumn{1}{|l|}{\cellcolor[HTML]{FFFFFF}\textbf{\begin{tabular}[c]{@{}l@{}}Windshield \\ Rails\end{tabular}}} &
  \multicolumn{1}{l|}{\cellcolor[HTML]{FFE8E7}2 1 0 0} &
  \multicolumn{1}{l|}{\cellcolor[HTML]{FFD5D5}4 0 0 1} &
  \multicolumn{1}{l|}{\cellcolor[HTML]{FFF0F0}1 1 0 0} &
  \multicolumn{1}{l|}{\cellcolor[HTML]{FFF0F0}0 2 0 0} &
  \multicolumn{1}{l|}{\cellcolor[HTML]{FFFFFF}NA} &
  \multicolumn{1}{l|}{\cellcolor[HTML]{FFF9F9}1 0 0 0} &
  \multicolumn{1}{l|}{\cellcolor[HTML]{FFF9F9}1 0 0 0} &
  \multicolumn{1}{l|}{\cellcolor[HTML]{FFB1B1}3 6 0 0} &
  \multicolumn{1}{l|}{\cellcolor[HTML]{FFE8E7}3 0 0 0} &
  \multicolumn{1}{l|}{\cellcolor[HTML]{FFC3C3}4 3 0 0} &
  \multicolumn{1}{l|}{\cellcolor[HTML]{FFBABA}7 1 0 0} &
  \multicolumn{1}{l|}{\cellcolor[HTML]{FFC3C3}3 4 0 0} &
  \multicolumn{1}{l|}{\cellcolor[HTML]{FFD5D5}4 1 0 0} &
  \multicolumn{1}{l|}{\cellcolor[HTML]{FFBABA}3 5 0 0} &
  \multicolumn{1}{l|}{\cellcolor[HTML]{FFDEDE}3 1 0 0} &
  \cellcolor[HTML]{ABDDC5}{\ul \textbf{1.08}} \\ \hline
\end{tabular}%
}
\caption{The highest \textit{Exposedness Score} found on any point of each part, for each vehicle, for points between 0\si{m} and 5\si{m} away from the camera, and for each of the four driving directions of that vehicle (south, east, north, and west, respectively).}
\label{tab:0-5Results}
\end{table}

\begin{table}[h]
\centering
\resizebox{\columnwidth}{!}{%
\begin{tabular}{|llllllllllllllllr|}
\hline
\rowcolor[HTML]{FFFFFF} 
\multicolumn{17}{|c|}{\cellcolor[HTML]{FFFFFF}\textbf{\begin{tabular}[c]{@{}c@{}}Highest registered exposedness index on any point between 5 and 25 meters from the camera that is found on the part when the vehicle is in each of the\\ four facings (South East North West).\end{tabular}}} \\ \hline
\rowcolor[HTML]{FFFFFF} 
\multicolumn{1}{|r|}{\cellcolor[HTML]{FFFFFF}\textbf{\begin{tabular}[c]{@{}r@{}}Vehicle \\ Type \\ \& Part\end{tabular}}} &
  \multicolumn{1}{c|}{\cellcolor[HTML]{FFFFFF}\textbf{\begin{tabular}[c]{@{}c@{}}Carryall\\ Car\end{tabular}}} &
  \multicolumn{1}{c|}{\cellcolor[HTML]{FFFFFF}\textbf{\begin{tabular}[c]{@{}c@{}}Double\\ Decker\\ Bus\end{tabular}}} &
  \multicolumn{1}{c|}{\cellcolor[HTML]{FFFFFF}\textbf{\begin{tabular}[c]{@{}c@{}}Four\\ Wheel\\ Truck\end{tabular}}} &
  \multicolumn{1}{c|}{\cellcolor[HTML]{FFFFFF}{\ul \textbf{Minibus}}} &
  \multicolumn{1}{c|}{\cellcolor[HTML]{FFFFFF}\textbf{\begin{tabular}[c]{@{}c@{}}Motor-\\ cycle\end{tabular}}} &
  \multicolumn{1}{c|}{\cellcolor[HTML]{FFFFFF}\textbf{\begin{tabular}[c]{@{}c@{}}Motor\\ Home\end{tabular}}} &
  \multicolumn{1}{c|}{\cellcolor[HTML]{FFFFFF}\textbf{\begin{tabular}[c]{@{}c@{}}Moving\\ Truck\end{tabular}}} &
  \multicolumn{1}{c|}{\cellcolor[HTML]{FFFFFF}\textbf{\begin{tabular}[c]{@{}c@{}}Panel\\ Van\end{tabular}}} &
  \multicolumn{1}{c|}{\cellcolor[HTML]{FFFFFF}\textbf{\begin{tabular}[c]{@{}c@{}}Pickup\\ Truck\end{tabular}}} &
  \multicolumn{1}{c|}{\cellcolor[HTML]{FFFFFF}{\ul \textbf{Sedan}}} &
  \multicolumn{1}{c|}{\cellcolor[HTML]{FFFFFF}\textbf{\begin{tabular}[c]{@{}c@{}}Single\\ Decker\\ Bus\end{tabular}}} &
  \multicolumn{1}{c|}{\cellcolor[HTML]{FFFFFF}\textbf{\begin{tabular}[c]{@{}c@{}}Smart\\ Car\end{tabular}}} &
  \multicolumn{1}{c|}{\cellcolor[HTML]{FFFFFF}\textbf{\begin{tabular}[c]{@{}c@{}}Station\\ Wagon\end{tabular}}} &
  \multicolumn{1}{c|}{\cellcolor[HTML]{FFFFFF}{\ul \textbf{SUV}}} &
  \multicolumn{1}{c|}{\cellcolor[HTML]{FFFFFF}\textbf{\begin{tabular}[c]{@{}c@{}}Topless \\ Conver-\\ tible\end{tabular}}} &
  \multicolumn{1}{c|}{\cellcolor[HTML]{FFFFFF}\textbf{\begin{tabular}[c]{@{}c@{}}Avg.\\ Highest\end{tabular}}} \\ \hline
\multicolumn{1}{|l|}{\cellcolor[HTML]{FFFFFF}\textbf{\begin{tabular}[c]{@{}l@{}}Back \\ Bumper\end{tabular}}} &
  \multicolumn{1}{l|}{\cellcolor[HTML]{FFFFFF}0 0 0 0} &
  \multicolumn{1}{l|}{\cellcolor[HTML]{FFF9F9}0 1 0 0} &
  \multicolumn{1}{l|}{\cellcolor[HTML]{FFDEDE}3 0 1 0} &
  \multicolumn{1}{l|}{\cellcolor[HTML]{FFA9A7}6 4 0 0} &
  \multicolumn{1}{l|}{\cellcolor[HTML]{FFFFFF}0 0 0 0} &
  \multicolumn{1}{l|}{\cellcolor[HTML]{FFF9F9}0 0 1 0} &
  \multicolumn{1}{l|}{\cellcolor[HTML]{FFF0F0}1 0 1 0} &
  \multicolumn{1}{l|}{\cellcolor[HTML]{FFFFFF}0 0 0 0} &
  \multicolumn{1}{l|}{\cellcolor[HTML]{FFDEDE}3 0 1 0} &
  \multicolumn{1}{l|}{\cellcolor[HTML]{FFE8E7}0 0 3 0} &
  \multicolumn{1}{l|}{\cellcolor[HTML]{FFFFFF}0 0 0 0} &
  \multicolumn{1}{l|}{\cellcolor[HTML]{FFF9F9}0 0 1 0} &
  \multicolumn{1}{l|}{\cellcolor[HTML]{FFCCCD}2 0 4 0} &
  \multicolumn{1}{l|}{\cellcolor[HTML]{FFF9F9}0 0 1 0} &
  \multicolumn{1}{l|}{\cellcolor[HTML]{FFE8E7}0 0 3 0} &
  \cellcolor[HTML]{EAF7F1}{\ul \textbf{0.6}} \\ \hline
\rowcolor[HTML]{FFFFFF} 
\multicolumn{1}{|l|}{\cellcolor[HTML]{FFFFFF}\textbf{\begin{tabular}[c]{@{}l@{}}Back \\ Central\\ Light\end{tabular}}} &
  \multicolumn{1}{l|}{\cellcolor[HTML]{FFFFFF}NA} &
  \multicolumn{1}{l|}{\cellcolor[HTML]{FFFFFF}0 0 0 0} &
  \multicolumn{1}{l|}{\cellcolor[HTML]{FFFFFF}NA} &
  \multicolumn{1}{l|}{\cellcolor[HTML]{FFF9F9}0 1 0 0} &
  \multicolumn{1}{l|}{\cellcolor[HTML]{FFFFFF}NA} &
  \multicolumn{1}{l|}{\cellcolor[HTML]{FFFFFF}0 0 0 0} &
  \multicolumn{1}{l|}{\cellcolor[HTML]{FFFFFF}NA} &
  \multicolumn{1}{l|}{\cellcolor[HTML]{FFFFFF}0 0 0 0} &
  \multicolumn{1}{l|}{\cellcolor[HTML]{FFFFFF}0 0 0 0} &
  \multicolumn{1}{l|}{\cellcolor[HTML]{FFFFFF}0 0 0 0} &
  \multicolumn{1}{l|}{\cellcolor[HTML]{FFF9F9}0 0 1 0} &
  \multicolumn{1}{l|}{\cellcolor[HTML]{FFFFFF}0 0 0 0} &
  \multicolumn{1}{l|}{\cellcolor[HTML]{FFF9F9}0 0 1 0} &
  \multicolumn{1}{l|}{\cellcolor[HTML]{FFFFFF}0 0 0 0} &
  \multicolumn{1}{l|}{\cellcolor[HTML]{FFFFFF}NA} &
  {\ul \textbf{0.05}} \\ \hline
\multicolumn{1}{|l|}{\cellcolor[HTML]{FFFFFF}\textbf{\begin{tabular}[c]{@{}l@{}}Back \\ Fenders\end{tabular}}} &
  \multicolumn{1}{l|}{\cellcolor[HTML]{FFCCCD}0 1 4 1} &
  \multicolumn{1}{l|}{\cellcolor[HTML]{FFCCCD}2 1 3 0} &
  \multicolumn{1}{l|}{\cellcolor[HTML]{FFFFFF}NA} &
  \multicolumn{1}{l|}{\cellcolor[HTML]{FFC3C3}3 0 4 0} &
  \multicolumn{1}{l|}{\cellcolor[HTML]{FFBABA}1 1 3 3} &
  \multicolumn{1}{l|}{\cellcolor[HTML]{FFD5D5}1 0 4 0} &
  \multicolumn{1}{l|}{\cellcolor[HTML]{FFF0F0}1 0 1 0} &
  \multicolumn{1}{l|}{\cellcolor[HTML]{FFC3C3}1 1 5 0} &
  \multicolumn{1}{l|}{\cellcolor[HTML]{FFB1B1}4 2 3 0} &
  \multicolumn{1}{l|}{\cellcolor[HTML]{FFC3C3}1 1 4 1} &
  \multicolumn{1}{l|}{\cellcolor[HTML]{FFE8E7}1 1 1 0} &
  \multicolumn{1}{l|}{\cellcolor[HTML]{FFB1B1}5 0 1 3} &
  \multicolumn{1}{l|}{\cellcolor[HTML]{FFC3C3}1 1 5 0} &
  \multicolumn{1}{l|}{\cellcolor[HTML]{FFE8E7}2 1 0 0} &
  \multicolumn{1}{l|}{\cellcolor[HTML]{FFC3C3}3 1 3 0} &
  \cellcolor[HTML]{C9EADA}{\ul \textbf{1.43}} \\ \hline
\rowcolor[HTML]{FFFFFF} 
\multicolumn{1}{|l|}{\cellcolor[HTML]{FFFFFF}\textbf{\begin{tabular}[c]{@{}l@{}}Back Low \\ Reflector\end{tabular}}} &
  \multicolumn{1}{l|}{\cellcolor[HTML]{FFFFFF}NA} &
  \multicolumn{1}{l|}{\cellcolor[HTML]{FFFFFF}0 0 0 0} &
  \multicolumn{1}{l|}{\cellcolor[HTML]{FFF9F9}0 0 1 0} &
  \multicolumn{1}{l|}{\cellcolor[HTML]{FFF9F9}0 1 0 0} &
  \multicolumn{1}{l|}{\cellcolor[HTML]{FFFFFF}NA} &
  \multicolumn{1}{l|}{\cellcolor[HTML]{FFFFFF}0 0 0 0} &
  \multicolumn{1}{l|}{\cellcolor[HTML]{FFFFFF}NA} &
  \multicolumn{1}{l|}{\cellcolor[HTML]{FFFFFF}0 0 0 0} &
  \multicolumn{1}{l|}{\cellcolor[HTML]{FFF0F0}2 0 0 0} &
  \multicolumn{1}{l|}{\cellcolor[HTML]{FFFFFF}0 0 0 0} &
  \multicolumn{1}{l|}{\cellcolor[HTML]{FFFFFF}0 0 0 0} &
  \multicolumn{1}{l|}{\cellcolor[HTML]{FFFFFF}0 0 0 0} &
  \multicolumn{1}{l|}{\cellcolor[HTML]{FFFFFF}0 0 0 0} &
  \multicolumn{1}{l|}{\cellcolor[HTML]{FFE8E7}0 0 3 0} &
  \multicolumn{1}{l|}{\cellcolor[HTML]{FFFFFF}0 0 0 0} &
  \cellcolor[HTML]{FDFEFE}{\ul \textbf{0.12}} \\ \hline
\rowcolor[HTML]{FFFFFF} 
\multicolumn{1}{|l|}{\cellcolor[HTML]{FFFFFF}{\ul \textbf{Back Plate}}} &
  \multicolumn{1}{l|}{\cellcolor[HTML]{FFFFFF}0 0 0 0} &
  \multicolumn{1}{l|}{\cellcolor[HTML]{FFFFFF}0 0 0 0} &
  \multicolumn{1}{l|}{\cellcolor[HTML]{FFF9F9}0 0 1 0} &
  \multicolumn{1}{l|}{\cellcolor[HTML]{FFFFFF}0 0 0 0} &
  \multicolumn{1}{l|}{\cellcolor[HTML]{FFFFFF}0 0 0 0} &
  \multicolumn{1}{l|}{\cellcolor[HTML]{FFF9F9}0 0 1 0} &
  \multicolumn{1}{l|}{\cellcolor[HTML]{FFFFFF}0 0 0 0} &
  \multicolumn{1}{l|}{\cellcolor[HTML]{FFFFFF}0 0 0 0} &
  \multicolumn{1}{l|}{\cellcolor[HTML]{FFF9F9}0 0 1 0} &
  \multicolumn{1}{l|}{\cellcolor[HTML]{FFF9F9}0 0 1 0} &
  \multicolumn{1}{l|}{\cellcolor[HTML]{FFFFFF}0 0 0 0} &
  \multicolumn{1}{l|}{\cellcolor[HTML]{FFFFFF}0 0 0 0} &
  \multicolumn{1}{l|}{\cellcolor[HTML]{FFFFFF}0 0 0 0} &
  \multicolumn{1}{l|}{\cellcolor[HTML]{FFF9F9}0 0 1 0} &
  \multicolumn{1}{l|}{\cellcolor[HTML]{FFFFFF}0 0 0 0} &
  \cellcolor[HTML]{FEFFFF}{\ul \textbf{0.08}} \\ \hline
\multicolumn{1}{|l|}{\cellcolor[HTML]{FFFFFF}\textbf{\begin{tabular}[c]{@{}l@{}}Back\\ Window\\ Rails\end{tabular}}} &
  \multicolumn{1}{l|}{\cellcolor[HTML]{FF9F9F}0 3 4 4} &
  \multicolumn{1}{l|}{\cellcolor[HTML]{FFE8E7}1 0 2 0} &
  \multicolumn{1}{l|}{\cellcolor[HTML]{FFFFFF}NA} &
  \multicolumn{1}{l|}{\cellcolor[HTML]{FFD5D5}1 0 4 0} &
  \multicolumn{1}{l|}{\cellcolor[HTML]{FFFFFF}NA} &
  \multicolumn{1}{l|}{\cellcolor[HTML]{FFF9F9}0 0 1 0} &
  \multicolumn{1}{l|}{\cellcolor[HTML]{FFFFFF}NA} &
  \multicolumn{1}{l|}{\cellcolor[HTML]{FFB1B1}0 0 4 5} &
  \multicolumn{1}{l|}{\cellcolor[HTML]{FFF0F0}0 0 2 0} &
  \multicolumn{1}{l|}{\cellcolor[HTML]{FFE8E7}0 1 2 0} &
  \multicolumn{1}{l|}{\cellcolor[HTML]{FFFFFF}0 0 0 0} &
  \multicolumn{1}{l|}{\cellcolor[HTML]{FFE8E7}0 0 0 3} &
  \multicolumn{1}{l|}{\cellcolor[HTML]{FFA9A7}3 1 4 2} &
  \multicolumn{1}{l|}{\cellcolor[HTML]{FFE8E7}1 0 2 0} &
  \multicolumn{1}{l|}{\cellcolor[HTML]{FFFFFF}NA} &
  \cellcolor[HTML]{E1F3EA}{\ul \textbf{0.83}} \\ \hline
\rowcolor[HTML]{FFFFFF} 
\multicolumn{1}{|l|}{\cellcolor[HTML]{FFFFFF}\textbf{\begin{tabular}[c]{@{}l@{}}Back\\ Window\end{tabular}}} &
  \multicolumn{1}{l|}{\cellcolor[HTML]{FFFFFF}NA} &
  \multicolumn{1}{l|}{\cellcolor[HTML]{FFFFFF}0 0 0 0} &
  \multicolumn{1}{l|}{\cellcolor[HTML]{FFFFFF}NA} &
  \multicolumn{1}{l|}{\cellcolor[HTML]{FFFFFF}0 0 0 0} &
  \multicolumn{1}{l|}{\cellcolor[HTML]{FFFFFF}NA} &
  \multicolumn{1}{l|}{\cellcolor[HTML]{FFD5D5}0 1 4 0} &
  \multicolumn{1}{l|}{\cellcolor[HTML]{FFFFFF}NA} &
  \multicolumn{1}{l|}{\cellcolor[HTML]{FFFFFF}NA} &
  \multicolumn{1}{l|}{\cellcolor[HTML]{FFF0F0}0 1 1 0} &
  \multicolumn{1}{l|}{\cellcolor[HTML]{FFCCCD}0 0 3 3} &
  \multicolumn{1}{l|}{\cellcolor[HTML]{FFF0F0}1 1 0 0} &
  \multicolumn{1}{l|}{\cellcolor[HTML]{FFFFFF}NA} &
  \multicolumn{1}{l|}{\cellcolor[HTML]{FFC3C3}1 0 5 1} &
  \multicolumn{1}{l|}{\cellcolor[HTML]{FFF9F9}0 0 1 0} &
  \multicolumn{1}{l|}{\cellcolor[HTML]{FFFFFF}NA} &
  \cellcolor[HTML]{F2FAF6}{\ul \textbf{0.38}} \\ \hline
\rowcolor[HTML]{FFFFFF} 
\multicolumn{1}{|l|}{\cellcolor[HTML]{FFFFFF}\textbf{\begin{tabular}[c]{@{}l@{}}Back\\ Window\\ Lower\\ Frame\end{tabular}}} &
  \multicolumn{1}{l|}{\cellcolor[HTML]{FFFFFF}NA} &
  \multicolumn{1}{l|}{\cellcolor[HTML]{FFFFFF}0 0 0 0} &
  \multicolumn{1}{l|}{\cellcolor[HTML]{FFFFFF}NA} &
  \multicolumn{1}{l|}{\cellcolor[HTML]{FFFFFF}0 0 0 0} &
  \multicolumn{1}{l|}{\cellcolor[HTML]{FFFFFF}NA} &
  \multicolumn{1}{l|}{\cellcolor[HTML]{FFFFFF}0 0 0 0} &
  \multicolumn{1}{l|}{\cellcolor[HTML]{FFFFFF}NA} &
  \multicolumn{1}{l|}{\cellcolor[HTML]{FFF9F9}0 0 1 0} &
  \multicolumn{1}{l|}{\cellcolor[HTML]{FFF9F9}0 0 1 0} &
  \multicolumn{1}{l|}{\cellcolor[HTML]{FFF9F9}0 0 1 0} &
  \multicolumn{1}{l|}{\cellcolor[HTML]{FFFFFF}0 0 0 0} &
  \multicolumn{1}{l|}{\cellcolor[HTML]{FFFFFF}NA} &
  \multicolumn{1}{l|}{\cellcolor[HTML]{FFF9F9}0 0 1 0} &
  \multicolumn{1}{l|}{\cellcolor[HTML]{FFF9F9}0 0 1 0} &
  \multicolumn{1}{l|}{\cellcolor[HTML]{FFFFFF}NA} &
  \cellcolor[HTML]{FEFFFF}{\ul \textbf{0.08}} \\ \hline
\multicolumn{1}{|l|}{\cellcolor[HTML]{FFFFFF}\textbf{\begin{tabular}[c]{@{}l@{}}Cowl\\ Cover\end{tabular}}} &
  \multicolumn{1}{l|}{\cellcolor[HTML]{FFFFFF}NA} &
  \multicolumn{1}{l|}{\cellcolor[HTML]{FF5756}5 0 7 7} &
  \multicolumn{1}{l|}{\cellcolor[HTML]{FF9596}2 0 7 3} &
  \multicolumn{1}{l|}{\cellcolor[HTML]{FFCCCD}2 0 1 3} &
  \multicolumn{1}{l|}{\cellcolor[HTML]{FFFFFF}NA} &
  \multicolumn{1}{l|}{\cellcolor[HTML]{FF8384}4 0 7 3} &
  \multicolumn{1}{l|}{\cellcolor[HTML]{FFE8E7}1 0 1 1} &
  \multicolumn{1}{l|}{\cellcolor[HTML]{FFBABA}3 1 1 3} &
  \multicolumn{1}{l|}{\cellcolor[HTML]{FFBABA}2 3 0 3} &
  \multicolumn{1}{l|}{\cellcolor[HTML]{FF9596}2 5 2 3} &
  \multicolumn{1}{l|}{\cellcolor[HTML]{FFFFFF}NA} &
  \multicolumn{1}{l|}{\cellcolor[HTML]{FF9596}4 2 3 3} &
  \multicolumn{1}{l|}{\cellcolor[HTML]{FFF9F9}0 0 0 1} &
  \multicolumn{1}{l|}{\cellcolor[HTML]{FFDEDE}1 0 0 3} &
  \multicolumn{1}{l|}{\cellcolor[HTML]{FFDEDE}2 0 0 2} &
  \cellcolor[HTML]{BEE5D2}{\ul \textbf{1.72}} \\ \hline
\multicolumn{1}{|l|}{\cellcolor[HTML]{FFFFFF}\textbf{\begin{tabular}[c]{@{}l@{}}Front \\ Bumper\end{tabular}}} &
  \multicolumn{1}{l|}{\cellcolor[HTML]{FFFFFF}NA} &
  \multicolumn{1}{l|}{\cellcolor[HTML]{FF2929}7 3 7 7} &
  \multicolumn{1}{l|}{\cellcolor[HTML]{FF9596}3 0 6 3} &
  \multicolumn{1}{l|}{\cellcolor[HTML]{FF7171}2 4 7 3} &
  \multicolumn{1}{l|}{\cellcolor[HTML]{FFFFFF}NA} &
  \multicolumn{1}{l|}{\cellcolor[HTML]{FF605F}7 0 7 4} &
  \multicolumn{1}{l|}{\cellcolor[HTML]{FF9596}3 2 4 3} &
  \multicolumn{1}{l|}{\cellcolor[HTML]{FF605F}4 3 7 4} &
  \multicolumn{1}{l|}{\cellcolor[HTML]{FF9596}4 2 3 3} &
  \multicolumn{1}{l|}{\cellcolor[HTML]{FF4444}4 5 5 7} &
  \multicolumn{1}{l|}{\cellcolor[HTML]{FF5756}7 1 4 7} &
  \multicolumn{1}{l|}{\cellcolor[HTML]{FF4D4D}7 0 6 7} &
  \multicolumn{1}{l|}{\cellcolor[HTML]{FF9596}4 1 3 4} &
  \multicolumn{1}{l|}{\cellcolor[HTML]{FF7171}7 2 4 3} &
  \multicolumn{1}{l|}{\cellcolor[HTML]{FF605F}7 3 5 3} &
  \cellcolor[HTML]{73C79E}{\ul \textbf{3.63}} \\ \hline
\multicolumn{1}{|l|}{\cellcolor[HTML]{FFFFFF}\textbf{\begin{tabular}[c]{@{}l@{}}Lower \\ Deflector\end{tabular}}} &
  \multicolumn{1}{l|}{\cellcolor[HTML]{FFFFFF}NA} &
  \multicolumn{1}{l|}{\cellcolor[HTML]{FF6868}4 3 3 7} &
  \multicolumn{1}{l|}{\cellcolor[HTML]{FFE8E7}0 0 3 0} &
  \multicolumn{1}{l|}{\cellcolor[HTML]{FFFFFF}NA} &
  \multicolumn{1}{l|}{\cellcolor[HTML]{FFFFFF}NA} &
  \multicolumn{1}{l|}{\cellcolor[HTML]{FFE8E7}1 2 0 0} &
  \multicolumn{1}{l|}{\cellcolor[HTML]{FFFFFF}0 0 0 0} &
  \multicolumn{1}{l|}{\cellcolor[HTML]{FFD5D5}1 1 3 0} &
  \multicolumn{1}{l|}{\cellcolor[HTML]{FFA9A7}3 1 3 3} &
  \multicolumn{1}{l|}{\cellcolor[HTML]{FF9F9F}3 2 3 3} &
  \multicolumn{1}{l|}{\cellcolor[HTML]{FFFFFF}NA} &
  \multicolumn{1}{l|}{\cellcolor[HTML]{FFBABA}4 0 2 2} &
  \multicolumn{1}{l|}{\cellcolor[HTML]{FFFFFF}NA} &
  \multicolumn{1}{l|}{\cellcolor[HTML]{FFBABA}2 1 2 3} &
  \multicolumn{1}{l|}{\cellcolor[HTML]{FFDEDE}3 0 0 1} &
  \cellcolor[HTML]{CCEBDC}{\ul \textbf{1.37}} \\ \hline
\multicolumn{1}{|l|}{\cellcolor[HTML]{FFFFFF}\textbf{\begin{tabular}[c]{@{}l@{}}Front\\ Doors\end{tabular}}} &
  \multicolumn{1}{l|}{\cellcolor[HTML]{FFFFFF}NA} &
  \multicolumn{1}{l|}{\cellcolor[HTML]{FFFFFF}NA} &
  \multicolumn{1}{l|}{\cellcolor[HTML]{FFB1B1}1 4 2 2} &
  \multicolumn{1}{l|}{\cellcolor[HTML]{FF9596}1 3 5 3} &
  \multicolumn{1}{l|}{\cellcolor[HTML]{FFFFFF}NA} &
  \multicolumn{1}{l|}{\cellcolor[HTML]{FFF9F9}1 0 0 0} &
  \multicolumn{1}{l|}{\cellcolor[HTML]{FFD5D5}1 1 2 1} &
  \multicolumn{1}{l|}{\cellcolor[HTML]{FFBABA}1 1 3 3} &
  \multicolumn{1}{l|}{\cellcolor[HTML]{FF8C8D}2 1 7 3} &
  \multicolumn{1}{l|}{\cellcolor[HTML]{FF8C8D}1 4 4 4} &
  \multicolumn{1}{l|}{\cellcolor[HTML]{FFC3C3}2 3 0 2} &
  \multicolumn{1}{l|}{\cellcolor[HTML]{FF7171}1 7 6 2} &
  \multicolumn{1}{l|}{\cellcolor[HTML]{FFB1B1}2 0 4 3} &
  \multicolumn{1}{l|}{\cellcolor[HTML]{FFBABA}1 0 4 3} &
  \multicolumn{1}{l|}{\cellcolor[HTML]{FFB1B1}2 0 3 4} &
  \cellcolor[HTML]{BAE3CF}{\ul \textbf{1.83}} \\ \hline
\multicolumn{1}{|l|}{\cellcolor[HTML]{FFFFFF}\textbf{\begin{tabular}[c]{@{}l@{}}Front \\ Fenders\end{tabular}}} &
  \multicolumn{1}{l|}{\cellcolor[HTML]{FF9596}1 1 7 3} &
  \multicolumn{1}{l|}{\cellcolor[HTML]{FF5756}2 7 7 3} &
  \multicolumn{1}{l|}{\cellcolor[HTML]{FFBABA}1 1 4 2} &
  \multicolumn{1}{l|}{\cellcolor[HTML]{FFB1B1}2 3 3 1} &
  \multicolumn{1}{l|}{\cellcolor[HTML]{FF2929}5 7 7 5} &
  \multicolumn{1}{l|}{\cellcolor[HTML]{FF9F9F}2 0 7 2} &
  \multicolumn{1}{l|}{\cellcolor[HTML]{FF9596}1 4 4 3} &
  \multicolumn{1}{l|}{\cellcolor[HTML]{FF4D4D}3 7 7 3} &
  \multicolumn{1}{l|}{\cellcolor[HTML]{FF7171}4 7 3 2} &
  \multicolumn{1}{l|}{\cellcolor[HTML]{FF4444}3 7 7 4} &
  \multicolumn{1}{l|}{\cellcolor[HTML]{FF8C8D}6 1 4 2} &
  \multicolumn{1}{l|}{\cellcolor[HTML]{FF8384}3 3 5 3} &
  \multicolumn{1}{l|}{\cellcolor[HTML]{FF4444}2 6 7 6} &
  \multicolumn{1}{l|}{\cellcolor[HTML]{FF4444}5 7 6 3} &
  \multicolumn{1}{l|}{\cellcolor[HTML]{FF3231}7 7 6 3} &
  \cellcolor[HTML]{63C092}{\ul \textbf{4.07}} \\ \hline
\multicolumn{1}{|l|}{\cellcolor[HTML]{FFFFFF}\textbf{\begin{tabular}[c]{@{}l@{}}Front\\ Plate\end{tabular}}} &
  \multicolumn{1}{l|}{\cellcolor[HTML]{FFB1B1}4 0 2 3} &
  \multicolumn{1}{l|}{\cellcolor[HTML]{FFBABA}2 0 3 3} &
  \multicolumn{1}{l|}{\cellcolor[HTML]{FFBABA}2 0 3 3} &
  \multicolumn{1}{l|}{\cellcolor[HTML]{FFD5D5}1 0 3 1} &
  \multicolumn{1}{l|}{\cellcolor[HTML]{FFFFFF}NA} &
  \multicolumn{1}{l|}{\cellcolor[HTML]{FF9596}4 7 1 0} &
  \multicolumn{1}{l|}{\cellcolor[HTML]{FFC3C3}1 0 3 3} &
  \multicolumn{1}{l|}{\cellcolor[HTML]{FFE8E7}1 0 1 1} &
  \multicolumn{1}{l|}{\cellcolor[HTML]{FFF0F0}1 0 0 1} &
  \multicolumn{1}{l|}{\cellcolor[HTML]{FFA9A7}3 0 2 5} &
  \multicolumn{1}{l|}{\cellcolor[HTML]{FFC3C3}2 0 2 3} &
  \multicolumn{1}{l|}{\cellcolor[HTML]{FFC3C3}4 0 0 3} &
  \multicolumn{1}{l|}{\cellcolor[HTML]{FFC3C3}3 0 1 3} &
  \multicolumn{1}{l|}{\cellcolor[HTML]{FFBABA}3 0 2 3} &
  \multicolumn{1}{l|}{\cellcolor[HTML]{FFF0F0}1 0 0 1} &
  \cellcolor[HTML]{C4E7D6}{\ul \textbf{1.58}} \\ \hline
\multicolumn{1}{|l|}{\cellcolor[HTML]{FFFFFF}\textbf{\begin{tabular}[c]{@{}l@{}}Front \\ Wheels\end{tabular}}} &
  \multicolumn{1}{l|}{\cellcolor[HTML]{FF3B3A}5 5 7 5} &
  \multicolumn{1}{l|}{\cellcolor[HTML]{FFB1B1}3 0 2 4} &
  \multicolumn{1}{l|}{\cellcolor[HTML]{FF7171}7 2 4 3} &
  \multicolumn{1}{l|}{\cellcolor[HTML]{FF7A7A}4 6 2 3} &
  \multicolumn{1}{l|}{\cellcolor[HTML]{FF4D4D}4 5 5 6} &
  \multicolumn{1}{l|}{\cellcolor[HTML]{FF8384}3 4 4 3} &
  \multicolumn{1}{l|}{\cellcolor[HTML]{FF4444}6 7 5 3} &
  \multicolumn{1}{l|}{\cellcolor[HTML]{FF4444}4 7 7 3} &
  \multicolumn{1}{l|}{\cellcolor[HTML]{FF5756}6 7 3 3} &
  \multicolumn{1}{l|}{\cellcolor[HTML]{FF605F}3 5 7 3} &
  \multicolumn{1}{l|}{\cellcolor[HTML]{FFDEDE}3 0 1 0} &
  \multicolumn{1}{l|}{\cellcolor[HTML]{FF8384}3 1 5 5} &
  \multicolumn{1}{l|}{\cellcolor[HTML]{FF3231}7 7 5 4} &
  \multicolumn{1}{l|}{\cellcolor[HTML]{FF3B3A}5 7 7 3} &
  \multicolumn{1}{l|}{\cellcolor[HTML]{FF3231}7 7 7 2} &
  \cellcolor[HTML]{57BB8A}{\ul \textbf{4.35}} \\ \hline
\multicolumn{1}{|l|}{\cellcolor[HTML]{FFFFFF}\textbf{\begin{tabular}[c]{@{}l@{}}Front \\ Windows\end{tabular}}} &
  \multicolumn{1}{l|}{\cellcolor[HTML]{FFFFFF}NA} &
  \multicolumn{1}{l|}{\cellcolor[HTML]{FF9596}1 7 0 4} &
  \multicolumn{1}{l|}{\cellcolor[HTML]{FFCCCD}1 0 3 2} &
  \multicolumn{1}{l|}{\cellcolor[HTML]{FFFFFF}0 0 0 0} &
  \multicolumn{1}{l|}{\cellcolor[HTML]{FFFFFF}NA} &
  \multicolumn{1}{l|}{\cellcolor[HTML]{FF9596}1 1 7 3} &
  \multicolumn{1}{l|}{\cellcolor[HTML]{FFB1B1}1 0 5 3} &
  \multicolumn{1}{l|}{\cellcolor[HTML]{FFC3C3}1 0 3 3} &
  \multicolumn{1}{l|}{\cellcolor[HTML]{FF9F9F}2 0 7 2} &
  \multicolumn{1}{l|}{\cellcolor[HTML]{FF5756}1 7 4 7} &
  \multicolumn{1}{l|}{\cellcolor[HTML]{FFB1B1}1 3 3 2} &
  \multicolumn{1}{l|}{\cellcolor[HTML]{FF8384}1 6 5 2} &
  \multicolumn{1}{l|}{\cellcolor[HTML]{FFB1B1}1 0 3 5} &
  \multicolumn{1}{l|}{\cellcolor[HTML]{FFC3C3}1 0 3 3} &
  \multicolumn{1}{l|}{\cellcolor[HTML]{FFFFFF}NA} &
  \cellcolor[HTML]{B7E2CD}{\ul \textbf{1.92}} \\ \hline
\multicolumn{1}{|l|}{\cellcolor[HTML]{FFFFFF}{\ul \textbf{Grill}}} &
  \multicolumn{1}{l|}{\cellcolor[HTML]{FFFFFF}NA} &
  \multicolumn{1}{l|}{\cellcolor[HTML]{FF8C8D}4 0 4 5} &
  \multicolumn{1}{l|}{\cellcolor[HTML]{FFBABA}2 0 3 3} &
  \multicolumn{1}{l|}{\cellcolor[HTML]{FFBABA}1 1 3 3} &
  \multicolumn{1}{l|}{\cellcolor[HTML]{FFFFFF}NA} &
  \multicolumn{1}{l|}{\cellcolor[HTML]{FF9F9F}4 0 4 3} &
  \multicolumn{1}{l|}{\cellcolor[HTML]{FFBABA}2 0 3 3} &
  \multicolumn{1}{l|}{\cellcolor[HTML]{FFA9A7}3 0 3 4} &
  \multicolumn{1}{l|}{\cellcolor[HTML]{FFC3C3}3 0 1 3} &
  \multicolumn{1}{l|}{\cellcolor[HTML]{FF9F9F}4 0 3 4} &
  \multicolumn{1}{l|}{\cellcolor[HTML]{FFBABA}3 0 1 4} &
  \multicolumn{1}{l|}{\cellcolor[HTML]{FFA9A7}4 0 2 4} &
  \multicolumn{1}{l|}{\cellcolor[HTML]{FF8C8D}4 0 2 7} &
  \multicolumn{1}{l|}{\cellcolor[HTML]{FFB1B1}4 0 2 3} &
  \multicolumn{1}{l|}{\cellcolor[HTML]{FFD5D5}3 0 1 1} &
  \cellcolor[HTML]{B3E0CA}{\ul \textbf{2.02}} \\ \hline
\multicolumn{1}{|l|}{\cellcolor[HTML]{FFFFFF}{\ul \textbf{Headlights}}} &
  \multicolumn{1}{l|}{\cellcolor[HTML]{FF8C8D}4 0 6 3} &
  \multicolumn{1}{l|}{\cellcolor[HTML]{FF4D4D}6 1 6 7} &
  \multicolumn{1}{l|}{\cellcolor[HTML]{FF8C8D}3 0 6 4} &
  \multicolumn{1}{l|}{\cellcolor[HTML]{FFC3C3}2 1 3 1} &
  \multicolumn{1}{l|}{\cellcolor[HTML]{FF605F}7 1 3 7} &
  \multicolumn{1}{l|}{\cellcolor[HTML]{FF6868}7 0 7 3} &
  \multicolumn{1}{l|}{\cellcolor[HTML]{FF8384}3 0 7 4} &
  \multicolumn{1}{l|}{\cellcolor[HTML]{FF8384}3 1 7 3} &
  \multicolumn{1}{l|}{\cellcolor[HTML]{FF9F9F}4 1 3 3} &
  \multicolumn{1}{l|}{\cellcolor[HTML]{FF6868}4 4 5 4} &
  \multicolumn{1}{l|}{\cellcolor[HTML]{FF8384}4 0 3 7} &
  \multicolumn{1}{l|}{\cellcolor[HTML]{FF9596}4 2 3 3} &
  \multicolumn{1}{l|}{\cellcolor[HTML]{FF5756}7 0 5 7} &
  \multicolumn{1}{l|}{\cellcolor[HTML]{FF6868}7 3 4 3} &
  \multicolumn{1}{l|}{\cellcolor[HTML]{FF605F}7 3 4 4} &
  \cellcolor[HTML]{70C59B}{\ul \textbf{3.73}} \\ \hline
\multicolumn{1}{|l|}{\cellcolor[HTML]{FFFFFF}{\ul \textbf{Hood}}} &
  \multicolumn{1}{l|}{\cellcolor[HTML]{FFBABA}1 1 2 4} &
  \multicolumn{1}{l|}{\cellcolor[HTML]{FFFFFF}NA} &
  \multicolumn{1}{l|}{\cellcolor[HTML]{FFBABA}2 0 3 3} &
  \multicolumn{1}{l|}{\cellcolor[HTML]{FFE8E7}1 0 1 1} &
  \multicolumn{1}{l|}{\cellcolor[HTML]{FFFFFF}NA} &
  \multicolumn{1}{l|}{\cellcolor[HTML]{FF9F9F}4 0 4 3} &
  \multicolumn{1}{l|}{\cellcolor[HTML]{FFC3C3}1 1 3 2} &
  \multicolumn{1}{l|}{\cellcolor[HTML]{FFC3C3}3 1 1 2} &
  \multicolumn{1}{l|}{\cellcolor[HTML]{FFD5D5}1 1 2 1} &
  \multicolumn{1}{l|}{\cellcolor[HTML]{FF9596}2 3 3 4} &
  \multicolumn{1}{l|}{\cellcolor[HTML]{FFFFFF}NA} &
  \multicolumn{1}{l|}{\cellcolor[HTML]{FFBABA}4 0 1 3} &
  \multicolumn{1}{l|}{\cellcolor[HTML]{FFCCCD}1 2 1 2} &
  \multicolumn{1}{l|}{\cellcolor[HTML]{FFC3C3}2 2 2 1} &
  \multicolumn{1}{l|}{\cellcolor[HTML]{FFC3C3}1 2 2 2} &
  \cellcolor[HTML]{C7E9D8}{\ul \textbf{1.48}} \\ \hline
\rowcolor[HTML]{FFFFFF} 
\multicolumn{1}{|l|}{\cellcolor[HTML]{FFFFFF}{\ul \textbf{Rear Doors}}} &
  \multicolumn{1}{l|}{\cellcolor[HTML]{FFFFFF}NA} &
  \multicolumn{1}{l|}{\cellcolor[HTML]{FFFFFF}NA} &
  \multicolumn{1}{l|}{\cellcolor[HTML]{FFFFFF}NA} &
  \multicolumn{1}{l|}{\cellcolor[HTML]{FFBABA}1 1 6 0} &
  \multicolumn{1}{l|}{\cellcolor[HTML]{FFFFFF}NA} &
  \multicolumn{1}{l|}{\cellcolor[HTML]{FFFFFF}NA} &
  \multicolumn{1}{l|}{\cellcolor[HTML]{FFFFFF}NA} &
  \multicolumn{1}{l|}{\cellcolor[HTML]{FFDEDE}1 1 2 0} &
  \multicolumn{1}{l|}{\cellcolor[HTML]{FFCCCD}1 1 4 0} &
  \multicolumn{1}{l|}{\cellcolor[HTML]{FF9F9F}1 0 3 7} &
  \multicolumn{1}{l|}{\cellcolor[HTML]{FFFFFF}NA} &
  \multicolumn{1}{l|}{\cellcolor[HTML]{FFFFFF}NA} &
  \multicolumn{1}{l|}{\cellcolor[HTML]{FFCCCD}2 0 3 1} &
  \multicolumn{1}{l|}{\cellcolor[HTML]{FFD5D5}1 0 3 1} &
  \multicolumn{1}{l|}{\cellcolor[HTML]{FFFFFF}NA} &
  \cellcolor[HTML]{E7F6EF}{\ul \textbf{0.67}} \\ \hline
\multicolumn{1}{|l|}{\cellcolor[HTML]{FFFFFF}\textbf{\begin{tabular}[c]{@{}l@{}}Rear\\ Wheels\end{tabular}}} &
  \multicolumn{1}{l|}{\cellcolor[HTML]{FFC3C3}1 0 4 2} &
  \multicolumn{1}{l|}{\cellcolor[HTML]{FF9596}4 5 3 0} &
  \multicolumn{1}{l|}{\cellcolor[HTML]{FF3B3A}7 7 7 1} &
  \multicolumn{1}{l|}{\cellcolor[HTML]{FF6868}7 7 3 0} &
  \multicolumn{1}{l|}{\cellcolor[HTML]{FFC3C3}2 1 1 3} &
  \multicolumn{1}{l|}{\cellcolor[HTML]{FFC3C3}3 0 4 0} &
  \multicolumn{1}{l|}{\cellcolor[HTML]{FF7A7A}7 0 7 1} &
  \multicolumn{1}{l|}{\cellcolor[HTML]{FF8C8D}7 1 4 1} &
  \multicolumn{1}{l|}{\cellcolor[HTML]{FF8C8D}4 4 5 0} &
  \multicolumn{1}{l|}{\cellcolor[HTML]{FFBABA}4 0 3 1} &
  \multicolumn{1}{l|}{\cellcolor[HTML]{FFCCCD}3 2 1 0} &
  \multicolumn{1}{l|}{\cellcolor[HTML]{FF9596}5 0 2 5} &
  \multicolumn{1}{l|}{\cellcolor[HTML]{FFA9A7}5 0 3 2} &
  \multicolumn{1}{l|}{\cellcolor[HTML]{FF9596}7 1 3 1} &
  \multicolumn{1}{l|}{\cellcolor[HTML]{FF9596}7 1 3 1} &
  \cellcolor[HTML]{91D3B2}{\ul \textbf{2.88}} \\ \hline
\multicolumn{1}{|l|}{\cellcolor[HTML]{FFFFFF}\textbf{\begin{tabular}[c]{@{}l@{}}Rear \\ Windows\end{tabular}}} &
  \multicolumn{1}{l|}{\cellcolor[HTML]{FFFFFF}NA} &
  \multicolumn{1}{l|}{\cellcolor[HTML]{FFBABA}1 4 3 0} &
  \multicolumn{1}{l|}{\cellcolor[HTML]{FFFFFF}NA} &
  \multicolumn{1}{l|}{\cellcolor[HTML]{FFF9F9}1 0 0 0} &
  \multicolumn{1}{l|}{\cellcolor[HTML]{FFFFFF}NA} &
  \multicolumn{1}{l|}{\cellcolor[HTML]{FFF0F0}1 0 1 0} &
  \multicolumn{1}{l|}{\cellcolor[HTML]{FFFFFF}NA} &
  \multicolumn{1}{l|}{\cellcolor[HTML]{FFFFFF}NA} &
  \multicolumn{1}{l|}{\cellcolor[HTML]{FFBABA}3 1 3 1} &
  \multicolumn{1}{l|}{\cellcolor[HTML]{FFBABA}1 0 3 4} &
  \multicolumn{1}{l|}{\cellcolor[HTML]{FFBABA}2 3 3 0} &
  \multicolumn{1}{l|}{\cellcolor[HTML]{FFFFFF}NA} &
  \multicolumn{1}{l|}{\cellcolor[HTML]{FFC3C3}1 0 5 1} &
  \multicolumn{1}{l|}{\cellcolor[HTML]{FFBABA}2 2 3 1} &
  \multicolumn{1}{l|}{\cellcolor[HTML]{FFFFFF}NA} &
  \cellcolor[HTML]{E1F3EA}{\ul \textbf{0.83}} \\ \hline
\multicolumn{1}{|l|}{\cellcolor[HTML]{FFFFFF}\textbf{\begin{tabular}[c]{@{}l@{}}Rocker\\ Panel\end{tabular}}} &
  \multicolumn{1}{l|}{\cellcolor[HTML]{FF7A7A}4 4 7 0} &
  \multicolumn{1}{l|}{\cellcolor[HTML]{FFFFFF}NA} &
  \multicolumn{1}{l|}{\cellcolor[HTML]{FFCCCD}2 0 4 0} &
  \multicolumn{1}{l|}{\cellcolor[HTML]{FFF9F9}0 0 0 1} &
  \multicolumn{1}{l|}{\cellcolor[HTML]{FFFFFF}NA} &
  \multicolumn{1}{l|}{\cellcolor[HTML]{FFD5D5}0 0 5 0} &
  \multicolumn{1}{l|}{\cellcolor[HTML]{FFF0F0}1 0 1 0} &
  \multicolumn{1}{l|}{\cellcolor[HTML]{FFFFFF}0 0 0 0} &
  \multicolumn{1}{l|}{\cellcolor[HTML]{FFCCCD}1 0 4 1} &
  \multicolumn{1}{l|}{\cellcolor[HTML]{FFE8E7}0 0 0 3} &
  \multicolumn{1}{l|}{\cellcolor[HTML]{FFFFFF}NA} &
  \multicolumn{1}{l|}{\cellcolor[HTML]{FFC3C3}0 3 3 1} &
  \multicolumn{1}{l|}{\cellcolor[HTML]{FFFFFF}0 0 0 0} &
  \multicolumn{1}{l|}{\cellcolor[HTML]{FFDEDE}0 0 4 0} &
  \multicolumn{1}{l|}{\cellcolor[HTML]{FFFFFF}0 0 0 0} &
  \cellcolor[HTML]{E2F3EB}{\ul \textbf{0.82}} \\ \hline
\multicolumn{1}{|l|}{\cellcolor[HTML]{FFFFFF}{\ul \textbf{Roof}}} &
  \multicolumn{1}{l|}{\cellcolor[HTML]{FF7171}2 7 5 2} &
  \multicolumn{1}{l|}{\cellcolor[HTML]{FFBABA}2 2 1 3} &
  \multicolumn{1}{l|}{\cellcolor[HTML]{FFB1B1}2 3 3 1} &
  \multicolumn{1}{l|}{\cellcolor[HTML]{FFCCCD}0 2 1 3} &
  \multicolumn{1}{l|}{\cellcolor[HTML]{FFFFFF}NA} &
  \multicolumn{1}{l|}{\cellcolor[HTML]{FFB1B1}1 0 4 4} &
  \multicolumn{1}{l|}{\cellcolor[HTML]{FFBABA}1 0 4 3} &
  \multicolumn{1}{l|}{\cellcolor[HTML]{FFCCCD}1 0 3 2} &
  \multicolumn{1}{l|}{\cellcolor[HTML]{FFC3C3}1 0 4 2} &
  \multicolumn{1}{l|}{\cellcolor[HTML]{FFA9A7}1 2 3 4} &
  \multicolumn{1}{l|}{\cellcolor[HTML]{FF7171}4 3 5 4} &
  \multicolumn{1}{l|}{\cellcolor[HTML]{FFB1B1}1 3 3 2} &
  \multicolumn{1}{l|}{\cellcolor[HTML]{FF9F9F}2 1 4 4} &
  \multicolumn{1}{l|}{\cellcolor[HTML]{FFDEDE}1 0 2 1} &
  \multicolumn{1}{l|}{\cellcolor[HTML]{FFFFFF}NA} &
  \cellcolor[HTML]{B4E1CB}{\ul \textbf{1.98}} \\ \hline
\multicolumn{1}{|l|}{\cellcolor[HTML]{FFFFFF}\textbf{\begin{tabular}[c]{@{}l@{}}Side\\ Mirrors\end{tabular}}} &
  \multicolumn{1}{l|}{\cellcolor[HTML]{FFFFFF}NA} &
  \multicolumn{1}{l|}{\cellcolor[HTML]{FF5756}7 0 7 5} &
  \multicolumn{1}{l|}{\cellcolor[HTML]{FF8384}4 2 5 3} &
  \multicolumn{1}{l|}{\cellcolor[HTML]{FF9596}4 0 1 7} &
  \multicolumn{1}{l|}{\cellcolor[HTML]{FF5756}7 6 3 3} &
  \multicolumn{1}{l|}{\cellcolor[HTML]{FF4444}7 0 7 7} &
  \multicolumn{1}{l|}{\cellcolor[HTML]{FF8384}4 0 3 7} &
  \multicolumn{1}{l|}{\cellcolor[HTML]{FF6868}5 0 5 7} &
  \multicolumn{1}{l|}{\cellcolor[HTML]{FF5756}7 0 5 7} &
  \multicolumn{1}{l|}{\cellcolor[HTML]{FF8384}4 0 5 5} &
  \multicolumn{1}{l|}{\cellcolor[HTML]{FF7171}7 1 6 2} &
  \multicolumn{1}{l|}{\cellcolor[HTML]{FFA9A7}3 2 2 3} &
  \multicolumn{1}{l|}{\cellcolor[HTML]{FF8384}5 0 3 6} &
  \multicolumn{1}{l|}{\cellcolor[HTML]{FF6868}5 1 4 7} &
  \multicolumn{1}{l|}{\cellcolor[HTML]{FF6868}7 0 3 7} &
  \cellcolor[HTML]{70C69C}{\ul \textbf{3.72}} \\ \hline
\rowcolor[HTML]{FFFFFF} 
\multicolumn{1}{|l|}{\cellcolor[HTML]{FFFFFF}{\ul \textbf{Tail Lights}}} &
  \multicolumn{1}{l|}{\cellcolor[HTML]{FFFFFF}0 0 0 0} &
  \multicolumn{1}{l|}{\cellcolor[HTML]{FFFFFF}0 0 0 0} &
  \multicolumn{1}{l|}{\cellcolor[HTML]{FFFFFF}0 0 0 0} &
  \multicolumn{1}{l|}{\cellcolor[HTML]{FFFFFF}0 0 0 0} &
  \multicolumn{1}{l|}{\cellcolor[HTML]{FFFFFF}0 0 0 0} &
  \multicolumn{1}{l|}{\cellcolor[HTML]{FFF9F9}0 0 1 0} &
  \multicolumn{1}{l|}{\cellcolor[HTML]{FFFFFF}0 0 0 0} &
  \multicolumn{1}{l|}{\cellcolor[HTML]{FFFFFF}0 0 0 0} &
  \multicolumn{1}{l|}{\cellcolor[HTML]{FFD5D5}4 0 1 0} &
  \multicolumn{1}{l|}{\cellcolor[HTML]{FFDEDE}0 0 4 0} &
  \multicolumn{1}{l|}{\cellcolor[HTML]{FFFFFF}0 0 0 0} &
  \multicolumn{1}{l|}{\cellcolor[HTML]{FFE8E7}0 0 0 3} &
  \multicolumn{1}{l|}{\cellcolor[HTML]{FFFFFF}0 0 0 0} &
  \multicolumn{1}{l|}{\cellcolor[HTML]{FFFFFF}0 0 0 0} &
  \multicolumn{1}{l|}{\cellcolor[HTML]{FFFFFF}0 0 0 0} &
  \cellcolor[HTML]{F9FDFB}{\ul \textbf{0.22}} \\ \hline
\multicolumn{1}{|l|}{\cellcolor[HTML]{FFFFFF}{\ul \textbf{Trunk}}} &
  \multicolumn{1}{l|}{\cellcolor[HTML]{FFFFFF}NA} &
  \multicolumn{1}{l|}{\cellcolor[HTML]{FFFFFF}NA} &
  \multicolumn{1}{l|}{\cellcolor[HTML]{FFFFFF}0 0 0 0} &
  \multicolumn{1}{l|}{\cellcolor[HTML]{FFF0F0}0 2 0 0} &
  \multicolumn{1}{l|}{\cellcolor[HTML]{FFFFFF}NA} &
  \multicolumn{1}{l|}{\cellcolor[HTML]{FFE8E7}0 0 3 0} &
  \multicolumn{1}{l|}{\cellcolor[HTML]{FFF9F9}0 0 1 0} &
  \multicolumn{1}{l|}{\cellcolor[HTML]{FFFFFF}0 0 0 0} &
  \multicolumn{1}{l|}{\cellcolor[HTML]{FFFFFF}NA} &
  \multicolumn{1}{l|}{\cellcolor[HTML]{FFE8E7}0 0 3 0} &
  \multicolumn{1}{l|}{\cellcolor[HTML]{FFFFFF}NA} &
  \multicolumn{1}{l|}{\cellcolor[HTML]{FFF9F9}0 0 1 0} &
  \multicolumn{1}{l|}{\cellcolor[HTML]{FFF9F9}0 0 1 0} &
  \multicolumn{1}{l|}{\cellcolor[HTML]{FFF9F9}0 0 1 0} &
  \multicolumn{1}{l|}{\cellcolor[HTML]{FFF0F0}0 0 2 0} &
  \cellcolor[HTML]{F8FDFB}{\ul \textbf{0.23}} \\ \hline
\multicolumn{1}{|l|}{\cellcolor[HTML]{FFFFFF}{\ul \textbf{Windshield}}} &
  \multicolumn{1}{l|}{\cellcolor[HTML]{FF605F}7 0 4 7} &
  \multicolumn{1}{l|}{\cellcolor[HTML]{FF3B3A}7 1 7 7} &
  \multicolumn{1}{l|}{\cellcolor[HTML]{FFC3C3}2 0 3 2} &
  \multicolumn{1}{l|}{\cellcolor[HTML]{FFC3C3}2 0 2 3} &
  \multicolumn{1}{l|}{\cellcolor[HTML]{FFFFFF}NA} &
  \multicolumn{1}{l|}{\cellcolor[HTML]{FF605F}4 3 7 4} &
  \multicolumn{1}{l|}{\cellcolor[HTML]{FFD5D5}1 0 1 3} &
  \multicolumn{1}{l|}{\cellcolor[HTML]{FFBABA}2 1 1 4} &
  \multicolumn{1}{l|}{\cellcolor[HTML]{FFCCCD}2 0 1 3} &
  \multicolumn{1}{l|}{\cellcolor[HTML]{FFA9A7}2 1 2 5} &
  \multicolumn{1}{l|}{\cellcolor[HTML]{FF7A7A}4 0 7 4} &
  \multicolumn{1}{l|}{\cellcolor[HTML]{FFA9A7}4 2 2 2} &
  \multicolumn{1}{l|}{\cellcolor[HTML]{FFB1B1}3 0 2 4} &
  \multicolumn{1}{l|}{\cellcolor[HTML]{FFC3C3}3 0 1 3} &
  \multicolumn{1}{l|}{\cellcolor[HTML]{FFC3C3}2 0 3 2} &
  \cellcolor[HTML]{A0D9BD}{\ul \textbf{2.48}} \\ \hline
\multicolumn{1}{|l|}{\cellcolor[HTML]{FFFFFF}\textbf{\begin{tabular}[c]{@{}l@{}}Windshield \\ Rails\end{tabular}}} &
  \multicolumn{1}{l|}{\cellcolor[HTML]{FF7171}3 3 3 7} &
  \multicolumn{1}{l|}{\cellcolor[HTML]{FF6868}5 1 7 4} &
  \multicolumn{1}{l|}{\cellcolor[HTML]{FFA9A7}1 1 7 1} &
  \multicolumn{1}{l|}{\cellcolor[HTML]{FFBABA}1 3 1 3} &
  \multicolumn{1}{l|}{\cellcolor[HTML]{FFFFFF}NA} &
  \multicolumn{1}{l|}{\cellcolor[HTML]{FFB1B1}1 3 3 2} &
  \multicolumn{1}{l|}{\cellcolor[HTML]{FFC3C3}1 0 3 3} &
  \multicolumn{1}{l|}{\cellcolor[HTML]{FF8C8D}3 3 4 3} &
  \multicolumn{1}{l|}{\cellcolor[HTML]{FF9F9F}2 0 6 3} &
  \multicolumn{1}{l|}{\cellcolor[HTML]{FF7A7A}3 2 5 5} &
  \multicolumn{1}{l|}{\cellcolor[HTML]{FF7A7A}3 2 7 3} &
  \multicolumn{1}{l|}{\cellcolor[HTML]{FFA9A7}1 2 5 2} &
  \multicolumn{1}{l|}{\cellcolor[HTML]{FF9F9F}3 0 4 4} &
  \multicolumn{1}{l|}{\cellcolor[HTML]{FFB1B1}1 1 4 3} &
  \multicolumn{1}{l|}{\cellcolor[HTML]{FF9F9F}2 1 5 3} &
  \cellcolor[HTML]{98D6B7}{\ul \textbf{2.7}} \\ \hline
\end{tabular}%
}
\caption{The highest \textit{Exposedness Score} found on any point of each part, for each vehicle, for points between 5\si{m} and 25\si{m} away from the camera, and for each of the four driving directions of that vehicle (south, east, north, and west, respectively).}
\label{tab:5-25Results}
\end{table}

\begin{table}[h]
\centering
\resizebox{\columnwidth}{!}{%
\begin{tabular}{|llllllllllllllllr|}
\hline
\rowcolor[HTML]{FFFFFF} 
\multicolumn{17}{|c|}{\cellcolor[HTML]{FFFFFF}\textbf{\begin{tabular}[c]{@{}c@{}}Highest registered exposedness index on any point between 25 and 75 meters from the camera that is found on the part when the vehicle is in each of the\\ four facings (South East North West).\end{tabular}}} \\ \hline
\rowcolor[HTML]{FFFFFF} 
\multicolumn{1}{|r|}{\cellcolor[HTML]{FFFFFF}\textbf{\begin{tabular}[c]{@{}r@{}}Vehicle \\ Type \\ \& Part\end{tabular}}} &
  \multicolumn{1}{c|}{\cellcolor[HTML]{FFFFFF}\textbf{\begin{tabular}[c]{@{}c@{}}Carryall\\ Car\end{tabular}}} &
  \multicolumn{1}{c|}{\cellcolor[HTML]{FFFFFF}\textbf{\begin{tabular}[c]{@{}c@{}}Double\\ Decker\\ Bus\end{tabular}}} &
  \multicolumn{1}{c|}{\cellcolor[HTML]{FFFFFF}\textbf{\begin{tabular}[c]{@{}c@{}}Four\\ Wheel\\ Truck\end{tabular}}} &
  \multicolumn{1}{c|}{\cellcolor[HTML]{FFFFFF}{\ul \textbf{Minibus}}} &
  \multicolumn{1}{c|}{\cellcolor[HTML]{FFFFFF}\textbf{\begin{tabular}[c]{@{}c@{}}Motor-\\ cycle\end{tabular}}} &
  \multicolumn{1}{c|}{\cellcolor[HTML]{FFFFFF}\textbf{\begin{tabular}[c]{@{}c@{}}Motor\\ Home\end{tabular}}} &
  \multicolumn{1}{c|}{\cellcolor[HTML]{FFFFFF}\textbf{\begin{tabular}[c]{@{}c@{}}Moving\\ Truck\end{tabular}}} &
  \multicolumn{1}{c|}{\cellcolor[HTML]{FFFFFF}\textbf{\begin{tabular}[c]{@{}c@{}}Panel\\ Van\end{tabular}}} &
  \multicolumn{1}{c|}{\cellcolor[HTML]{FFFFFF}\textbf{\begin{tabular}[c]{@{}c@{}}Pickup\\ Truck\end{tabular}}} &
  \multicolumn{1}{c|}{\cellcolor[HTML]{FFFFFF}{\ul \textbf{Sedan}}} &
  \multicolumn{1}{c|}{\cellcolor[HTML]{FFFFFF}\textbf{\begin{tabular}[c]{@{}c@{}}Single\\ Decker\\ Bus\end{tabular}}} &
  \multicolumn{1}{c|}{\cellcolor[HTML]{FFFFFF}\textbf{\begin{tabular}[c]{@{}c@{}}Smart\\ Car\end{tabular}}} &
  \multicolumn{1}{c|}{\cellcolor[HTML]{FFFFFF}\textbf{\begin{tabular}[c]{@{}c@{}}Station\\ Wagon\end{tabular}}} &
  \multicolumn{1}{c|}{\cellcolor[HTML]{FFFFFF}{\ul \textbf{SUV}}} &
  \multicolumn{1}{c|}{\cellcolor[HTML]{FFFFFF}\textbf{\begin{tabular}[c]{@{}c@{}}Topless \\ Conver-\\ tible\end{tabular}}} &
  \multicolumn{1}{c|}{\cellcolor[HTML]{FFFFFF}\textbf{\begin{tabular}[c]{@{}c@{}}Avg.\\ Highest\end{tabular}}} \\ \hline
\multicolumn{1}{|l|}{\cellcolor[HTML]{FFFFFF}\textbf{\begin{tabular}[c]{@{}l@{}}Back \\ Bumper\end{tabular}}} &
  \multicolumn{1}{l|}{\cellcolor[HTML]{FFFFFF}0 0 0 0} &
  \multicolumn{1}{l|}{\cellcolor[HTML]{FFFFFF}0 0 0 0} &
  \multicolumn{1}{l|}{\cellcolor[HTML]{FFC3C3}0 0 0 7} &
  \multicolumn{1}{l|}{\cellcolor[HTML]{FF7A7A}7 0 3 5} &
  \multicolumn{1}{l|}{\cellcolor[HTML]{FFFFFF}0 0 0 0} &
  \multicolumn{1}{l|}{\cellcolor[HTML]{FFCCCD}2 0 0 4} &
  \multicolumn{1}{l|}{\cellcolor[HTML]{FFF0F0}1 0 1 0} &
  \multicolumn{1}{l|}{\cellcolor[HTML]{FFF0F0}0 0 0 2} &
  \multicolumn{1}{l|}{\cellcolor[HTML]{FFC3C3}0 0 0 7} &
  \multicolumn{1}{l|}{\cellcolor[HTML]{FFCCCD}3 0 3 0} &
  \multicolumn{1}{l|}{\cellcolor[HTML]{FFCCCD}1 5 0 0} &
  \multicolumn{1}{l|}{\cellcolor[HTML]{FFF0F0}2 0 0 0} &
  \multicolumn{1}{l|}{\cellcolor[HTML]{FFE8E7}0 0 0 3} &
  \multicolumn{1}{l|}{\cellcolor[HTML]{FFFFFF}0 0 0 0} &
  \multicolumn{1}{l|}{\cellcolor[HTML]{FFE8E7}0 0 0 3} &
  \cellcolor[HTML]{E6F5EE}{\ul \textbf{0.38}} \\ \hline
\rowcolor[HTML]{FFFFFF} 
\multicolumn{1}{|l|}{\cellcolor[HTML]{FFFFFF}\textbf{\begin{tabular}[c]{@{}l@{}}Back \\ Central\\ Light\end{tabular}}} &
  \multicolumn{1}{l|}{\cellcolor[HTML]{FFFFFF}NA} &
  \multicolumn{1}{l|}{\cellcolor[HTML]{FFFFFF}0 0 0 0} &
  \multicolumn{1}{l|}{\cellcolor[HTML]{FFFFFF}NA} &
  \multicolumn{1}{l|}{\cellcolor[HTML]{FFF9F9}1 0 0 0} &
  \multicolumn{1}{l|}{\cellcolor[HTML]{FFFFFF}NA} &
  \multicolumn{1}{l|}{\cellcolor[HTML]{FFFFFF}0 0 0 0} &
  \multicolumn{1}{l|}{\cellcolor[HTML]{FFFFFF}NA} &
  \multicolumn{1}{l|}{\cellcolor[HTML]{FFFFFF}0 0 0 0} &
  \multicolumn{1}{l|}{\cellcolor[HTML]{FFF9F9}1 0 0 0} &
  \multicolumn{1}{l|}{\cellcolor[HTML]{FFFFFF}0 0 0 0} &
  \multicolumn{1}{l|}{\cellcolor[HTML]{FFFFFF}0 0 0 0} &
  \multicolumn{1}{l|}{\cellcolor[HTML]{FFFFFF}0 0 0 0} &
  \multicolumn{1}{l|}{\cellcolor[HTML]{FFDEDE}0 0 3 1} &
  \multicolumn{1}{l|}{\cellcolor[HTML]{FFFFFF}0 0 0 0} &
  \multicolumn{1}{l|}{\cellcolor[HTML]{FFFFFF}NA} &
  \cellcolor[HTML]{FCFEFD}{\ul \textbf{0.08}} \\ \hline
\multicolumn{1}{|l|}{\cellcolor[HTML]{FFFFFF}\textbf{\begin{tabular}[c]{@{}l@{}}Back \\ Fenders\end{tabular}}} &
  \multicolumn{1}{l|}{\cellcolor[HTML]{FFC3C3}0 0 0 7} &
  \multicolumn{1}{l|}{\cellcolor[HTML]{FFB1B1}2 0 3 4} &
  \multicolumn{1}{l|}{\cellcolor[HTML]{FFFFFF}NA} &
  \multicolumn{1}{l|}{\cellcolor[HTML]{FFBABA}4 0 1 3} &
  \multicolumn{1}{l|}{\cellcolor[HTML]{FFE8E7}0 0 3 0} &
  \multicolumn{1}{l|}{\cellcolor[HTML]{FFC3C3}4 0 0 3} &
  \multicolumn{1}{l|}{\cellcolor[HTML]{FFDEDE}1 0 0 3} &
  \multicolumn{1}{l|}{\cellcolor[HTML]{FFC3C3}0 0 5 2} &
  \multicolumn{1}{l|}{\cellcolor[HTML]{FF7171}2 0 7 7} &
  \multicolumn{1}{l|}{\cellcolor[HTML]{FF8C8D}6 0 6 1} &
  \multicolumn{1}{l|}{\cellcolor[HTML]{FF8C8D}1 7 1 4} &
  \multicolumn{1}{l|}{\cellcolor[HTML]{FFCCCD}0 0 5 1} &
  \multicolumn{1}{l|}{\cellcolor[HTML]{FFCCCD}2 0 3 1} &
  \multicolumn{1}{l|}{\cellcolor[HTML]{FFF9F9}0 0 1 0} &
  \multicolumn{1}{l|}{\cellcolor[HTML]{FFD5D5}0 0 3 2} &
  \cellcolor[HTML]{B9E3CF}{\ul \textbf{1}} \\ \hline
\rowcolor[HTML]{FFFFFF} 
\multicolumn{1}{|l|}{\cellcolor[HTML]{FFFFFF}\textbf{\begin{tabular}[c]{@{}l@{}}Back Low \\ Reflector\end{tabular}}} &
  \multicolumn{1}{l|}{\cellcolor[HTML]{FFFFFF}NA} &
  \multicolumn{1}{l|}{\cellcolor[HTML]{FFFFFF}0 0 0 0} &
  \multicolumn{1}{l|}{\cellcolor[HTML]{FFE8E7}0 0 0 3} &
  \multicolumn{1}{l|}{\cellcolor[HTML]{FFFFFF}0 0 0 0} &
  \multicolumn{1}{l|}{\cellcolor[HTML]{FFFFFF}NA} &
  \multicolumn{1}{l|}{\cellcolor[HTML]{FFFFFF}0 0 0 0} &
  \multicolumn{1}{l|}{\cellcolor[HTML]{FFFFFF}NA} &
  \multicolumn{1}{l|}{\cellcolor[HTML]{FFFFFF}0 0 0 0} &
  \multicolumn{1}{l|}{\cellcolor[HTML]{FFFFFF}0 0 0 0} &
  \multicolumn{1}{l|}{\cellcolor[HTML]{FFF0F0}2 0 0 0} &
  \multicolumn{1}{l|}{\cellcolor[HTML]{FFFFFF}0 0 0 0} &
  \multicolumn{1}{l|}{\cellcolor[HTML]{FFFFFF}0 0 0 0} &
  \multicolumn{1}{l|}{\cellcolor[HTML]{FFFFFF}0 0 0 0} &
  \multicolumn{1}{l|}{\cellcolor[HTML]{FFFFFF}0 0 0 0} &
  \multicolumn{1}{l|}{\cellcolor[HTML]{FFFFFF}0 0 0 0} &
  {\ul \textbf{0.03}} \\ \hline
\rowcolor[HTML]{FFFFFF} 
\multicolumn{1}{|l|}{\cellcolor[HTML]{FFFFFF}{\ul \textbf{Back Plate}}} &
  \multicolumn{1}{l|}{\cellcolor[HTML]{FFFFFF}0 0 0 0} &
  \multicolumn{1}{l|}{\cellcolor[HTML]{FFFFFF}0 0 0 0} &
  \multicolumn{1}{l|}{\cellcolor[HTML]{FFFFFF}0 0 0 0} &
  \multicolumn{1}{l|}{\cellcolor[HTML]{FFF0F0}1 0 0 1} &
  \multicolumn{1}{l|}{\cellcolor[HTML]{FFFFFF}0 0 0 0} &
  \multicolumn{1}{l|}{\cellcolor[HTML]{FFE8E7}3 0 0 0} &
  \multicolumn{1}{l|}{\cellcolor[HTML]{FFFFFF}0 0 0 0} &
  \multicolumn{1}{l|}{\cellcolor[HTML]{FFFFFF}0 0 0 0} &
  \multicolumn{1}{l|}{\cellcolor[HTML]{FFFFFF}0 0 0 0} &
  \multicolumn{1}{l|}{\cellcolor[HTML]{FFF9F9}1 0 0 0} &
  \multicolumn{1}{l|}{\cellcolor[HTML]{FFFFFF}0 0 0 0} &
  \multicolumn{1}{l|}{\cellcolor[HTML]{FFFFFF}0 0 0 0} &
  \multicolumn{1}{l|}{\cellcolor[HTML]{FF9F9F}4 0 3 4} &
  \multicolumn{1}{l|}{\cellcolor[HTML]{FFFFFF}0 0 0 0} &
  \multicolumn{1}{l|}{\cellcolor[HTML]{FFDEDE}2 0 1 1} &
  \cellcolor[HTML]{F0F9F5}{\ul \textbf{0.25}} \\ \hline
\multicolumn{1}{|l|}{\cellcolor[HTML]{FFFFFF}\textbf{\begin{tabular}[c]{@{}l@{}}Back\\ Window\\ Rails\end{tabular}}} &
  \multicolumn{1}{l|}{\cellcolor[HTML]{FFD5D5}3 0 2 0} &
  \multicolumn{1}{l|}{\cellcolor[HTML]{FFDEDE}0 0 0 4} &
  \multicolumn{1}{l|}{\cellcolor[HTML]{FFFFFF}NA} &
  \multicolumn{1}{l|}{\cellcolor[HTML]{FFB1B1}2 0 1 6} &
  \multicolumn{1}{l|}{\cellcolor[HTML]{FFFFFF}NA} &
  \multicolumn{1}{l|}{\cellcolor[HTML]{FFFFFF}0 0 0 0} &
  \multicolumn{1}{l|}{\cellcolor[HTML]{FFFFFF}NA} &
  \multicolumn{1}{l|}{\cellcolor[HTML]{FFDEDE}0 0 0 4} &
  \multicolumn{1}{l|}{\cellcolor[HTML]{FFE8E7}1 0 1 1} &
  \multicolumn{1}{l|}{\cellcolor[HTML]{FFE8E7}1 0 2 0} &
  \multicolumn{1}{l|}{\cellcolor[HTML]{FF4444}7 7 3 4} &
  \multicolumn{1}{l|}{\cellcolor[HTML]{FFDEDE}0 0 0 4} &
  \multicolumn{1}{l|}{\cellcolor[HTML]{FF7A7A}7 0 7 1} &
  \multicolumn{1}{l|}{\cellcolor[HTML]{FFFFFF}0 0 0 0} &
  \multicolumn{1}{l|}{\cellcolor[HTML]{FFFFFF}NA} &
  \cellcolor[HTML]{D5EEE2}{\ul \textbf{0.62}} \\ \hline
\multicolumn{1}{|l|}{\cellcolor[HTML]{FFFFFF}\textbf{\begin{tabular}[c]{@{}l@{}}Back\\ Window\end{tabular}}} &
  \multicolumn{1}{l|}{\cellcolor[HTML]{FFFFFF}NA} &
  \multicolumn{1}{l|}{\cellcolor[HTML]{FFC3C3}2 0 0 5} &
  \multicolumn{1}{l|}{\cellcolor[HTML]{FFFFFF}NA} &
  \multicolumn{1}{l|}{\cellcolor[HTML]{FFF0F0}1 0 0 1} &
  \multicolumn{1}{l|}{\cellcolor[HTML]{FFFFFF}NA} &
  \multicolumn{1}{l|}{\cellcolor[HTML]{FFF9F9}1 0 0 0} &
  \multicolumn{1}{l|}{\cellcolor[HTML]{FFFFFF}NA} &
  \multicolumn{1}{l|}{\cellcolor[HTML]{FFFFFF}NA} &
  \multicolumn{1}{l|}{\cellcolor[HTML]{FFE8E7}1 0 1 1} &
  \multicolumn{1}{l|}{\cellcolor[HTML]{FFB1B1}3 0 3 3} &
  \multicolumn{1}{l|}{\cellcolor[HTML]{FFDEDE}0 4 0 0} &
  \multicolumn{1}{l|}{\cellcolor[HTML]{FFFFFF}NA} &
  \multicolumn{1}{l|}{\cellcolor[HTML]{FFD5D5}1 0 1 3} &
  \multicolumn{1}{l|}{\cellcolor[HTML]{FFFFFF}0 0 0 0} &
  \multicolumn{1}{l|}{\cellcolor[HTML]{FFFFFF}NA} &
  \cellcolor[HTML]{F1FAF5}{\ul \textbf{0.23}} \\ \hline
\rowcolor[HTML]{FFFFFF} 
\multicolumn{1}{|l|}{\cellcolor[HTML]{FFFFFF}\textbf{\begin{tabular}[c]{@{}l@{}}Back\\ Window\\ Lower\\ Frame\end{tabular}}} &
  \multicolumn{1}{l|}{\cellcolor[HTML]{FFFFFF}NA} &
  \multicolumn{1}{l|}{\cellcolor[HTML]{FFFFFF}0 0 0 0} &
  \multicolumn{1}{l|}{\cellcolor[HTML]{FFFFFF}NA} &
  \multicolumn{1}{l|}{\cellcolor[HTML]{FFF0F0}1 0 0 1} &
  \multicolumn{1}{l|}{\cellcolor[HTML]{FFFFFF}NA} &
  \multicolumn{1}{l|}{\cellcolor[HTML]{FFFFFF}0 0 0 0} &
  \multicolumn{1}{l|}{\cellcolor[HTML]{FFFFFF}NA} &
  \multicolumn{1}{l|}{\cellcolor[HTML]{FFFFFF}0 0 0 0} &
  \multicolumn{1}{l|}{\cellcolor[HTML]{FFFFFF}0 0 0 0} &
  \multicolumn{1}{l|}{\cellcolor[HTML]{FFF0F0}1 0 1 0} &
  \multicolumn{1}{l|}{\cellcolor[HTML]{FFFFFF}0 0 0 0} &
  \multicolumn{1}{l|}{\cellcolor[HTML]{FFFFFF}NA} &
  \multicolumn{1}{l|}{\cellcolor[HTML]{FFF9F9}0 0 1 0} &
  \multicolumn{1}{l|}{\cellcolor[HTML]{FFFFFF}0 0 0 0} &
  \multicolumn{1}{l|}{\cellcolor[HTML]{FFFFFF}NA} &
  \cellcolor[HTML]{FDFFFE}{\ul \textbf{0.07}} \\ \hline
\multicolumn{1}{|l|}{\cellcolor[HTML]{FFFFFF}\textbf{\begin{tabular}[c]{@{}l@{}}Cowl\\ Cover\end{tabular}}} &
  \multicolumn{1}{l|}{\cellcolor[HTML]{FFFFFF}NA} &
  \multicolumn{1}{l|}{\cellcolor[HTML]{FFC3C3}3 0 4 0} &
  \multicolumn{1}{l|}{\cellcolor[HTML]{FFFFFF}0 0 0 0} &
  \multicolumn{1}{l|}{\cellcolor[HTML]{FFF0F0}2 0 0 0} &
  \multicolumn{1}{l|}{\cellcolor[HTML]{FFFFFF}NA} &
  \multicolumn{1}{l|}{\cellcolor[HTML]{FF9F9F}4 0 7 0} &
  \multicolumn{1}{l|}{\cellcolor[HTML]{FFFFFF}0 0 0 0} &
  \multicolumn{1}{l|}{\cellcolor[HTML]{FFF0F0}2 0 0 0} &
  \multicolumn{1}{l|}{\cellcolor[HTML]{FFF9F9}1 0 0 0} &
  \multicolumn{1}{l|}{\cellcolor[HTML]{FFE8E7}3 0 0 0} &
  \multicolumn{1}{l|}{\cellcolor[HTML]{FFFFFF}NA} &
  \multicolumn{1}{l|}{\cellcolor[HTML]{FFE8E7}0 0 3 0} &
  \multicolumn{1}{l|}{\cellcolor[HTML]{FFDEDE}3 0 1 0} &
  \multicolumn{1}{l|}{\cellcolor[HTML]{FFDEDE}3 0 1 0} &
  \multicolumn{1}{l|}{\cellcolor[HTML]{FFE8E7}2 0 1 0} &
  \cellcolor[HTML]{D2EDE0}{\ul \textbf{0.67}} \\ \hline
\multicolumn{1}{|l|}{\cellcolor[HTML]{FFFFFF}\textbf{\begin{tabular}[c]{@{}l@{}}Front \\ Bumper\end{tabular}}} &
  \multicolumn{1}{l|}{\cellcolor[HTML]{FFFFFF}NA} &
  \multicolumn{1}{l|}{\cellcolor[HTML]{FFBABA}3 0 5 0} &
  \multicolumn{1}{l|}{\cellcolor[HTML]{FFE8E7}3 0 0 0} &
  \multicolumn{1}{l|}{\cellcolor[HTML]{FFCCCD}4 0 2 0} &
  \multicolumn{1}{l|}{\cellcolor[HTML]{FFFFFF}NA} &
  \multicolumn{1}{l|}{\cellcolor[HTML]{FFA9A7}4 0 6 0} &
  \multicolumn{1}{l|}{\cellcolor[HTML]{FFF9F9}1 0 0 0} &
  \multicolumn{1}{l|}{\cellcolor[HTML]{FFC3C3}7 0 0 0} &
  \multicolumn{1}{l|}{\cellcolor[HTML]{FFE8E7}3 0 0 0} &
  \multicolumn{1}{l|}{\cellcolor[HTML]{FFBABA}7 0 1 0} &
  \multicolumn{1}{l|}{\cellcolor[HTML]{FFCCCD}3 0 3 0} &
  \multicolumn{1}{l|}{\cellcolor[HTML]{FFFFFF}0 0 0 0} &
  \multicolumn{1}{l|}{\cellcolor[HTML]{FFBABA}7 0 1 0} &
  \multicolumn{1}{l|}{\cellcolor[HTML]{FFA9A7}7 0 3 0} &
  \multicolumn{1}{l|}{\cellcolor[HTML]{FFB1B1}7 0 2 0} &
  \cellcolor[HTML]{A2DABF}{\ul \textbf{1.32}} \\ \hline
\multicolumn{1}{|l|}{\cellcolor[HTML]{FFFFFF}\textbf{\begin{tabular}[c]{@{}l@{}}Lower \\ Deflector\end{tabular}}} &
  \multicolumn{1}{l|}{\cellcolor[HTML]{FFFFFF}NA} &
  \multicolumn{1}{l|}{\cellcolor[HTML]{FFD5D5}2 0 3 0} &
  \multicolumn{1}{l|}{\cellcolor[HTML]{FFFFFF}0 0 0 0} &
  \multicolumn{1}{l|}{\cellcolor[HTML]{FFFFFF}NA} &
  \multicolumn{1}{l|}{\cellcolor[HTML]{FFFFFF}NA} &
  \multicolumn{1}{l|}{\cellcolor[HTML]{FFFFFF}0 0 0 0} &
  \multicolumn{1}{l|}{\cellcolor[HTML]{FFF9F9}1 0 0 0} &
  \multicolumn{1}{l|}{\cellcolor[HTML]{FFF0F0}2 0 0 0} &
  \multicolumn{1}{l|}{\cellcolor[HTML]{FFF9F9}1 0 0 0} &
  \multicolumn{1}{l|}{\cellcolor[HTML]{FFD5D5}4 0 1 0} &
  \multicolumn{1}{l|}{\cellcolor[HTML]{FFFFFF}NA} &
  \multicolumn{1}{l|}{\cellcolor[HTML]{FFFFFF}0 0 0 0} &
  \multicolumn{1}{l|}{\cellcolor[HTML]{FFFFFF}NA} &
  \multicolumn{1}{l|}{\cellcolor[HTML]{FFE8E7}3 0 0 0} &
  \multicolumn{1}{l|}{\cellcolor[HTML]{FFDEDE}3 0 1 0} &
  \cellcolor[HTML]{E9F6F0}{\ul \textbf{0.35}} \\ \hline
\multicolumn{1}{|l|}{\cellcolor[HTML]{FFFFFF}\textbf{\begin{tabular}[c]{@{}l@{}}Front\\ Doors\end{tabular}}} &
  \multicolumn{1}{l|}{\cellcolor[HTML]{FFFFFF}NA} &
  \multicolumn{1}{l|}{\cellcolor[HTML]{FFFFFF}NA} &
  \multicolumn{1}{l|}{\cellcolor[HTML]{FFD5D5}3 0 2 0} &
  \multicolumn{1}{l|}{\cellcolor[HTML]{FFBABA}3 0 1 4} &
  \multicolumn{1}{l|}{\cellcolor[HTML]{FFFFFF}NA} &
  \multicolumn{1}{l|}{\cellcolor[HTML]{FFF9F9}0 0 1 0} &
  \multicolumn{1}{l|}{\cellcolor[HTML]{FFFFFF}0 0 0 0} &
  \multicolumn{1}{l|}{\cellcolor[HTML]{FFDEDE}4 0 0 0} &
  \multicolumn{1}{l|}{\cellcolor[HTML]{FFCCCD}4 0 1 1} &
  \multicolumn{1}{l|}{\cellcolor[HTML]{FFF0F0}1 0 1 0} &
  \multicolumn{1}{l|}{\cellcolor[HTML]{FFBABA}3 0 2 3} &
  \multicolumn{1}{l|}{\cellcolor[HTML]{FFE8E7}3 0 0 0} &
  \multicolumn{1}{l|}{\cellcolor[HTML]{FFF0F0}1 0 1 0} &
  \multicolumn{1}{l|}{\cellcolor[HTML]{FFCCCD}3 0 2 1} &
  \multicolumn{1}{l|}{\cellcolor[HTML]{FFD5D5}4 0 1 0} &
  \cellcolor[HTML]{D0ECDF}{\ul \textbf{0.68}} \\ \hline
\multicolumn{1}{|l|}{\cellcolor[HTML]{FFFFFF}\textbf{\begin{tabular}[c]{@{}l@{}}Front \\ Fenders\end{tabular}}} &
  \multicolumn{1}{l|}{\cellcolor[HTML]{FFFFFF}0 0 0 0} &
  \multicolumn{1}{l|}{\cellcolor[HTML]{FFC3C3}3 0 4 0} &
  \multicolumn{1}{l|}{\cellcolor[HTML]{FFFFFF}0 0 0 0} &
  \multicolumn{1}{l|}{\cellcolor[HTML]{FFC3C3}6 0 1 0} &
  \multicolumn{1}{l|}{\cellcolor[HTML]{FFDEDE}0 0 4 0} &
  \multicolumn{1}{l|}{\cellcolor[HTML]{FFF9F9}0 0 1 0} &
  \multicolumn{1}{l|}{\cellcolor[HTML]{FFFFFF}0 0 0 0} &
  \multicolumn{1}{l|}{\cellcolor[HTML]{FFDEDE}4 0 0 0} &
  \multicolumn{1}{l|}{\cellcolor[HTML]{FFCCCD}3 0 3 0} &
  \multicolumn{1}{l|}{\cellcolor[HTML]{FFD5D5}4 0 1 0} &
  \multicolumn{1}{l|}{\cellcolor[HTML]{FFF0F0}2 0 0 0} &
  \multicolumn{1}{l|}{\cellcolor[HTML]{FFF0F0}0 0 2 0} &
  \multicolumn{1}{l|}{\cellcolor[HTML]{FFDEDE}3 0 1 0} &
  \multicolumn{1}{l|}{\cellcolor[HTML]{FFC3C3}6 0 1 0} &
  \multicolumn{1}{l|}{\cellcolor[HTML]{FFBABA}7 0 1 0} &
  \cellcolor[HTML]{BDE5D1}{\ul \textbf{0.95}} \\ \hline
\multicolumn{1}{|l|}{\cellcolor[HTML]{FFFFFF}\textbf{\begin{tabular}[c]{@{}l@{}}Front\\ Plate\end{tabular}}} &
  \multicolumn{1}{l|}{\cellcolor[HTML]{FFFFFF}0 0 0 0} &
  \multicolumn{1}{l|}{\cellcolor[HTML]{FFBABA}3 0 5 0} &
  \multicolumn{1}{l|}{\cellcolor[HTML]{FFFFFF}0 0 0 0} &
  \multicolumn{1}{l|}{\cellcolor[HTML]{FFCCCD}4 0 2 0} &
  \multicolumn{1}{l|}{\cellcolor[HTML]{FFFFFF}NA} &
  \multicolumn{1}{l|}{\cellcolor[HTML]{FFA9A7}7 0 3 0} &
  \multicolumn{1}{l|}{\cellcolor[HTML]{FFF0F0}2 0 0 0} &
  \multicolumn{1}{l|}{\cellcolor[HTML]{FFFFFF}0 0 0 0} &
  \multicolumn{1}{l|}{\cellcolor[HTML]{FFFFFF}0 0 0 0} &
  \multicolumn{1}{l|}{\cellcolor[HTML]{FFD5D5}4 0 1 0} &
  \multicolumn{1}{l|}{\cellcolor[HTML]{FFDEDE}3 0 1 0} &
  \multicolumn{1}{l|}{\cellcolor[HTML]{FFFFFF}0 0 0 0} &
  \multicolumn{1}{l|}{\cellcolor[HTML]{FFDEDE}4 0 0 0} &
  \multicolumn{1}{l|}{\cellcolor[HTML]{FFB1B1}6 0 3 0} &
  \multicolumn{1}{l|}{\cellcolor[HTML]{FFDEDE}3 0 1 0} &
  \cellcolor[HTML]{C3E7D5}{\ul \textbf{0.87}} \\ \hline
\multicolumn{1}{|l|}{\cellcolor[HTML]{FFFFFF}\textbf{\begin{tabular}[c]{@{}l@{}}Front \\ Wheels\end{tabular}}} &
  \multicolumn{1}{l|}{\cellcolor[HTML]{FFF0F0}0 0 2 0} &
  \multicolumn{1}{l|}{\cellcolor[HTML]{FF9F9F}4 0 3 4} &
  \multicolumn{1}{l|}{\cellcolor[HTML]{FFCCCD}3 0 3 0} &
  \multicolumn{1}{l|}{\cellcolor[HTML]{FFA9A7}4 0 2 4} &
  \multicolumn{1}{l|}{\cellcolor[HTML]{FFD5D5}0 0 5 0} &
  \multicolumn{1}{l|}{\cellcolor[HTML]{FF8C8D}5 0 7 1} &
  \multicolumn{1}{l|}{\cellcolor[HTML]{FFE8E7}3 0 0 0} &
  \multicolumn{1}{l|}{\cellcolor[HTML]{FFDEDE}4 0 0 0} &
  \multicolumn{1}{l|}{\cellcolor[HTML]{FFDEDE}4 0 0 0} &
  \multicolumn{1}{l|}{\cellcolor[HTML]{FFDEDE}3 0 1 0} &
  \multicolumn{1}{l|}{\cellcolor[HTML]{FFC3C3}3 0 3 1} &
  \multicolumn{1}{l|}{\cellcolor[HTML]{FFC3C3}0 0 7 0} &
  \multicolumn{1}{l|}{\cellcolor[HTML]{FFBABA}7 0 1 0} &
  \multicolumn{1}{l|}{\cellcolor[HTML]{FFBABA}6 0 2 0} &
  \multicolumn{1}{l|}{\cellcolor[HTML]{FFC3C3}5 0 2 0} &
  \cellcolor[HTML]{96D5B6}{\ul \textbf{1.48}} \\ \hline
\multicolumn{1}{|l|}{\cellcolor[HTML]{FFFFFF}\textbf{\begin{tabular}[c]{@{}l@{}}Front \\ Windows\end{tabular}}} &
  \multicolumn{1}{l|}{\cellcolor[HTML]{FFFFFF}NA} &
  \multicolumn{1}{l|}{\cellcolor[HTML]{FFE8E7}3 0 0 0} &
  \multicolumn{1}{l|}{\cellcolor[HTML]{FFDEDE}3 0 1 0} &
  \multicolumn{1}{l|}{\cellcolor[HTML]{FFBABA}1 0 3 4} &
  \multicolumn{1}{l|}{\cellcolor[HTML]{FFFFFF}NA} &
  \multicolumn{1}{l|}{\cellcolor[HTML]{FFA9A7}5 0 3 2} &
  \multicolumn{1}{l|}{\cellcolor[HTML]{FFC3C3}0 0 2 5} &
  \multicolumn{1}{l|}{\cellcolor[HTML]{FFF0F0}2 0 0 0} &
  \multicolumn{1}{l|}{\cellcolor[HTML]{FFCCCD}3 0 2 1} &
  \multicolumn{1}{l|}{\cellcolor[HTML]{FFDEDE}3 0 1 0} &
  \multicolumn{1}{l|}{\cellcolor[HTML]{FFF9F9}0 0 0 1} &
  \multicolumn{1}{l|}{\cellcolor[HTML]{FFF0F0}2 0 0 0} &
  \multicolumn{1}{l|}{\cellcolor[HTML]{FFF0F0}1 0 1 0} &
  \multicolumn{1}{l|}{\cellcolor[HTML]{FFE8E7}2 0 1 0} &
  \multicolumn{1}{l|}{\cellcolor[HTML]{FFFFFF}NA} &
  \cellcolor[HTML]{D3EDE0}{\ul \textbf{0.65}} \\ \hline
\multicolumn{1}{|l|}{\cellcolor[HTML]{FFFFFF}{\ul \textbf{Grill}}} &
  \multicolumn{1}{l|}{\cellcolor[HTML]{FFFFFF}NA} &
  \multicolumn{1}{l|}{\cellcolor[HTML]{FFBABA}3 0 5 0} &
  \multicolumn{1}{l|}{\cellcolor[HTML]{FFF0F0}2 0 0 0} &
  \multicolumn{1}{l|}{\cellcolor[HTML]{FFCCCD}4 0 2 0} &
  \multicolumn{1}{l|}{\cellcolor[HTML]{FFFFFF}NA} &
  \multicolumn{1}{l|}{\cellcolor[HTML]{FFB1B1}5 0 4 0} &
  \multicolumn{1}{l|}{\cellcolor[HTML]{FFE8E7}3 0 0 0} &
  \multicolumn{1}{l|}{\cellcolor[HTML]{FFD5D5}5 0 0 0} &
  \multicolumn{1}{l|}{\cellcolor[HTML]{FFE8E7}3 0 0 0} &
  \multicolumn{1}{l|}{\cellcolor[HTML]{FFBABA}7 0 1 0} &
  \multicolumn{1}{l|}{\cellcolor[HTML]{FFDEDE}2 0 2 0} &
  \multicolumn{1}{l|}{\cellcolor[HTML]{FFFFFF}0 0 0 0} &
  \multicolumn{1}{l|}{\cellcolor[HTML]{FFB1B1}7 0 2 0} &
  \multicolumn{1}{l|}{\cellcolor[HTML]{FFB1B1}7 0 2 0} &
  \multicolumn{1}{l|}{\cellcolor[HTML]{FFF9F9}0 0 1 0} &
  \cellcolor[HTML]{B1E0C9}{\ul \textbf{1.12}} \\ \hline
\multicolumn{1}{|l|}{\cellcolor[HTML]{FFFFFF}{\ul \textbf{Headlights}}} &
  \multicolumn{1}{l|}{\cellcolor[HTML]{FFFFFF}0 0 0 0} &
  \multicolumn{1}{l|}{\cellcolor[HTML]{FFC3C3}3 0 4 0} &
  \multicolumn{1}{l|}{\cellcolor[HTML]{FFE8E7}3 0 0 0} &
  \multicolumn{1}{l|}{\cellcolor[HTML]{FFBABA}7 0 1 0} &
  \multicolumn{1}{l|}{\cellcolor[HTML]{FFC3C3}0 0 7 0} &
  \multicolumn{1}{l|}{\cellcolor[HTML]{FF9F9F}4 0 7 0} &
  \multicolumn{1}{l|}{\cellcolor[HTML]{FFE8E7}3 0 0 0} &
  \multicolumn{1}{l|}{\cellcolor[HTML]{FFDEDE}4 0 0 0} &
  \multicolumn{1}{l|}{\cellcolor[HTML]{FFE8E7}3 0 0 0} &
  \multicolumn{1}{l|}{\cellcolor[HTML]{FFD5D5}4 0 1 0} &
  \multicolumn{1}{l|}{\cellcolor[HTML]{FFD5D5}2 0 3 0} &
  \multicolumn{1}{l|}{\cellcolor[HTML]{FFD5D5}0 0 5 0} &
  \multicolumn{1}{l|}{\cellcolor[HTML]{FFB1B1}7 0 2 0} &
  \multicolumn{1}{l|}{\cellcolor[HTML]{FFA9A7}7 0 3 0} &
  \multicolumn{1}{l|}{\cellcolor[HTML]{FFCCCD}4 0 2 0} &
  \cellcolor[HTML]{9AD6B9}{\ul \textbf{1.43}} \\ \hline
\multicolumn{1}{|l|}{\cellcolor[HTML]{FFFFFF}{\ul \textbf{Hood}}} &
  \multicolumn{1}{l|}{\cellcolor[HTML]{FFFFFF}0 0 0 0} &
  \multicolumn{1}{l|}{\cellcolor[HTML]{FFFFFF}NA} &
  \multicolumn{1}{l|}{\cellcolor[HTML]{FFF9F9}1 0 0 0} &
  \multicolumn{1}{l|}{\cellcolor[HTML]{FFDEDE}3 0 1 0} &
  \multicolumn{1}{l|}{\cellcolor[HTML]{FFFFFF}NA} &
  \multicolumn{1}{l|}{\cellcolor[HTML]{FF9F9F}4 0 7 0} &
  \multicolumn{1}{l|}{\cellcolor[HTML]{FFF9F9}1 0 0 0} &
  \multicolumn{1}{l|}{\cellcolor[HTML]{FFDEDE}4 0 0 0} &
  \multicolumn{1}{l|}{\cellcolor[HTML]{FFF9F9}1 0 0 0} &
  \multicolumn{1}{l|}{\cellcolor[HTML]{FFD5D5}4 0 1 0} &
  \multicolumn{1}{l|}{\cellcolor[HTML]{FFFFFF}NA} &
  \multicolumn{1}{l|}{\cellcolor[HTML]{FFFFFF}0 0 0 0} &
  \multicolumn{1}{l|}{\cellcolor[HTML]{FFE8E7}3 0 0 0} &
  \multicolumn{1}{l|}{\cellcolor[HTML]{FFDEDE}3 0 1 0} &
  \multicolumn{1}{l|}{\cellcolor[HTML]{FFDEDE}3 0 1 0} &
  \cellcolor[HTML]{D4EEE1}{\ul \textbf{0.63}} \\ \hline
\rowcolor[HTML]{FFFFFF} 
\multicolumn{1}{|l|}{\cellcolor[HTML]{FFFFFF}{\ul \textbf{Rear Doors}}} &
  \multicolumn{1}{l|}{\cellcolor[HTML]{FFFFFF}NA} &
  \multicolumn{1}{l|}{\cellcolor[HTML]{FFFFFF}NA} &
  \multicolumn{1}{l|}{\cellcolor[HTML]{FFFFFF}NA} &
  \multicolumn{1}{l|}{\cellcolor[HTML]{FFDEDE}3 0 0 1} &
  \multicolumn{1}{l|}{\cellcolor[HTML]{FFFFFF}NA} &
  \multicolumn{1}{l|}{\cellcolor[HTML]{FFFFFF}NA} &
  \multicolumn{1}{l|}{\cellcolor[HTML]{FFFFFF}NA} &
  \multicolumn{1}{l|}{\cellcolor[HTML]{FFC3C3}3 0 4 0} &
  \multicolumn{1}{l|}{\cellcolor[HTML]{FFE8E7}3 0 0 0} &
  \multicolumn{1}{l|}{\cellcolor[HTML]{FFDEDE}4 0 0 0} &
  \multicolumn{1}{l|}{\cellcolor[HTML]{FFFFFF}NA} &
  \multicolumn{1}{l|}{\cellcolor[HTML]{FFFFFF}NA} &
  \multicolumn{1}{l|}{\cellcolor[HTML]{FFCCCD}2 0 3 1} &
  \multicolumn{1}{l|}{\cellcolor[HTML]{FFDEDE}2 0 1 1} &
  \multicolumn{1}{l|}{\cellcolor[HTML]{FFFFFF}NA} &
  \cellcolor[HTML]{E4F4EC}{\ul \textbf{0.42}} \\ \hline
\multicolumn{1}{|l|}{\cellcolor[HTML]{FFFFFF}\textbf{\begin{tabular}[c]{@{}l@{}}Rear\\ Wheels\end{tabular}}} &
  \multicolumn{1}{l|}{\cellcolor[HTML]{FFDEDE}2 0 2 0} &
  \multicolumn{1}{l|}{\cellcolor[HTML]{FF6868}7 0 3 7} &
  \multicolumn{1}{l|}{\cellcolor[HTML]{FF6868}7 0 6 4} &
  \multicolumn{1}{l|}{\cellcolor[HTML]{FF4444}7 0 7 7} &
  \multicolumn{1}{l|}{\cellcolor[HTML]{FFD5D5}3 0 2 0} &
  \multicolumn{1}{l|}{\cellcolor[HTML]{FFBABA}4 0 2 2} &
  \multicolumn{1}{l|}{\cellcolor[HTML]{FF605F}7 0 4 7} &
  \multicolumn{1}{l|}{\cellcolor[HTML]{FF5756}7 0 5 7} &
  \multicolumn{1}{l|}{\cellcolor[HTML]{FF6868}5 0 5 7} &
  \multicolumn{1}{l|}{\cellcolor[HTML]{FF8C8D}3 0 6 4} &
  \multicolumn{1}{l|}{\cellcolor[HTML]{FFB1B1}3 0 3 3} &
  \multicolumn{1}{l|}{\cellcolor[HTML]{FF7A7A}7 0 2 6} &
  \multicolumn{1}{l|}{\cellcolor[HTML]{FF8384}3 0 4 7} &
  \multicolumn{1}{l|}{\cellcolor[HTML]{FF4D4D}6 0 7 7} &
  \multicolumn{1}{l|}{\cellcolor[HTML]{FF6868}6 0 6 5} &
  \cellcolor[HTML]{57BB8A}{\ul \textbf{2.35}} \\ \hline
\multicolumn{1}{|l|}{\cellcolor[HTML]{FFFFFF}\textbf{\begin{tabular}[c]{@{}l@{}}Rear \\ Windows\end{tabular}}} &
  \multicolumn{1}{l|}{\cellcolor[HTML]{FFFFFF}NA} &
  \multicolumn{1}{l|}{\cellcolor[HTML]{FFCCCD}2 0 0 4} &
  \multicolumn{1}{l|}{\cellcolor[HTML]{FFFFFF}NA} &
  \multicolumn{1}{l|}{\cellcolor[HTML]{FF9596}5 0 3 4} &
  \multicolumn{1}{l|}{\cellcolor[HTML]{FFFFFF}NA} &
  \multicolumn{1}{l|}{\cellcolor[HTML]{FFC3C3}3 0 1 3} &
  \multicolumn{1}{l|}{\cellcolor[HTML]{FFFFFF}NA} &
  \multicolumn{1}{l|}{\cellcolor[HTML]{FFFFFF}NA} &
  \multicolumn{1}{l|}{\cellcolor[HTML]{FFF0F0}1 0 1 0} &
  \multicolumn{1}{l|}{\cellcolor[HTML]{FFA9A7}2 0 1 7} &
  \multicolumn{1}{l|}{\cellcolor[HTML]{FF3231}2 7 7 7} &
  \multicolumn{1}{l|}{\cellcolor[HTML]{FFFFFF}NA} &
  \multicolumn{1}{l|}{\cellcolor[HTML]{FFBABA}1 0 4 3} &
  \multicolumn{1}{l|}{\cellcolor[HTML]{FFF9F9}0 0 1 0} &
  \multicolumn{1}{l|}{\cellcolor[HTML]{FFFFFF}NA} &
  \cellcolor[HTML]{D9F0E5}{\ul \textbf{0.57}} \\ \hline
\multicolumn{1}{|l|}{\cellcolor[HTML]{FFFFFF}\textbf{\begin{tabular}[c]{@{}l@{}}Rocker\\ Panel\end{tabular}}} &
  \multicolumn{1}{l|}{\cellcolor[HTML]{FFFFFF}0 0 0 0} &
  \multicolumn{1}{l|}{\cellcolor[HTML]{FFFFFF}NA} &
  \multicolumn{1}{l|}{\cellcolor[HTML]{FFC3C3}7 0 0 0} &
  \multicolumn{1}{l|}{\cellcolor[HTML]{FFFFFF}0 0 0 0} &
  \multicolumn{1}{l|}{\cellcolor[HTML]{FFFFFF}NA} &
  \multicolumn{1}{l|}{\cellcolor[HTML]{FFF9F9}0 0 0 1} &
  \multicolumn{1}{l|}{\cellcolor[HTML]{FFDEDE}0 0 0 4} &
  \multicolumn{1}{l|}{\cellcolor[HTML]{FFDEDE}0 0 4 0} &
  \multicolumn{1}{l|}{\cellcolor[HTML]{FFCCCD}3 0 1 2} &
  \multicolumn{1}{l|}{\cellcolor[HTML]{FFF0F0}2 0 0 0} &
  \multicolumn{1}{l|}{\cellcolor[HTML]{FFFFFF}NA} &
  \multicolumn{1}{l|}{\cellcolor[HTML]{FFF9F9}1 0 0 0} &
  \multicolumn{1}{l|}{\cellcolor[HTML]{FFF9F9}0 0 0 1} &
  \multicolumn{1}{l|}{\cellcolor[HTML]{FFFFFF}0 0 0 0} &
  \multicolumn{1}{l|}{\cellcolor[HTML]{FFFFFF}0 0 0 0} &
  \cellcolor[HTML]{ECF8F2}{\ul \textbf{0.3}} \\ \hline
\multicolumn{1}{|l|}{\cellcolor[HTML]{FFFFFF}{\ul \textbf{Roof}}} &
  \multicolumn{1}{l|}{\cellcolor[HTML]{FFD5D5}2 0 3 0} &
  \multicolumn{1}{l|}{\cellcolor[HTML]{FFB1B1}3 0 3 3} &
  \multicolumn{1}{l|}{\cellcolor[HTML]{FFFFFF}0 0 0 0} &
  \multicolumn{1}{l|}{\cellcolor[HTML]{FFC3C3}2 0 2 3} &
  \multicolumn{1}{l|}{\cellcolor[HTML]{FFFFFF}NA} &
  \multicolumn{1}{l|}{\cellcolor[HTML]{FF9596}7 0 4 1} &
  \multicolumn{1}{l|}{\cellcolor[HTML]{FFDEDE}3 0 1 0} &
  \multicolumn{1}{l|}{\cellcolor[HTML]{FFDEDE}1 0 1 2} &
  \multicolumn{1}{l|}{\cellcolor[HTML]{FFE8E7}1 0 1 1} &
  \multicolumn{1}{l|}{\cellcolor[HTML]{FFDEDE}2 0 0 2} &
  \multicolumn{1}{l|}{\cellcolor[HTML]{FF8C8D}1 7 2 3} &
  \multicolumn{1}{l|}{\cellcolor[HTML]{FFCCCD}2 0 4 0} &
  \multicolumn{1}{l|}{\cellcolor[HTML]{FF9F9F}2 0 6 3} &
  \multicolumn{1}{l|}{\cellcolor[HTML]{FFB1B1}3 0 3 3} &
  \multicolumn{1}{l|}{\cellcolor[HTML]{FFFFFF}NA} &
  \cellcolor[HTML]{BBE4D0}{\ul \textbf{0.98}} \\ \hline
\multicolumn{1}{|l|}{\cellcolor[HTML]{FFFFFF}\textbf{\begin{tabular}[c]{@{}l@{}}Side\\ Mirrors\end{tabular}}} &
  \multicolumn{1}{l|}{\cellcolor[HTML]{FFFFFF}NA} &
  \multicolumn{1}{l|}{\cellcolor[HTML]{FFBABA}3 0 4 1} &
  \multicolumn{1}{l|}{\cellcolor[HTML]{FFE8E7}3 0 0 0} &
  \multicolumn{1}{l|}{\cellcolor[HTML]{FFBABA}3 0 2 3} &
  \multicolumn{1}{l|}{\cellcolor[HTML]{FFFFFF}0 0 0 0} &
  \multicolumn{1}{l|}{\cellcolor[HTML]{FF8C8D}5 0 7 1} &
  \multicolumn{1}{l|}{\cellcolor[HTML]{FFE8E7}3 0 0 0} &
  \multicolumn{1}{l|}{\cellcolor[HTML]{FFCCCD}6 0 0 0} &
  \multicolumn{1}{l|}{\cellcolor[HTML]{FF9596}6 0 3 3} &
  \multicolumn{1}{l|}{\cellcolor[HTML]{FFF9F9}1 0 0 0} &
  \multicolumn{1}{l|}{\cellcolor[HTML]{FFD5D5}1 0 4 0} &
  \multicolumn{1}{l|}{\cellcolor[HTML]{FFC3C3}0 0 7 0} &
  \multicolumn{1}{l|}{\cellcolor[HTML]{FFF0F0}2 0 0 0} &
  \multicolumn{1}{l|}{\cellcolor[HTML]{FFBABA}6 0 2 0} &
  \multicolumn{1}{l|}{\cellcolor[HTML]{FFCCCD}4 0 2 0} &
  \cellcolor[HTML]{A8DCC3}{\ul \textbf{1.23}} \\ \hline
\rowcolor[HTML]{FFFFFF} 
\multicolumn{1}{|l|}{\cellcolor[HTML]{FFFFFF}{\ul \textbf{Tail Lights}}} &
  \multicolumn{1}{l|}{\cellcolor[HTML]{FFFFFF}0 0 0 0} &
  \multicolumn{1}{l|}{\cellcolor[HTML]{FFFFFF}0 0 0 0} &
  \multicolumn{1}{l|}{\cellcolor[HTML]{FFFFFF}0 0 0 0} &
  \multicolumn{1}{l|}{\cellcolor[HTML]{FFF0F0}2 0 0 0} &
  \multicolumn{1}{l|}{\cellcolor[HTML]{FFFFFF}0 0 0 0} &
  \multicolumn{1}{l|}{\cellcolor[HTML]{FFFFFF}0 0 0 0} &
  \multicolumn{1}{l|}{\cellcolor[HTML]{FFFFFF}0 0 0 0} &
  \multicolumn{1}{l|}{\cellcolor[HTML]{FFFFFF}0 0 0 0} &
  \multicolumn{1}{l|}{\cellcolor[HTML]{FFFFFF}0 0 0 0} &
  \multicolumn{1}{l|}{\cellcolor[HTML]{FFCCCD}3 0 3 0} &
  \multicolumn{1}{l|}{\cellcolor[HTML]{FFFFFF}0 0 0 0} &
  \multicolumn{1}{l|}{\cellcolor[HTML]{FFFFFF}0 0 0 0} &
  \multicolumn{1}{l|}{\cellcolor[HTML]{FFFFFF}0 0 0 0} &
  \multicolumn{1}{l|}{\cellcolor[HTML]{FFFFFF}0 0 0 0} &
  \multicolumn{1}{l|}{\cellcolor[HTML]{FFF0F0}0 0 0 2} &
  \cellcolor[HTML]{F8FDFA}{\ul \textbf{0.13}} \\ \hline
\rowcolor[HTML]{FFFFFF} 
\multicolumn{1}{|l|}{\cellcolor[HTML]{FFFFFF}{\ul \textbf{Trunk}}} &
  \multicolumn{1}{l|}{\cellcolor[HTML]{FFFFFF}NA} &
  \multicolumn{1}{l|}{\cellcolor[HTML]{FFFFFF}NA} &
  \multicolumn{1}{l|}{\cellcolor[HTML]{FFFFFF}0 0 0 0} &
  \multicolumn{1}{l|}{\cellcolor[HTML]{FFBABA}3 0 0 5} &
  \multicolumn{1}{l|}{\cellcolor[HTML]{FFFFFF}NA} &
  \multicolumn{1}{l|}{\cellcolor[HTML]{FFFFFF}0 0 0 0} &
  \multicolumn{1}{l|}{\cellcolor[HTML]{FFE8E7}3 0 0 0} &
  \multicolumn{1}{l|}{\cellcolor[HTML]{FFFFFF}0 0 0 0} &
  \multicolumn{1}{l|}{\cellcolor[HTML]{FFFFFF}NA} &
  \multicolumn{1}{l|}{\cellcolor[HTML]{FFD5D5}1 0 1 3} &
  \multicolumn{1}{l|}{\cellcolor[HTML]{FFFFFF}NA} &
  \multicolumn{1}{l|}{\cellcolor[HTML]{FFF9F9}1 0 0 0} &
  \multicolumn{1}{l|}{\cellcolor[HTML]{FFE8E7}1 0 1 1} &
  \multicolumn{1}{l|}{\cellcolor[HTML]{FFFFFF}0 0 0 0} &
  \multicolumn{1}{l|}{\cellcolor[HTML]{FFD5D5}3 0 1 1} &
  \cellcolor[HTML]{F0F9F5}{\ul \textbf{0.25}} \\ \hline
\multicolumn{1}{|l|}{\cellcolor[HTML]{FFFFFF}{\ul \textbf{Windshield}}} &
  \multicolumn{1}{l|}{\cellcolor[HTML]{FFC3C3}0 0 7 0} &
  \multicolumn{1}{l|}{\cellcolor[HTML]{FFA9A7}3 0 7 0} &
  \multicolumn{1}{l|}{\cellcolor[HTML]{FFF9F9}1 0 0 0} &
  \multicolumn{1}{l|}{\cellcolor[HTML]{FFCCCD}4 0 2 0} &
  \multicolumn{1}{l|}{\cellcolor[HTML]{FFFFFF}NA} &
  \multicolumn{1}{l|}{\cellcolor[HTML]{FF9F9F}4 0 7 0} &
  \multicolumn{1}{l|}{\cellcolor[HTML]{FFD5D5}3 0 2 0} &
  \multicolumn{1}{l|}{\cellcolor[HTML]{FFDEDE}4 0 0 0} &
  \multicolumn{1}{l|}{\cellcolor[HTML]{FFDEDE}3 0 1 0} &
  \multicolumn{1}{l|}{\cellcolor[HTML]{FFDEDE}3 0 1 0} &
  \multicolumn{1}{l|}{\cellcolor[HTML]{FFCCCD}3 0 3 0} &
  \multicolumn{1}{l|}{\cellcolor[HTML]{FFE8E7}0 0 3 0} &
  \multicolumn{1}{l|}{\cellcolor[HTML]{FFD5D5}3 0 2 0} &
  \multicolumn{1}{l|}{\cellcolor[HTML]{FFC3C3}6 0 1 0} &
  \multicolumn{1}{l|}{\cellcolor[HTML]{FFDEDE}3 0 1 0} &
  \cellcolor[HTML]{A5DBC0}{\ul \textbf{1.28}} \\ \hline
\multicolumn{1}{|l|}{\cellcolor[HTML]{FFFFFF}\textbf{\begin{tabular}[c]{@{}l@{}}Windshield \\ Rails\end{tabular}}} &
  \multicolumn{1}{l|}{\cellcolor[HTML]{FFD5D5}0 0 5 0} &
  \multicolumn{1}{l|}{\cellcolor[HTML]{FFFFFF}0 0 0 0} &
  \multicolumn{1}{l|}{\cellcolor[HTML]{FFFFFF}0 0 0 0} &
  \multicolumn{1}{l|}{\cellcolor[HTML]{FFCCCD}4 0 2 0} &
  \multicolumn{1}{l|}{\cellcolor[HTML]{FFFFFF}NA} &
  \multicolumn{1}{l|}{\cellcolor[HTML]{FFF0F0}1 0 0 1} &
  \multicolumn{1}{l|}{\cellcolor[HTML]{FFF9F9}1 0 0 0} &
  \multicolumn{1}{l|}{\cellcolor[HTML]{FFF0F0}2 0 0 0} &
  \multicolumn{1}{l|}{\cellcolor[HTML]{FFD5D5}3 0 2 0} &
  \multicolumn{1}{l|}{\cellcolor[HTML]{FFDEDE}3 0 1 0} &
  \multicolumn{1}{l|}{\cellcolor[HTML]{FFDEDE}3 0 1 0} &
  \multicolumn{1}{l|}{\cellcolor[HTML]{FFE8E7}0 0 3 0} &
  \multicolumn{1}{l|}{\cellcolor[HTML]{FFDEDE}3 0 1 0} &
  \multicolumn{1}{l|}{\cellcolor[HTML]{FFD5D5}4 0 1 0} &
  \multicolumn{1}{l|}{\cellcolor[HTML]{FFDEDE}3 0 1 0} &
  \cellcolor[HTML]{CDEBDC}{\ul \textbf{0.73}} \\ \hline
\end{tabular}%
}
\caption{The highest \textit{Exposedness Score} found on any point of each part, for each vehicle, for points between 25\si{m} and 75\si{m} away from the camera, and for each of the four driving directions of that vehicle (south, east, north, and west, respectively).}
\label{tab:25-75Results}
\end{table}

\section{IDs used during data generation of this study}\label{secA2}
Each code is made up of 9 parts, divided by vertical lines ($|$). For example, the following code:

\begin{lstlisting}[breaklines=true]
    SMALL,US|Carryall|S_1,S_2,S_3|1|Carryall|S_1,S_2,S_3,S_4|A-1,2;B-1,2,3|0|5
\end{lstlisting}

Can be divided into nine segments:
\begin{enumerate}
    \item \textbf{Setup ID: }Determines the positions used (divided by vehizle size) and the driving direction.
    \item \textbf{Target Vehicle Model: }Determines the 3D model of the \textit{Target Vehicle} used for the simulation, based on the type.
    \item \textbf{Target Vehicle Positions: }Determines the positions that the \textit{Target Vehicle} can be in, prefixed by an indication of the vehicle size (S, M, L, or XL).
    \item \textbf{Number of Vehicles in Simulation: }Determines the number of vehicles for this \textit{Simulation Run}, including the \textit{Target Vehicle} and any \textit{Blocking Vehicles}. This value should never be lower than 1 or higher than 8.
    \item \textbf{Blocking Vehicle(s) Model: }Determines the 3D model of the \textit{Blocking Vehicle} used for the simulation, based on their type. The available options for this element are the same as for the Target Vehicle Model.
    \item \textbf{Blocking Vehicle(s) Positions: }Determines the positions that the \textit{Blocking Vehicle(s)} can be in, prefixed by an indication of the vehicle size (S, M, L, or XL).
    \item \textbf{Camera Positions and Directions: }The \textit{Camera Positions} and \textit{Camera Directions} for this \textit{Simulation Run}. During the simulation, the program will take \textit{Data Captures} at each \textit{Camera Position} and \textit{Camera Direction}, at five different heights. The structure of each specified position and direction is "[Camera Position 1]-[Camera Direction 1],[Camera Direction 2];[Camera Position 2]-[Camera Direction 3]...", separated by semicolons (;).
    \item \textbf{Minimum Recording Distance:} The minimum distance from the camera before a point is recorded in a \textit{Data Capture}. Points closer to the camera will not be recorded. The minimum number is 0, the maximum number is the Maximum Recording Distance.
    \item \textbf{Maximum Recording Distance:} The maximum distance from the camera for a point to be recorded in a \textit{Data Capture}. Points farther to the camera than this distance will not be recorded. The minimum value is the Minimum Recording Distance.
\end{enumerate}

The following is a list of all the IDs used in the generation of data for this study:







\pagebreak
\end{appendices}

\bibliography{sn-bibliography}

\end{document}